\documentstyle[12pt,epsf]{report}
\begin{document}
\baselineskip=24pt

\thispagestyle{empty}

\begin{center}

{\LARGE \bf 
Universality in the Vibrational Spectra of Amorphous Systems}

\vskip 4cm

{\large \sl A thesis submitted in partial fulfillment\\ of the requirements
for the degree of \\ Doctor of Philosophy.}

\vskip 2cm

{\Large \bf Gurpreet Singh Matharoo}\\

\vskip 2.5cm


\begin{figure}[htp]
\epsfxsize=1in
\centerline{\epsfbox{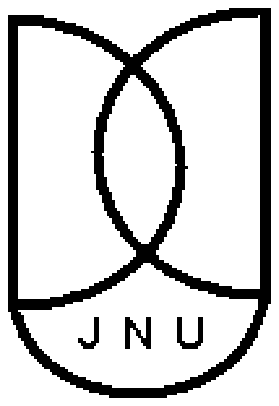}}
\label{jnulogo}
\end{figure}

{\large \bf School of Physical Sciences\\Jawaharlal Nehru University\\
New Delhi - 110 067 \\

\vspace{0.5in}

September 2005}
\end{center}

\newpage

\pagenumbering{roman}
\vspace{0.5in}
\begin{center}
{\huge \bf {Dedication\ldots}}
\end{center}

\begin{center}
\vspace{3.3in}  
{\large \sl This thesis is dedicated to the memory of \\
       my beloved father\\}
{\large \bf Mr. Shivcharan Singh Matharoo.}
\end{center}

\newpage

\begin{center}
{\large \bf DECLARATION}
\end{center}

\hrule

\vskip 2cm

I hereby declare that the work reported in this thesis is entirely
original and has been carried out by me in the School of Physical Sciences,
Jawaharlal Nehru University, New Delhi. The work has been supervised by
Dr. Subir Kumar Sarkar.  
I further declare that it has not formed the basis for the award of any
degree, diploma, associateship or similar title of any university or
institution.

\vskip 1cm

\noindent September, 2005 \hfill {\bf Gurpreet Singh Matharoo}\\

\vskip 1cm

\noindent {\bf Prof. H. B. Bohidar} \hfill {\bf Dr. Subir Kumar Sarkar}\\
Dean \hfill Thesis Supervisor\\
%
\noindent School of Physical Sciences       \hfill{ School of Physical Sciences}\\     
Jawaharlal Nehru University                  \hfill{Jawaharlal Nehru University}\\
New Delhi - 110067                             \hfill{New Delhi - 110067}

\newpage

\begin{center}
{\Huge\bf Acknowledgements}
\end{center}

It would not have been possible for me without my advisor Dr. Subir
Kumar Sarkar to complete this thesis. He not only taught me Physics
and the art of computer programming but also to do things sincerely
and in a truthful manner. I still have a very long way to
go and I do hope that he supports me in the same manner in the 
years to come.

I had the privilege to work with Prof. Akhilesh Pandey in the first
part of the thesis. He not only helped me with theoretical ideas but also
corrected me at many places in my computer programs.

I would like to thank Prof. Shankar P. Das for his concern
and valuable suggestions and discussions from time to time. I have
benefitted from them on many occasions.

Prof. Sanjay Puri has been a combination of teacher and a senior friend
to me. I have not only benefited from him in Statistical Mechanics 
but also as a player on the cricket field.

My teachers at Delhi University, namely, Prof. S. K.Muthu, Prof. K. L. Baluja 
and Dr. A. K. Kavathekar have been a constant source of encouragement to me
for almost seven years now. 

In SPS, I have spent a lot of quality time with Vikas, Anjana, Charan, Madhusudan
Niladri, Navendu, and Ruchi. They have helped me a lot in tackling various
problems.

Manju Chattopadhyay, Ashish Tyagi, Sanjeev Kumar and Bhavna Vidhani are
some very important names in my life. I would like to thank all the friends who have written about me in my 
autograph book, their comments will remain as a treasure with me all
the time.

I would also like to thank my mother for everything.

The acknowledgment will be incomplete if I dont mention the name of 
my best friend Robin. Finally, thanks Vinayak! for being there with me
all the time.
\vskip1.0cm
\hrule

\vskip 2cm

\newpage

\begin{center}
{\large \bf List Of Publications}\\
\end{center}
\vskip1.0in
\hrule
\vskip1.2cm

\begin{enumerate}

\item{\it Universality in the Vibrational Spectra of Single-Component
Amorphous Clusters. }\\ 
Subir K. Sarkar, Gurpreet S. Matharoo and Akhilesh Pandey,\\ 
{\bf PHYSICAL REVIEW LETTER} 92, 215503 (2004).

\item{\it Vibrational Spectra of Amorphous Clusters : Universal Aspects. }\\ 
Gurpreet S. Matharoo, Subir K. Sarkar and Akhilesh Pandey,\\  
{\bf PHYSICAL REVIEW B} 72, 075401 (2005). 

\item{\it Vibrational Spectra of Simple Single-Component Systems:
Evolution with Disorder and Passage to Quasi-Universality. }\\
Subir K. Sarkar and Gurpreet S. Matharoo.\\
(to be submitted). 
\vskip1.2cm
\hrule
\vskip1.0in
{\bf Note:} Chapters 2 to 4 of this thesis are based primarily on the
papers 1 and 2 listed above. Chapter 5 is based on the preprint
listed as paper 3.

\end{enumerate}

\tableofcontents
\begin{description}
\item[] Bibliography~~.~~.~~.~~.~~.~~.~~.~~.~~.~~.~~.~~.~~.~~.~~.~~.~~.~~.~~.~~.~~.~~.~~.~~.~~.
~~.~~.~~.~~.~~.~~.~~.~~.~~~$87$
\end{description}

\chapter{Introduction}
\pagenumbering{arabic}
\setcounter{page}{1}

~~~~~The concept of universality has been familiar in physics for several
decades now [1] and has received
widespread usage in many areas. In this thesis we examine the applicability
of this concept to the domain of amorphous systems [2-22]. In particular we
examine the theme of universality in the context of the vibrational spectra
of such  systems. We consider both clusters and bulk systems.
Amorphosity immediately introduces some severe complications
that have been the subject of a lot of attention for many years now [23-65].
Thus, even within the harmonic approximation, which we use
consistently everywhere in this thesis, computation of the vibrational
spectrum poses serious difficulties in any analytical scheme and
invariably various kinds of approximations are used. These approximations
may be with respect to both the equilibrium structure of the disordered
system and the actual computation of the normal modes around this
structure. Our computation, which is entirely numerical in nature,
however makes no assumptions beyond the nature of the potential
that describes the interaction amongst the constituent particles of
the system. This is due to the fact that any candidate configuration
for a solid state structure, amorphous or crystalline, must
correspond to a local minimum of the potential energy function which
can be computed very accurately using standard numerical techniques.
These minima are also called `inherent structures'. For amorphous
states the positional arrangement of the constituent units is highly
disordered. The vibrational spectrum corresponding to a particular
stable configuration is derived from the solutions of an eigenvalue
problem in which the Hessian matrix for that configuration is
involved. When the configuration is disordered, it is clear that the
corresponding Hessian matrix will have substantial amount of randomness
in it. Thus, it is not surprising that ideas from the theory of
Random Matrices have been used in recent times to examine the
vibrational spectra of disordered systems [32-34,55,61,66-68]. Infact,
this is a key
element in our analysis of the universal aspects of the vibrational
spectra of disordered systems. Random Matrix theories have
been used extensively [69-78] for the analysis of statistical fluctuations
of the spectra of complex quantum systems such as complex nuclei,
atoms, molecules, quantum chaotic systems, disordered mesoscopic
systems etc. Study of the spectra of these systems
has revealed that although the smooth part of the density of
states will be system dependent, the fluctuations around the mean
densities are universal and there are only three universality classes.
In contrast, we look for universality both in the spectral
fluctuations and in the density of states.

We have studied both clusters and bulk systems while investigating
amorphous states. We have varied the nature of interaction amongst
the particles of the system under consideration in order to reveal
the possible presence of universality (i.e. independence of the
potential). For clusters, the number of particles is varied to
investigate the effect of finite size on various properties.
To study bulk amorphous systems, we actually use periodic crystals
with as large a primitive cell as possible. For any finite disorder,
the number of particles in the primitive cell should ideally be made
arbitrarily large. However, in practice, this is limited by the
available computational resources. We should note here that due to
the periodic nature of our approximation to the bulk disordered system,
the analysis of spectral fluctuations becomes a lot more subtle
and has not been reported in this thesis. We report results only on:
(1) Universality in the density of states for clusters,
(2) Universality in spectral fluctuations in the case of clusters, and
(3) Universality in the density of states for bulk amorphous systems.
In the following two paragraphs, we provide a summary of the
results reported in the subsequent chapters.

Chapter 2 describes the study of universality in both density of states
and statistical fluctuations in clusters made up of only one type of
particle. We have used four different kinds of potentials. Of these,
two belong to the `sum over pairs' category and the other two contain
many body interactions and hence are suitable for metals. The two
primary results are: (1) The spectral fluctuations are described
by the Gaussian Orthogonal Ensemble of random matrices to an extraordinarily
high degree of accuracy and (2) Over a large central region of the vibrational
spectrum, the density of states is described by the same functional
form, containing one scale of frequency, in all cases. In chapter 3
we extend the studies of chapter 2 to binary systems in order to see
how general are the results obtained  for single component systems.
We find that, atleast for the limited class of binary systems that we
have studied, results for single component systems are reproduced. Thus,
additional complexity does not cause any essential change in the results.
In chapter 4  we examine the question of possible universality of the
density of states over the {\it entire} spectrum rather than over only a
limited central part of it. From our analysis of the data, it seems
that universality of the form of the density of states does not extend
over the entire spectrum for the potentials and system sizes
investigated as far as clusters are concerned. This question is again
taken up, for bulk systems, in the next chapter.

In chapter 5  we show that there are indeed certain limiting situations
in which the normalized density of states assume a universal shape
{\it over the entire spectrum}.
Here  we study systems made up of point particles that interact via
two body forces only. The generic shape of this two body potential
has a minimum at a certain distance at which the potential is
negative. For larger distances  the potential rapidly rises zero.
At shorter distances also  the potential rapidly rises either to a
finite value or to an infinite value. The critical parameter for our
purposes is how fast the curvature of this potential changes around
the minimum. We introduce a parameter that quantifies the rapidity
of this variation and we find that when this parameter becomes very
large, the vibrational spectrum for completely disordered states assumes
a shape that does not depend on the specific functional form of the
potential that is tuned to achieve the very fast variation of the
curvature around the minimum. We study the implications of these
observations in understanding the quasi-universal shape that has been
observed recently for the vibrational spectra of molecular glasses [65].
The methodology that is used for the investigation of the bulk amorphous
systems has also been used in this chapter to study how the nature of
the vibrational spectrum changes as one goes from crystalline state
to completely amorphous states. For example, this gives a detailed
picture of how excess vibrational modes develop in both the low
frequency and high frequency domains of the spectrum. The accumulation
in the low frequency domain is the reason behind the existence of
what are called `boson peaks'[21-22, 45-65]. Although there are model
theoretical calculations that study the evolution of the vibrational spectrum
with increasing disorder, our calculation is distinguished by the
fact that we do not need to use any specific model of disorder. Once
the choice of potential is made, there is no need to make any further
approximation.

\chapter{Universality of Statistical Fluctuations : Single-Component Amorphous Clusters}

     In this chapter we explore the theme of universality in the vibrational
spectra of amorphous clusters made up of only one kind of constituent units
(atoms or molecules) [67-68].
All our clusters are three dimensional and to check for universality,
we use different kinds of potentials available in the literature [79-82].
More specifically, they are Lennard-Jones, Morse, Sutton - Chen and Gupta
potentials. The explicit expressions of these four types are given by

{\bf Lennard-Jones potential}:
\begin{equation}
V = 4 \sum\limits_{i<j} \left [\left( {1 \over r_{ij}} \right)^{12} -
\left( {1 \over r_{ij}} \right)^6\right ]
\end{equation}
with the factor of 4 omitted while calculating vibrational frequencies.

{\bf Morse potential}:
\begin{equation}
V = \sum\limits_{i<j} \left [\exp({-2\alpha(r_{ij}-1)})
- 2 \exp({-\alpha(r_{ij}-1)})\right]
\end{equation}
with the value of $\alpha = 6$.

{\bf Sutton - Chen potential}:
\begin{equation}
V = (1/2) \sum_{i=1}^N\sum\limits_{j \neq i} \left ( 1 \over r_{ij} \right)^{9} -\beta \sum_{i=1}^N
\sqrt{\sum\limits_{j \neq i} \left( 1 \over r_{ij} \right)^{6}}
\end{equation}
We take the value of $\beta$ to be equal to 39.432 which corresponds to nickel.\newline
{\bf Gupta potential}:
\begin{equation}
V= \sum_{i=1}^N (A \sum\limits_{j \neq i}\exp(-p(r_{ij} -1))) - \sum_{i=1}^N \sqrt{\sum\limits_{j \neq i}
\exp(-2q(r_{ij}-1)}
\end{equation}
The values of A, p and q are material specific and we have used the values applicable to
nickel or vanadium.

In all the expressions given above  $r_{ij}$ is the distance between the particles 
labelled i and j. N is the total number of particles in the cluster. As far as the 
functional forms are concerned, Lennard-Jones and Morse are of the type 
`sum over pairs' whereas the Sutton - Chen and Gupta potentials, which are often used
to describe metallic clusters, contain many body interactions in them. Choice of
these two distinct families of potentials is deliberate since universality is a primary
theme here and we want to explore as many qualitatively distinct kinds of potentials as 
possible.

In order to compute the vibrational spectra of amorphous clusters the first step that
needs to be taken is the generation of the stable geometries around which vibration takes
place. Obviously, any such geometry will correspond to a local minimum of the potential 
energy function in the configuration space. Configurations corresponding to
these minima are also called ``inherent structures''.
Each such inherent structure is a candidate for the geometry of a (atleast) metastable cluster at 
sufficiently low temperature. There are many methods available to generate these inherent structures.
The method we have used is called ``Homotopy Minimization'' [83]. In this method if we need to generate
a local minimum of a function $V$, we actually perform minimization of a sequence of functions of
the form $\left(\theta V + (1-\theta) U\right)$. Here $U$ is an appropriately selected simple function.
$\theta$ is a parameter that changes its values from 0 to 1 in a finite number of steps (about 20).
In the first step of the homotopy minimization the initial guess of the configuration is a random
distribution of the elements of the cluster within a sphere of suitable radius. If the radius is too 
large, the process of minimization is very slow and if it is too small, it may cause numerical 
instability due to high gradients that may be generated. As the value of $\theta$ keeps changing
subsequently, the initial guess at any stage for the configuration is the one that minimized the function
$\left(\theta V + (1-\theta) U\right)$ for the previous value of $\theta$. In all our calculations the number of 
particles is never below 100. For such a large number of particles  the method of homotopy minimization
leads, in one trial, to one of the higher energy local minima and the number of such minima is 
essentially limitless for the system sizes we utilize. But since the actual number of minima we 
generate is relatively small, the range of energies for the local minima produced is rather narrow. 

After generating an inherent structure  it is straightforward to compute the corresponding vibrational 
spectrum in the harmonic approximation. For this purpose  the Hessian matrix of the potential energy
function is computed for the configuration at hand and the subsequent steps for calculating the frequencies
for various normal modes of oscillations are standard. This process involves solving an eigenvalue problem
where the eigenvalues are proportional to the squares of the frequencies. In this chapter we 
denote the square of the frequency of a mode by $\lambda$ and analysis of the density as well as 
fluctuations will be presented only for $\lambda$. One could equally well choose
the frequency itself. But this would have made no difference to the kind of analysis that we present
in this chapter.

  Let us denote the elements of the eigenvalue spectrum for a particular local minimum (arranged in 
increasing order) by $\lambda(i)$, with i=1,2,...,$3N$. Since any potential that we deal with describes
the interactions of particles amongst themselves only, the conditions of translational and rotational
invariance must obviously be satisfied. This also implies that the six lowest frequency modes will have
$\lambda = 0$. For the remaining $\left(3N - 6\right)$ values of $\lambda$, characterization is done in terms of an
analysis of the mean local density and fluctuations around it. Consider the quantity
$\Delta_{i} \equiv \left(\lambda(i+1) - \lambda(i)\right)$. This raw value of the nearest neighbor spacing, when considered
as a function of i, will have a smooth component and a rapidly fluctuating component. The inverse of
the smooth component is called the local Density of States (DOS). This is a physical quantity that can
be measured through various experimental techniques and plays a direct role in determining various 
thermodynamic properties of the system. Here we denote this DOS function by $g(\lambda)$.
It turns out that for a precise numerical analysis of the nature of fluctuations, a precise 
knowledge of $g(\lambda)$ is required. However, it is rarely the case that one has an analytical
knowledge of the $g(\lambda)$ function. For our present examples also  this is the situation.
Hence, we can determine $g(\lambda)$ only numerically in the present analysis. The procedure
employed for extracting the $g(\lambda)$ function from the numerically computed eigenvalue 
spectrum is described in the next paragraph.

Let the number of eigenvalues less than or equal to $\lambda$ be represented by $H(\lambda)$.
Now plot the eigenvalue $\lambda$ along the x - axis and the corresponding eigenvalue number
(in the ordered spectrum) along the y - axis. By definition, this plot is monotonically 
increasing and $H(\lambda)$ is a staircase function that passes through all the points
of this plot. Let $S(\lambda)$ be the smooth best fit function for this plot. For any
finite eigenvalue spectrum $S(\lambda)$ is not actually uniquely defined. But for our purposes
an approximation maybe made as follows. Generate a suitable function space by combining 
various elementary functions and using a small number of parameters. The actual construction of
the function space is done through the inspection of the data and by trial and error.
Now this best fit function $S(\lambda)$ maybe generated by varying the parameters in the
function space and performing a least square fit. Since the best fit $S(\lambda)$ function
is known analytically now, we can simply take its derivative and that constitutes our
numerically determined DOS function for $\lambda$.  

The first key result of this chapter is the following observation. If 
we exclude about 10\% to 15\% of all the eigenvalues at both the low
frequency and high frequency ends,\

$S(\lambda)$ can be approximated quite accurately by the functional form
$D(\lambda) \equiv a-b\exp(-c\lambda)$ for the rest of the spectrum.
This observation is applicable to all the four potentials we have studied
and also for every system size $(\geq 100)$.

How well does the function $D(\lambda)$ approximate $S(\lambda)$ ?
This question can be answered by defining a misfit function 
$m(i)\equiv i- D(\lambda(i))$. In the range in which we do the fitting
(i.e. excluding 10\% to 15\% at each end) let $m_{max}$ be the maximum
absolute value of the misfit function as calculated with the best fit
values of the parameters a,b and c. 
Then, if we divide $m_{max}$ by the total number of eigenvalues within the
range of fit, we get an index that can be considered to be an appropriate
measure of the accuracy of the approximation. For all the cases that are 
reported in this chapter, the value of this index stays around or below 0.01.
In figures 2.1, 2.3, 2.5 and 2.7 we show how closely the best fit $D(\lambda)$
function approximates the integrated density of states. We have taken
one sample local minimum with the largest cluster size for each of the
four types of potentials. An alternative demonstration is given in
figures 2.2, 2.4, 2.6 and 2.8 where we plot the misfit function for the same
four local minima.

Before we proceed to perform an analysis of the statistical fluctuations of the
eigenvalue spectra, a general comment regarding the density of states function
is in order. This is regarding the evolution of the $g(\lambda)$ function with
the number of particles in the cluster for a given potential. The existence of 
thermodynamic limit would suggest that $g(\lambda)$ will have the structure $Nf(\lambda)$
when $N$ is sufficiently large. Here $f(\lambda)$ does not depend on the cluster size
but should be a function of energy per particle. Given the structure of the $g(\lambda)$
function that we are using, $f(\lambda)$ will have the form $(bc/N)exp(-c\lambda)$.
Thus, $c$ and $(bc/N)$ define the scales of $\lambda$ and the normalized density of states
- both these quantities being, in general, a function of the energy of the local minimum in question.
In our case, for a given potential and a given number of particles, the local minima are 
generated only in a rather narrow range of energy. If we make the standard assumption that the 
density of states function evolves smoothly with the energy of the local minimum, the
implication of 

\begin{figure}[htp]
\vskip+2cm
\epsfxsize=3in
\centerline{\epsfbox{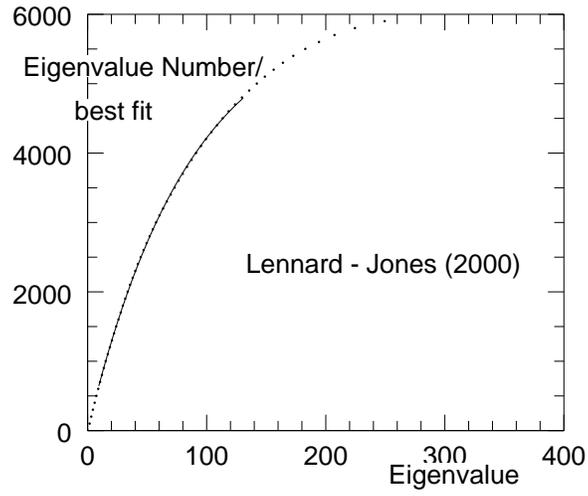}}
\caption{\sl Best fit function for Lennard-Jones potential.}
\label{fig201}
\end{figure}
       
\begin{figure}[htp]
\vskip+2cm
\epsfxsize=3in
\centerline{\epsfbox{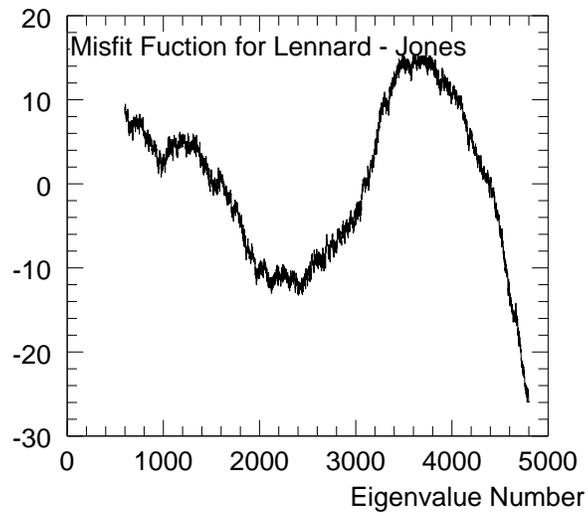}}
\caption{\sl Misfit function for Lennard-Jones potential.}
\label{fig205}
\end{figure}

\begin{figure}[htp]
\vskip+2cm
\epsfxsize=3in
\centerline{\epsfbox{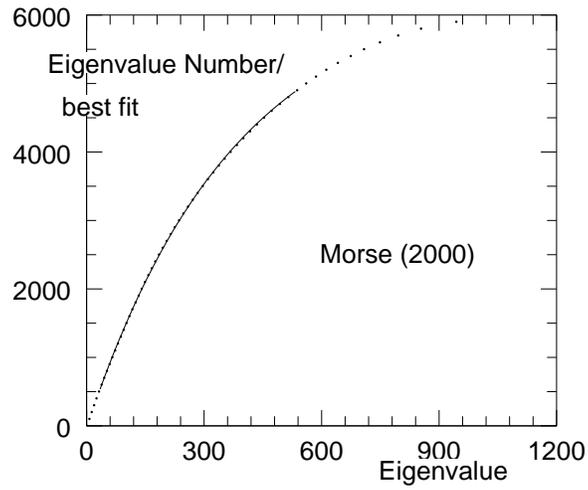}}
\caption{\sl Best fit function for Morse potential.}
\label{fig202}
\end{figure}

\begin{figure}[htp]
\vskip+2cm
\epsfxsize=3in
\centerline{\epsfbox{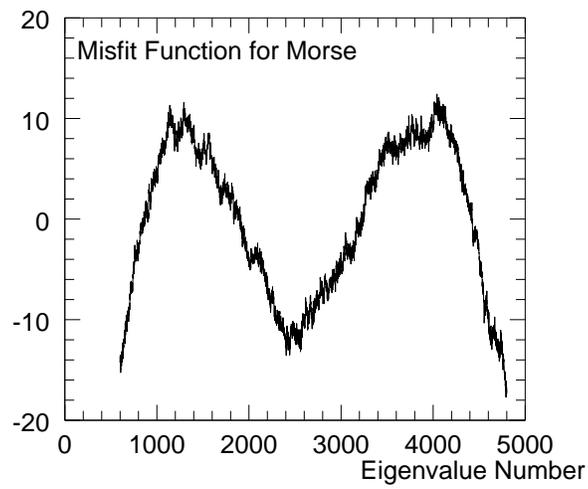}}
\caption{\sl Misfit function for Morse potential.}
\label{fig206}
\end{figure}

\begin{figure}[htp]
\vskip+2cm
\epsfxsize=3in
\centerline{\epsfbox{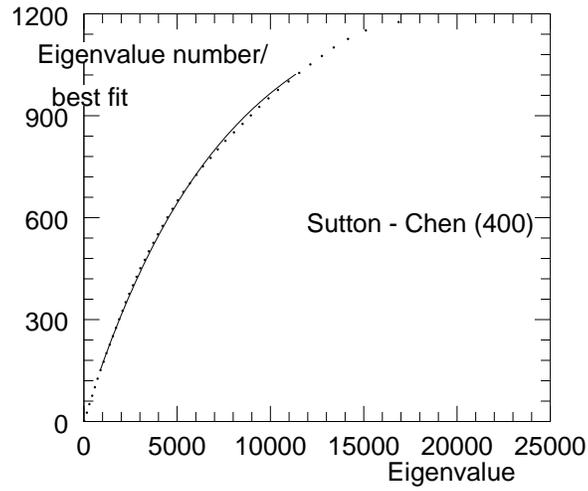}}
\caption{\sl Best fit function for Sutton - Chen potential.}
\label{fig203}
\end{figure}

\begin{figure}[htp]
\vskip+2cm
\epsfxsize=3in
\centerline{\epsfbox{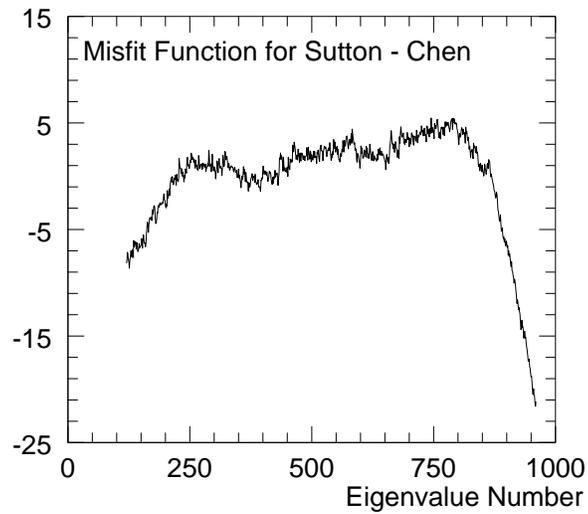}}
\caption{\sl Misfit function for Sutton - Chen potential.}
\label{fig207}
\end{figure}

\begin{figure}[htp]
\vskip+2cm
\epsfxsize=3in
\centerline{\epsfbox{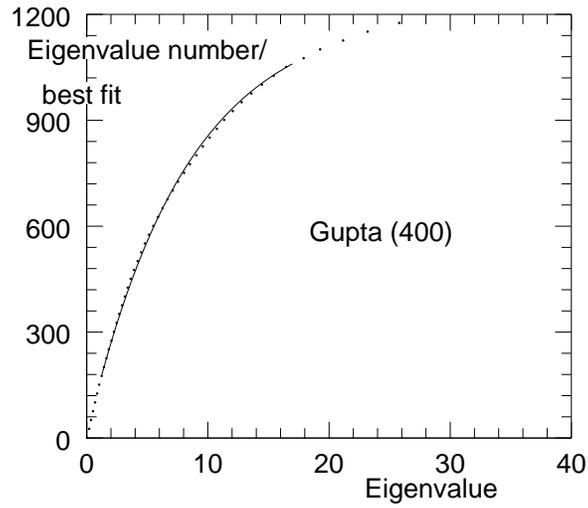}}
\caption{\sl Best fit function for Gupta potential.}
\label{fig204}
\end{figure}

\begin{figure}[htp]
\vskip+2cm
\epsfxsize=3in
\centerline{\epsfbox{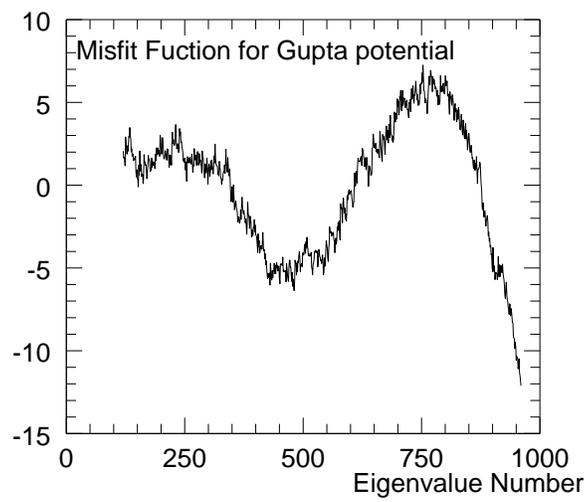}}
\caption{\sl Misfit function for Gupta potential.}
\label{fig208}
\end{figure}

\newpage
the narrowness of the range of the energies produced is that both 
$c$ and $bc/N$ should also be in a correspondingly narrow range. We have computed the mean
and the standard deviation of these two parameters and the data is displayed in { \bf Table 2.1}.
\begin {center}
{\bf Table 2.1}
\end{center}
\begin{center}
\begin{tabular}{|c|c|c|c|c|c|c|} \hline

{\bf Potential} &{\bf Number of}
& \multicolumn {2}{|c|}{\bf c} 
& \multicolumn {2}{|c|}{\bf cb/N}
& {\bf s}\\ \cline{3-7}

&{\bf Particles}&{\bf Average} &{\bf Standard} & {\bf Average}&{\bf Standard}&{\bf Variance}\\
&&&{\bf Deviation}&& {\bf Deviation}& \\ \hline

&200
&1.60x10$^{-2}$&6.80x10$^{-4}$&1.56x10$^{-2}$&4.02x10$^{-4}$&0.2844\\ \cline{2-7}
\bf LJ&500
&1.38x10$^{-2}$&4.49x10$^{-4}$&1.37x10$^{-2}$&2.18x10$^{-4}$&0.2844\\ \cline{2-7}
&1000
&1.26x10$^{-2}$&2.51x10$^{-4}$&1.27x10$^{-2}$&1.34x10$^{-4}$&0.2852\\ \cline{2-7}
&2000
&1.17x10$^{-2}$&1.51x10$^{-4}$&1.20x10$^{-2}$&8.19x10$^{-5}$&0.2854\\ \hline
&200
&3.63x10$^{-3}$&1.43x10$^{-4}$&3.65x10$^{-3}$&9.59x10$^{-5}$&0.2864\\ \cline{2-7}
\bf Morse&500
&3.21x10$^{-3}$&7.97x10$^{-5}$&3.29x10$^{-3}$&5.07x10$^{-5}$&0.2844\\ \cline{2-7}
&1000
&2.97x10$^{-3}$&5.50x10$^{-5}$&3.09x10$^{-3}$&3.29x10$^{-5}$&0.2857\\ \cline{2-7}
&2000
&2.78x10$^{-3}$&3.66x10$^{-5}$&2.95x10$^{-3}$&2.26x10$^{-5}$&0.2845\\ \hline
&100
&1.74x10$^{-4}$&6.76x10$^{-6}$&1.71x10$^{-4}$&6.44x10$^{-6}$&0.2908\\ \cline{2-7}
\bf SC&200
&1.65x10$^{-4}$&4.19x10$^{-6}$&1.65x10$^{-4}$&4.34x10$^{-6}$&0.2882\\ \cline{2-7}
&300
&1.60x10$^{-4}$&3.37x10$^{-6}$&1.62x10$^{-4}$&4.25x10$^{-6}$&0.2833\\ \cline{2-7}
&400
&1.58x10$^{-4}$&2.70x10$^{-6}$&1.61x10$^{-4}$&4.45x10$^{-6}$&0.2856\\ \hline
&100
&1.96x10$^{-2}$&8.76x10$^{-4}$&1.86x10$^{-2}$&8.53x10$^{-4}$&0.2942\\ \cline{2-7}
\bf Gupta&200
&1.71x10$^{-2}$&6.14x10$^{-4}$&1.63x10$^{-2}$&3.83x10$^{-4}$&0.2838\\ \cline{2-7}
&400
&1.49x10$^{-2}$&7.39x10$^{-4}$&1.46x10$^{-2}$&2.48x10$^{-4}$&0.2841\\ \hline

\end{tabular}
\end{center}

\vspace{0.2in}

Inspection of this table shows that the standard deviation decreases with increase in the
value of $N$ in all the cases. Simultaneously, the mean values appear to go to a non - zero
limit. This suggests the existence of a well defined density of states function as the 
number of particles in the cluster becomes very large. For the algorithm that we are using to
generate the local minima, we have no direct control over their energies. Thus, the apparent
convergence of $c$ as well as $bc/N$ suggests that the energy per particle goes to some
limiting value in the large $N$ limit. However, we have no {\it{ a priori}} control over this
limiting value of the energy per particle.

In order to analyze the statistical fluctuations of the eigenvalue spectra, we use procedures
that are standard in the theory of Random Matrices [69-72]. The first step is to convert the spectrum
of eigenvalues into what is called an ``unfolded'' spectrum. This is done by using the map 
$s(i)\equiv S(\lambda(i))$. Given the definition of the $S(\lambda)$ function that has been 
provided earlier, it is obvious that the average nearest neighbor spacing of the transformed
eigenvalues will be unity everywhere in the spectrum. This standardization of the scale of 
spacing makes it possible to compare the nature of fluctuations between different parts of the
same spectrum or two entirely different spectra. One benefit of the process of unfolding is 
that the statistics of the analysis of the fluctuations can be significantly improved
by combining the data from many different unfolded spectra if there is reason to believe 
that the nature of the statistical fluctuations is the same for all the spectra. However,
it should be kept in mind that the transformation from the raw spectrum to the unfolded 
spectrum is implemented by using the $D(\lambda)$ function with the best fit values of a,b and
c specific to that local minimum. To summarize the procedure:
\newline(1) Take the ensemble of local minima
generated for a particular potential with a specific \hskip+10.0in number of particles. 
\newline(2) Generate the
eigenvalue spectra for all the minima, and \newline (3) Unfold all the spectra. \newline
On this collection of the unfolded spectra, we perform the following kinds of analysis of
statistical fluctuations: \newline
(1) Calculate the distribution ($p(s)$) of the nearest neighbor spacing
$(s)$. \newline
(2) Define the variable $n(r)$ to be the number of levels within a window of length of $r$ placed
randomly in the spectrum. Then we calculate variance [$\Sigma^{2}(r)$], skewness
[$\gamma_{1}(r)$]
and excess [$\gamma_{2}(r)$] parameters of the distribution of the variable $n(r)$.

Before we present the data on the statistical analysis, we should 
reiterate that the approximation $D(\lambda)$ to the $S(\lambda)$ function holds good only 
over a large central region of the spectrum. We consistently take this region to be 
what is left after excluding the lowest 10\% and the highest 20\% of each spectrum.
This is called `region II'. We refer to the spectral regions below and above this
range as `region I' and `region III', respectively. Analysis of the spectral 
fluctuations for these two relatively small regions are performed separately whenever
they contain adequate number of eigenmodes. We now proceed to present the results
of the analysis of the statistical fluctuations for the Lennard-Jones, Morse, 
Sutton - Chen and Gupta (for nickel) potentials. The largest cluster sizes that we have used
with these four potentials are 2000, 2000, 400, and 400, respectively. The reason 
behind the largest system size being much smaller for Gupta and Sutton - Chen
potentials is the presence of the many body terms. This causes a very large
increase in the requirement of the computational power. As a result, we are 
constrained to use much smaller system sizes and fewer local minima as compared
to Morse and Lennard-Jones potentials (which are of the form of sum over pairs).
\vskip 0.2in

{ \huge \underline{Region II}}

In this section we present the results of the analysis of statistical 
fluctuations in the large central region of the spectrum for all the 
four types of potentials. Analysis of individual spectra suggests that
in every case  the fluctuations have the characteristics of the
Gaussian Orthogonal Ensemble (GOE) of random matrices. However, the
quality of the statistics is unsatisfactory owing to the relatively
small number of levels in the region II of each spectrum -- not 
exceeding about 4000 in any case. Since all the spectra individually
have the features of GOE, we are able to improve statistics by combining 
the data for all the spectra obtained for a given number of particles 
with a particular potential.

First of all  we present the data for the distribution $p(s)$ of the 
normalized nearest neighbor spacings $(s)$ for the largest cluster size
generated with each potential in figures 2.9, 2.10, 2.11 and 2.12. In each
of the figures  we also show the prediction from the Wigner's surmise
$[p(s) = (\pi s/2)\exp(-\pi s^{2}/4)]$ as well as the exact prediction
for the GOE [69].
\begin{figure}[htp]
\vskip+0.5cm
\epsfxsize=7.5in
\centerline{\epsfbox{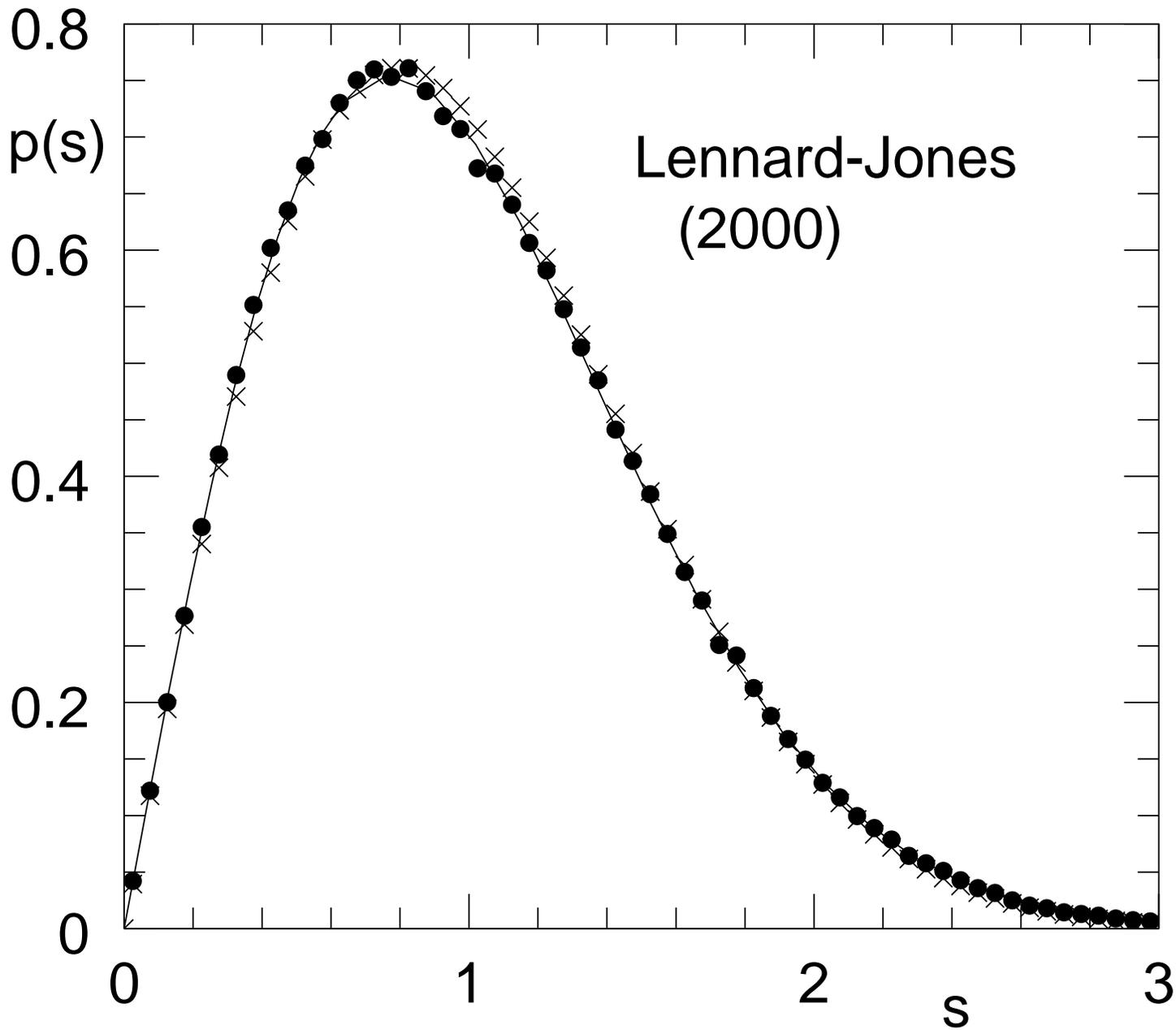}}
\caption{\sl Probability density [$p(s)$] for normalized nearest neighbor
spacing ($s$) for Lennard-Jones potential.
Filled circles: Our data. 
Crosses: Wigner's surmise for GOE.
Continuous line: Exact prediction for the GOE.}
\label{fig209}
\end{figure}

\begin{figure}[htp]
\vskip+0.5cm
\epsfxsize=7.5in
\centerline{\epsfbox{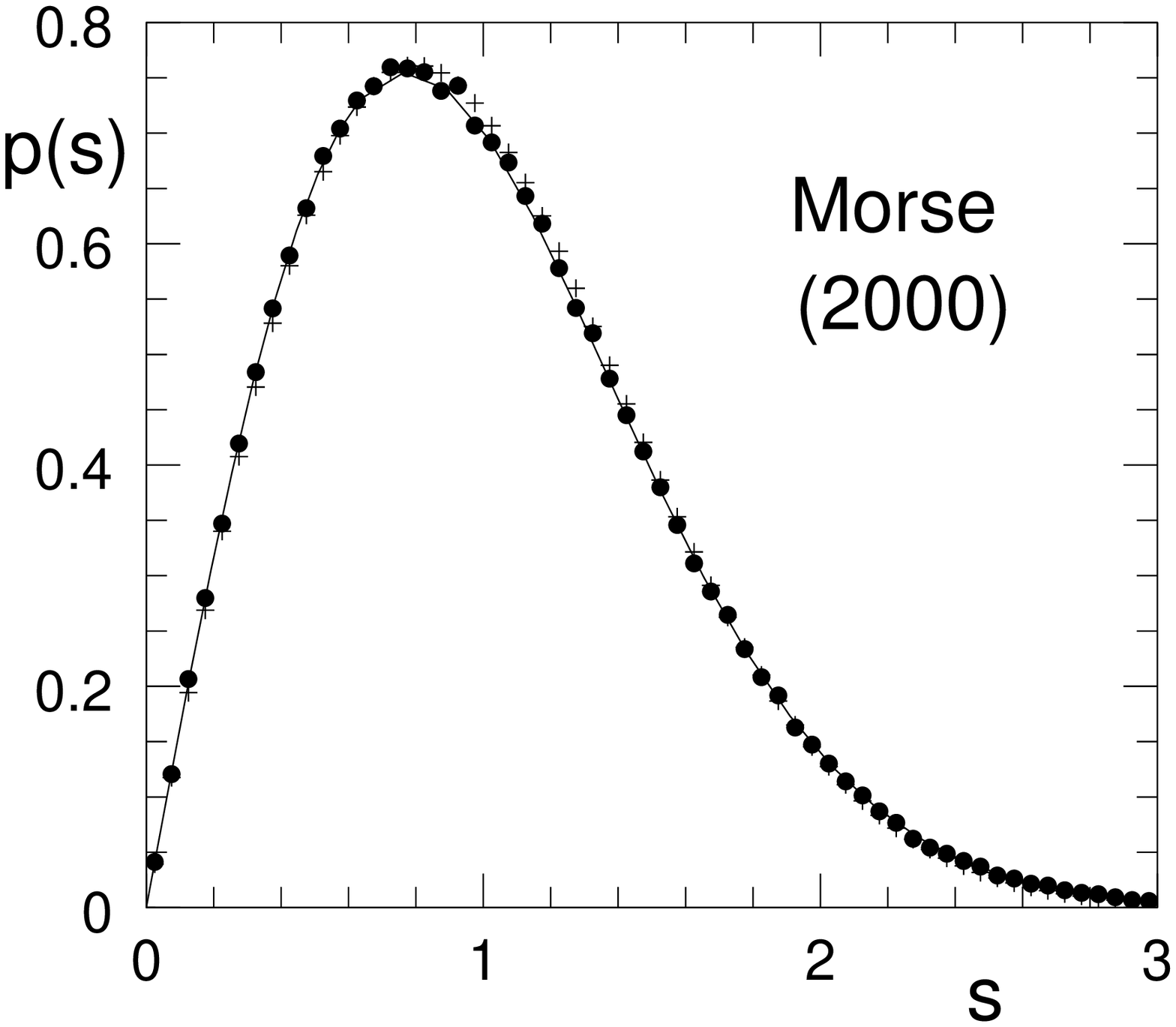}}
\caption{\sl Probability density [$p(s)$] for normalized nearest neighbor
spacing ($s$) for Morse potential.
Filled circles: Our data.
Crosses: Wigner's surmise for GOE.
Continuous line: Exact prediction for the GOE.}
\label{fig210}
\end{figure}

\begin{figure}[htp]
\vskip+0.5cm
\epsfxsize=7.5in
\centerline{\epsfbox{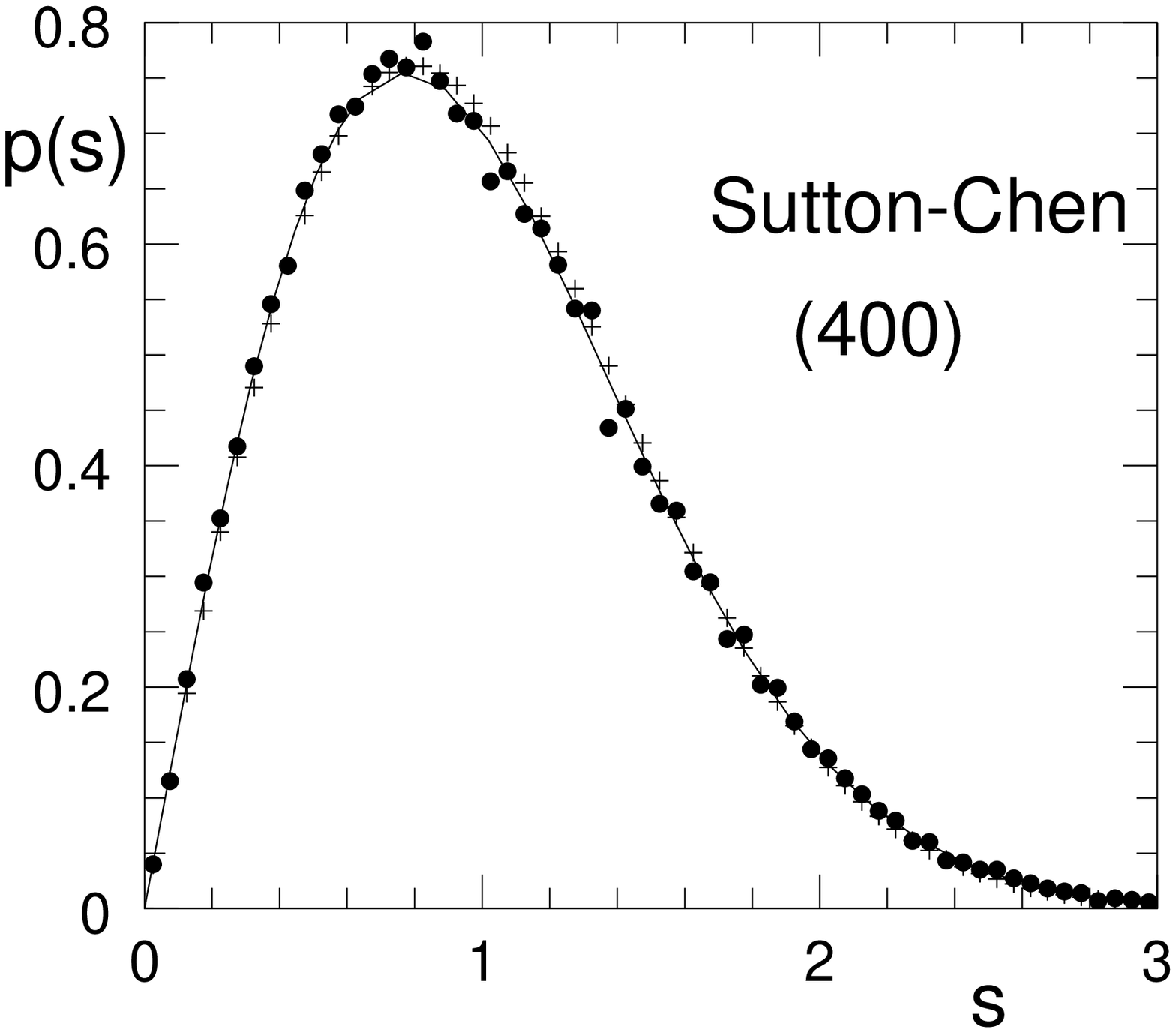}}
\caption{\sl Probability density [$p(s)$] for normalized nearest neighbor
spacing ($s$) for Sutton - Chen potential.
Filled circles: Our data.
Crosses: Wigner's surmise for GOE.
Continuous line: Exact prediction for the GOE.}
\label{fig211}
\end{figure}

\begin{figure}[htp]
\vskip+0.5cm
\epsfxsize=7.5in
\centerline{\epsfbox{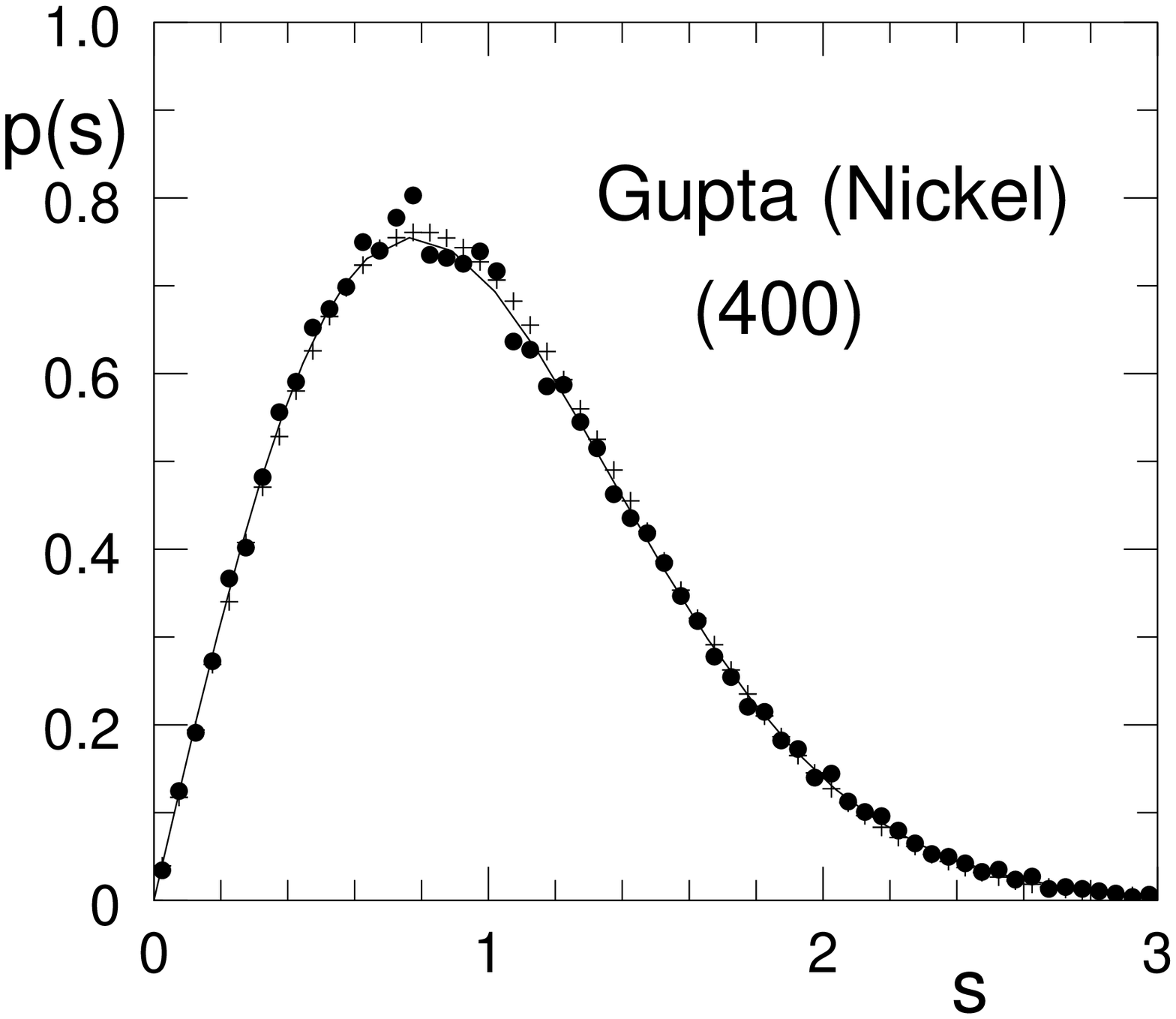}}
\caption{\sl Probability density [$p(s)$] for normalized nearest neighbor
spacing ($s$) for Gupta potential.
Filled circles: Our data.
Crosses: Wigner's surmise for GOE.
Continuous line: Exact prediction for the GOE.}
\label{fig212}
\end{figure}

The Wigner surmise is an approximate but highly accurate formula for
$p(s)$. Its deviation from the exact prediction is typically of the
order of 1\% or less and it has been extensively used in the literature.
Part of the reason behind this extensive use of a formula that is known
to be approximate is that the level of statistics in most applications 
is simply not large enough to be able to meaningfully  distinguish 
between the approximate formula and the exact result. However, in 
our case  the ensembles for Lennard-Jones potential and 
Morse potential are large enough (of the order of one million levels)
to make it possible to verify in a statistically significant manner 
whether our data follows the Wigner's surmise or the exact result.
The plots for $p(s)$ in figures 2.9 and 2.10 for Lennard-Jones and Morse
potential show clearly that our data agrees with the exact predictions
for GOE rather than the Wigner surmise.

For the Sutton - Chen and the Gupta potential ( figures 2.11 and 2.12),
the data for $p(s)$ show a significantly higher level of scatter around 
the two predictions but it turns out that even for these two cases, 
the data points fall essentially within the range of permissible statistical 
fluctuations. For example, the absolute number of data points in a bin near
the peak of the $p(s)$ curve are about 12000, 19000, 1200 and 2300 for the
Lennard-Jones, Morse, Sutton - Chen and Gupta potential, respectively.
From this, the level of allowed statistical fluctuation can be computed
and we find that the extent of scatter in the actual data is within 
permissible limits.

A single valued indicator that follows from the $p(s)$ function is the
variance of the nearest neighbor spacing. We have calculated this parameter
also for every ensemble and the results are available in the last column of
{\bf Table 2.1}. From the exact prediction for the GOE, this variance 
should be 0.286 whereas the Wigner's surmise leads to the value of 0.273.
Inspection of {\bf Table 2.1} shows that the computed values of the 
variance are in much closer agreement with the exact GOE prediction.
\newpage
We should point out here that the distribution of the nearest neighbor spacing
is not too sensitive with respect to the size of the cluster. Even for a system
with only about 100 particles, there is very good agreement between the computed
$p(s)$ function and the predictions (see figures 2.13, 2.14, 2.15 and 2.16).

The significance of this observation is the following:\newline
Any attempt at verifying these predictions in laboratory experiments or ab - initio
calculations will involve only relatively small clusters (of the order of 100 or so) 
since bigger sizes are not feasible presently. Thus, it is important that the GOE
properties should be observable even for systems of such small sizes. 
Our results for the clusters containing only 100 particles suggest that this is
indeed the case.

Now we consider the computation of $\Sigma^{2}(r)$, which is the variance 
of $n(r)$, the number of levels within a window of length $r$ placed 
randomly in the spectrum. For a credible computation of this characteristic,
special care must be taken while unfolding the spectrum. As we have mentioned
earlier, the function $D(\lambda)$ with appropriately chosen values of a, b, 
and c does provide a rather close approximation to the ideal smooth function
$S(\lambda)$. But an examination of the misfit functions (see figures 2.2,
2.4, 2.6 and 2.8) shows that it is still not close enough as far as the 
computation of $\Sigma^{2}(r)$ is concerned. Specifically, the misfit 
functions have an amplitude of fluctuations (around zero) of the order
of 1\% of the range of the fit and large scale systematic mismatch is
apparent. Since  the value of $\Sigma^{2}(r)$ for GOE stays around unity,
even for fairly large values of $r$, this level of systematic mismatch
renders any calculation of variance for large values of $r$ unacceptable.
For GOE  the spectrum is extremely rigid and $\Sigma^{2}(r)$ grows only as
$ln(r)$. On the other hand, the error in the computation of $\Sigma^{2}(r)$
which is caused by the systematic mismatch that we have mentioned earlier
grows proportional to $r^{2}$ and the constant of proportionality grows with
the extent of the mismatch. Hence, it is quite essential that the amplitude 
of mismatch should be brought down to the lowest possible level and ideally
the mismatch function should reflect
\begin{figure}[htp]
\vskip+1.0cm
\epsfxsize=3.0in
\centerline{\epsfbox{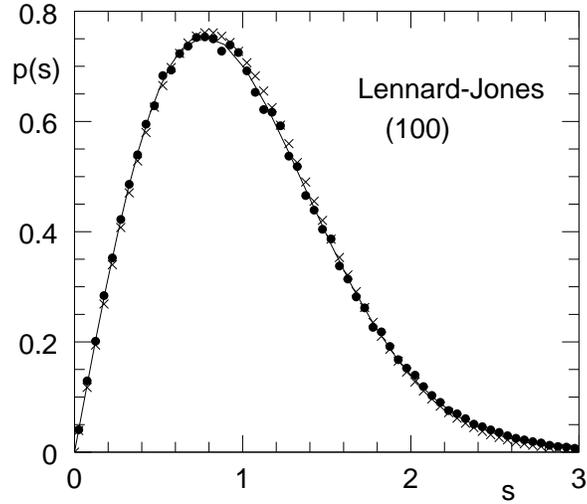}}
\caption{\sl Probability density [$p(s)$] for normalized nearest neighbor
spacing ($s$) for Lennard-Jones potential with $N = 100$.
Filled circles: Our data.
Crosses: Wigner's surmise for GOE.
Continuous line: Exact prediction for the GOE.}
\label{fig213}
\end{figure}

\begin{figure}[htp]
\vskip+0.5cm
\epsfxsize=3.0in
\centerline{\epsfbox{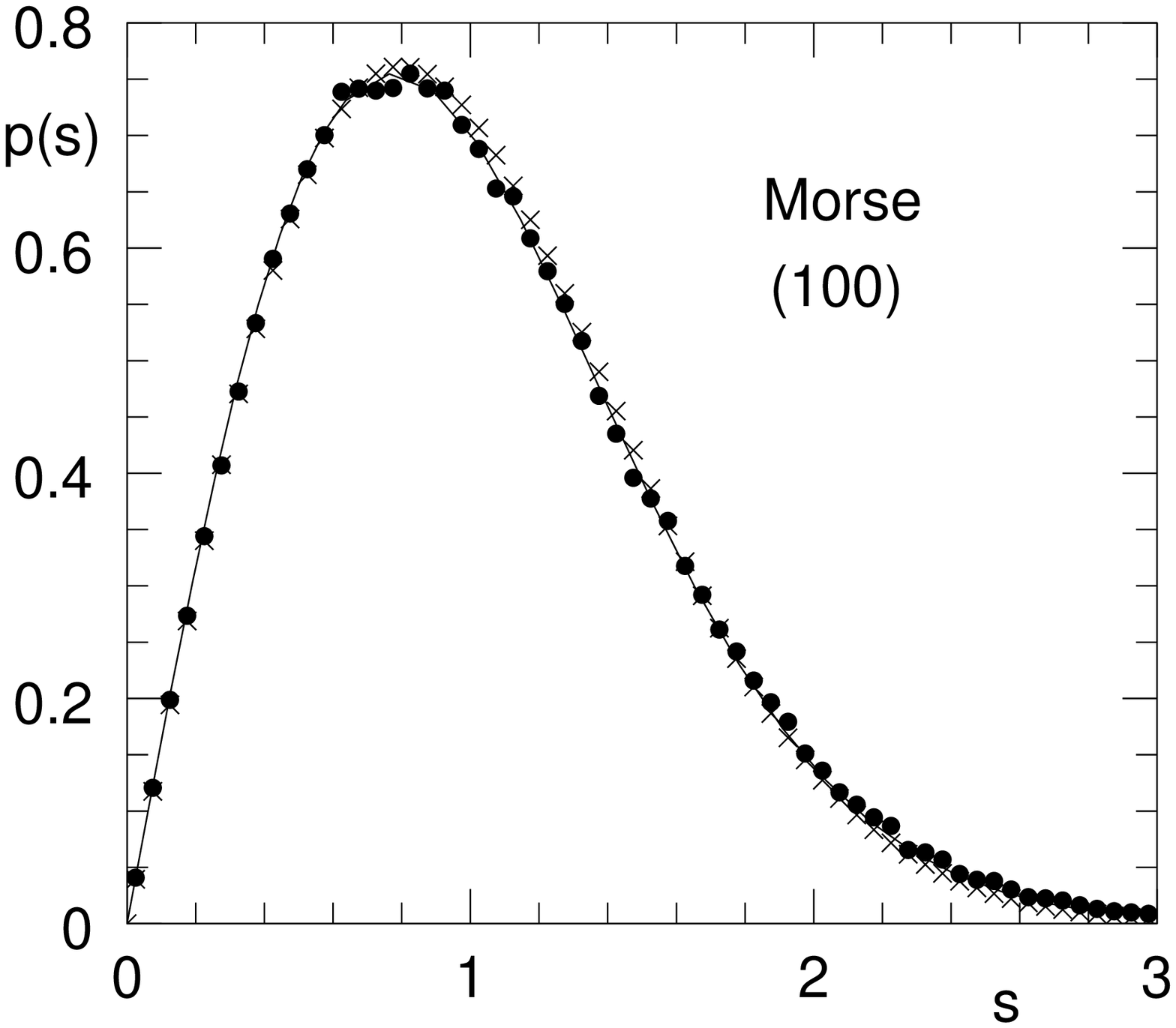}}
\caption{\sl Probability density [$p(s)$] for normalized nearest neighbor
spacing ($s$) for Morse potential with $N = 100$. 
Filled circles: Our data.
Crosses: Wigner's surmise for GOE.
Continuous line: Exact prediction for the GOE.}
\label{fig214}
\end{figure}

\begin{figure}[htp]
\vskip+0.5cm
\epsfxsize=3.0in
\centerline{\epsfbox{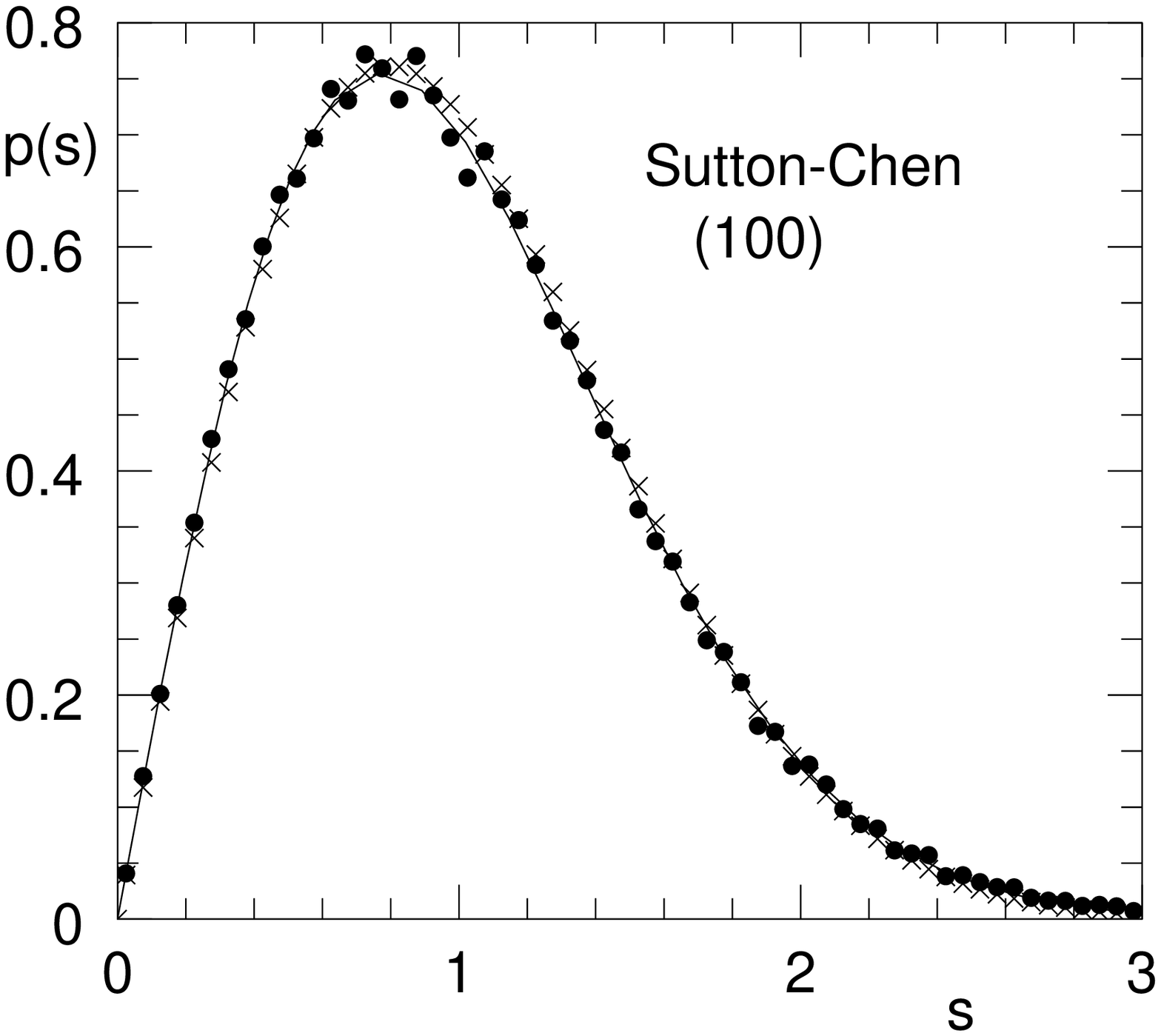}}
\caption{\sl Probability density [$p(s)$] for normalized nearest neighbor
spacing ($s$)for Sutton - Chen potential with $N = 100$.
Filled circles: Our data.
Crosses: Wigner's surmise for GOE.
Continuous line: Exact prediction for the GOE.}
\label{fig215}
\end{figure}

\begin{figure}[htp]
\vskip+0.5cm
\epsfxsize=3.0in
\centerline{\epsfbox{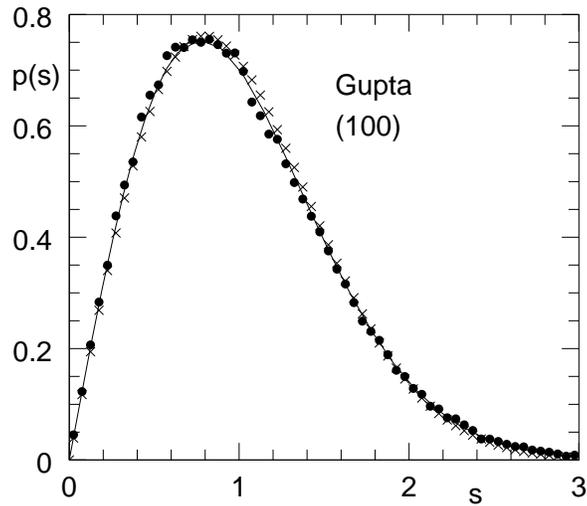}}
\caption{\sl Probability density [$p(s)$] for normalized nearest neighbor
spacing ($s$)for Gupta potential with $N=100$.
Filled circles: Our data.
Crosses: Wigner's surmise for GOE.
Continuous line: Exact prediction for the GOE.}
\label{fig216}
\end{figure}

\vspace{1in}
 only the statistical fluctuations which
are inherent to GOE. In our analysis  we have effected a reduction in the 
amplitude of the mismatch function by adding a correction term to the
principal  fitting function $D(\lambda)$. To do this  we first of all
identify the regions of the mismatch function where the behavior is
rather irregular. Since there is no simple way to get a smooth fit through
these regions they are eliminated from further consideration.
For the remaining relatively smooth regions ( which we call `accepted regions')
,which are typically 2 or 3
in number, we separately construct quadratic fits to the mismatch function.
Adding these quadratic functions to the principal fitting function 
$D(\lambda)$, we get the final approximation to $S(\lambda)$ for each
of the accepted regions separately. Thus, the approximation to $S(\lambda)$
which is actually used for unfolding is somewhat different for each of the 
accepted  regions within the same spectrum. We should mention here that
all the data for the normalized nearest neighbor spacing that are reported
in this thesis are generated from spectra that have been unfolded by using
the combined unfolding function -- although the $p(s)$ function would not
be altered recognizably even if we omitted the quadratic correction term
in the unfolding function.
For the calculation of $\Sigma^{2}(r)$ for a given value of
$r$, we use the ensemble of $n(r)$ computed for all the accepted regions of all
the spectra for a given cluster size and a given potential. This same 
ensemble is also used for calculation of the skewness [$\gamma_{1}(r)$] and
excess [$\gamma_{2}(r)$] parameters which are defined as
\begin{equation}
\gamma_{1}(r) = (1/M) \sum_{j=1}^M((n_{j} - \overline{n})/\sigma)^{3}
\end{equation}
\begin{equation}
\gamma_{2}(r) = (1/M) \sum_{j=1}^M((n_{j} - \overline{n})/\sigma)^{4} - 3 
\end{equation}
where  $M$ is the number of data points in the ensemble for $n(r)$.
$\sigma$ and $\overline{n}$ are the standard deviation and mean of $n(r)$,
respectively.

Data for $\Sigma^{2}(r)$ is displayed for the four potentials in figures 
2.17, 2.18, 2.19 and 2.20. In each figure  we combine the data for several
cluster sizes to exhibit the trend of dependence on 

\begin{figure}[htp]
\vskip+0.5cm
\epsfxsize=3in
\centerline{\epsfbox{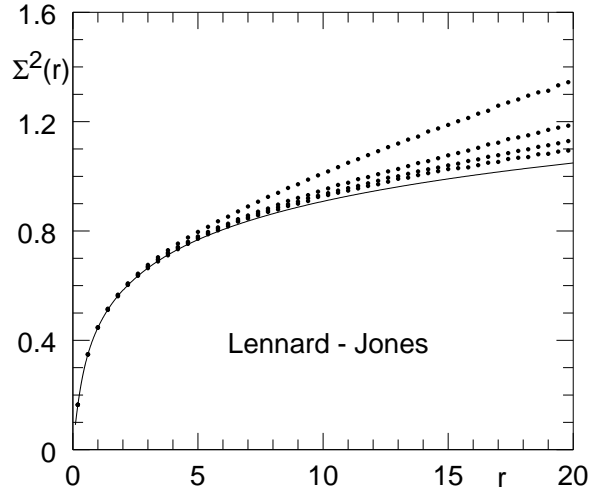}}
\caption{\sl Variance of the number of levels in interval of length $r$
plotted as a function of $r$ for Lennard-Jones potential.
Continuous line: Prediction for the GOE.
Filled Circles: Our data.
Number of particles $N$ = 200, 500, 1000 and 2000 from top to bottom.}
\label{fig217}
\end{figure}
       
\begin{figure}[htp]
\vskip+0.5cm
\epsfxsize=3in
\centerline{\epsfbox{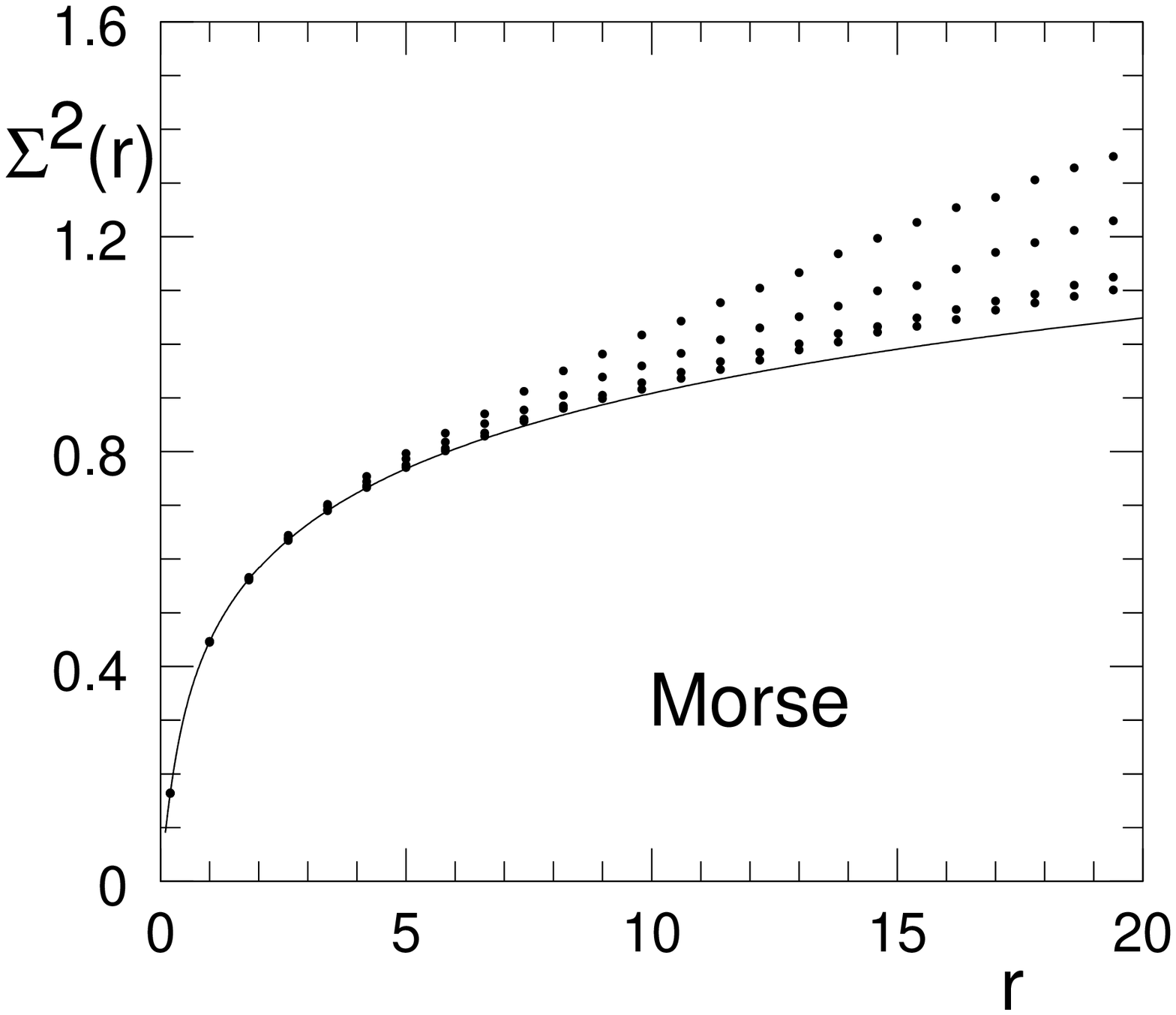}}
\caption{\sl Variance of the number of levels in interval of length $r$
plotted as a function of $r$ for Morse potential.
Continuous line: Prediction for the GOE.
Filled Circles: Our data.
Number of particles $N$ = 200, 500, 1000 and 2000 from top to bottom.}
\label{fig218}
\end{figure}

\begin{figure}[htp]
\vskip+0.5cm
\epsfxsize=3in
\centerline{\epsfbox{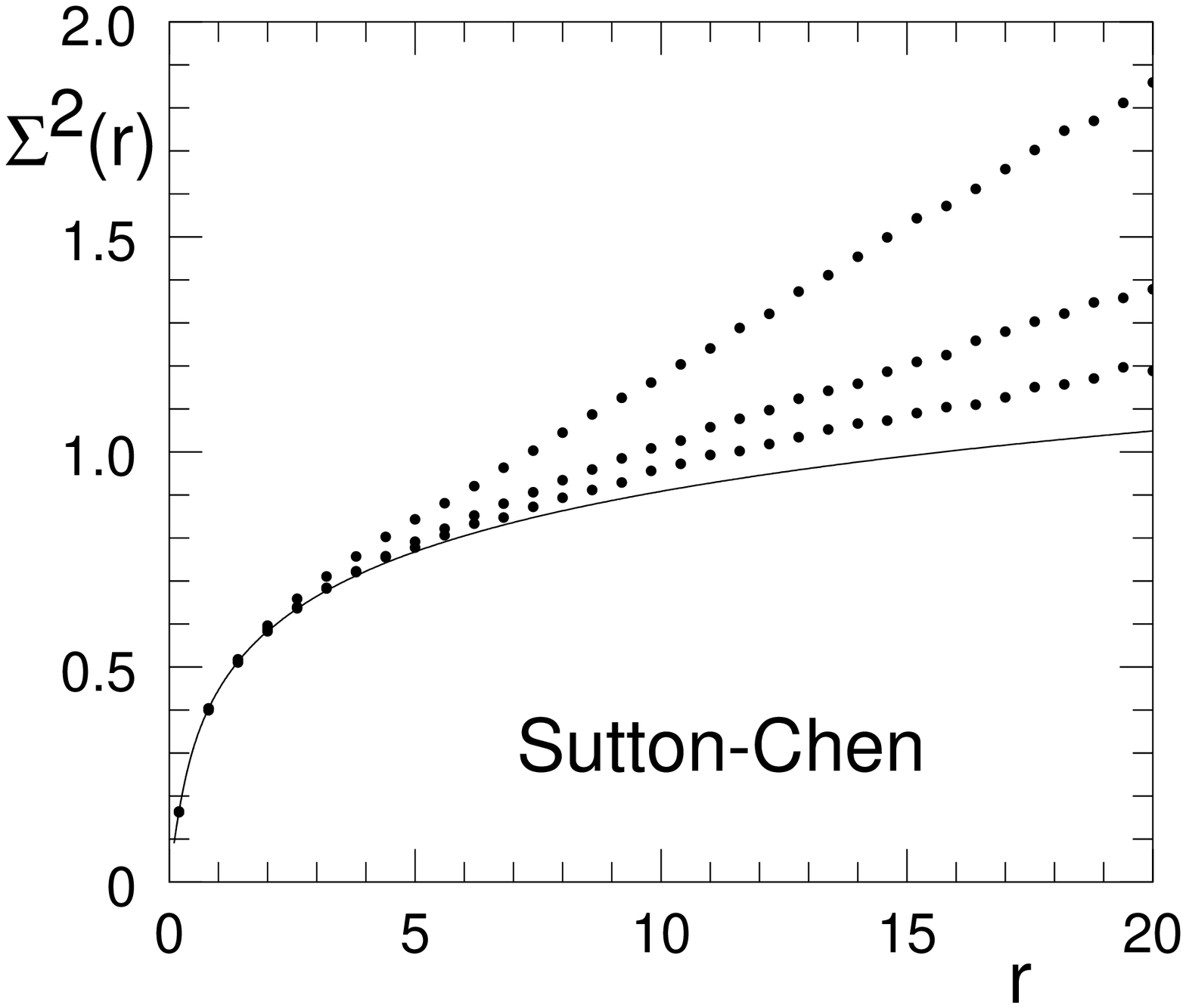}}
\caption{\sl Variance of the number of levels in interval of length $r$
plotted as a function of $r$ for Sutton - Chen potential.
Continuous line: Prediction for the GOE.
Filled Circles: Our data.
Number of particles $N$ = 100, 200  and 400 from top to bottom.}
\label{fig219}
\end{figure}

\begin{figure}[htp]
\vskip+0.5cm
\epsfxsize=3in
\centerline{\epsfbox{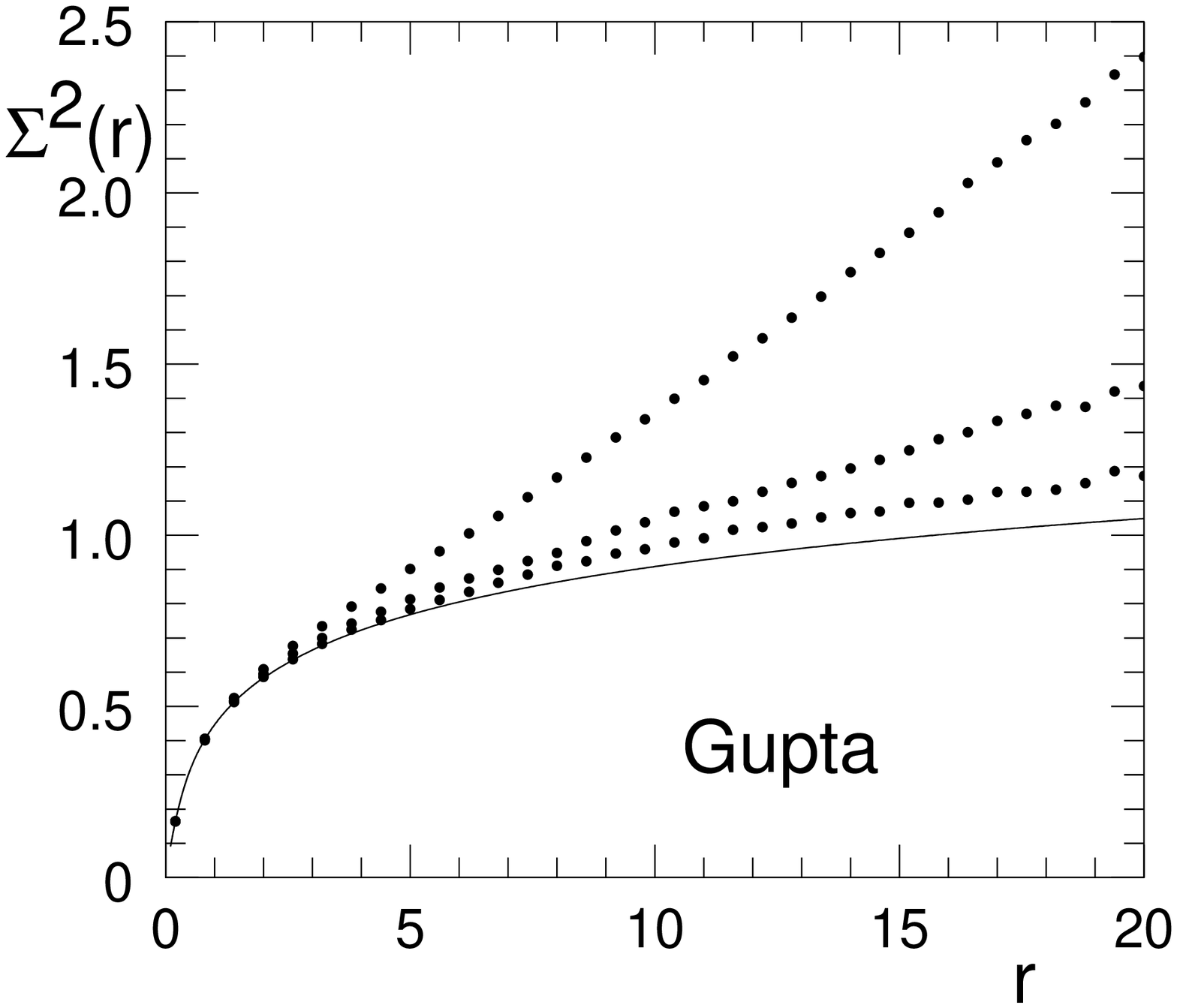}}
\caption{\sl Variance of the number of levels in interval of length $r$
plotted as a function of $r$ for Gupta potential.
Continuous line: Prediction for the GOE.
Filled Circles: Our data.
Number of particles $N$ = 100, 200, and 400 from top to bottom.}
\label{fig220}
\end{figure}

\begin{figure}[htp]
\vskip+0.5cm
\epsfxsize=3in
\centerline{\epsfbox{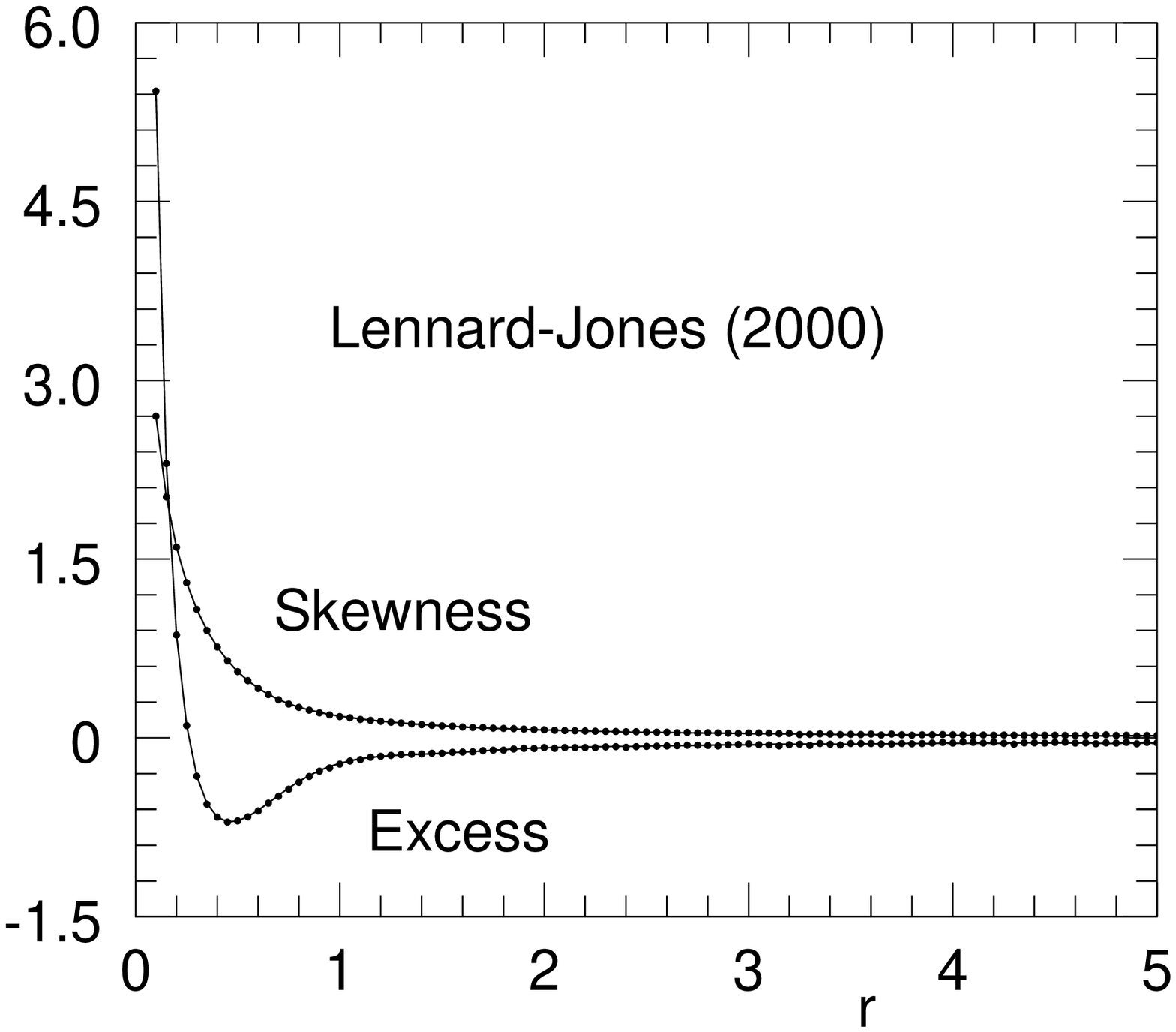}}
\caption{\sl Skewness and Excess parameters of the distribution of $n(r)$,
 the number of levels in interval of length $r$,
plotted as a function of $r$ for Lennard-Jones potential with $N=2000$.
Continuous lines: Predictions for the GOE.
Filled Circles: Our data.}
\label{fig221}
\end{figure}
       
\begin{figure}[htp]
\vskip+0.5cm
\epsfxsize=3in
\centerline{\epsfbox{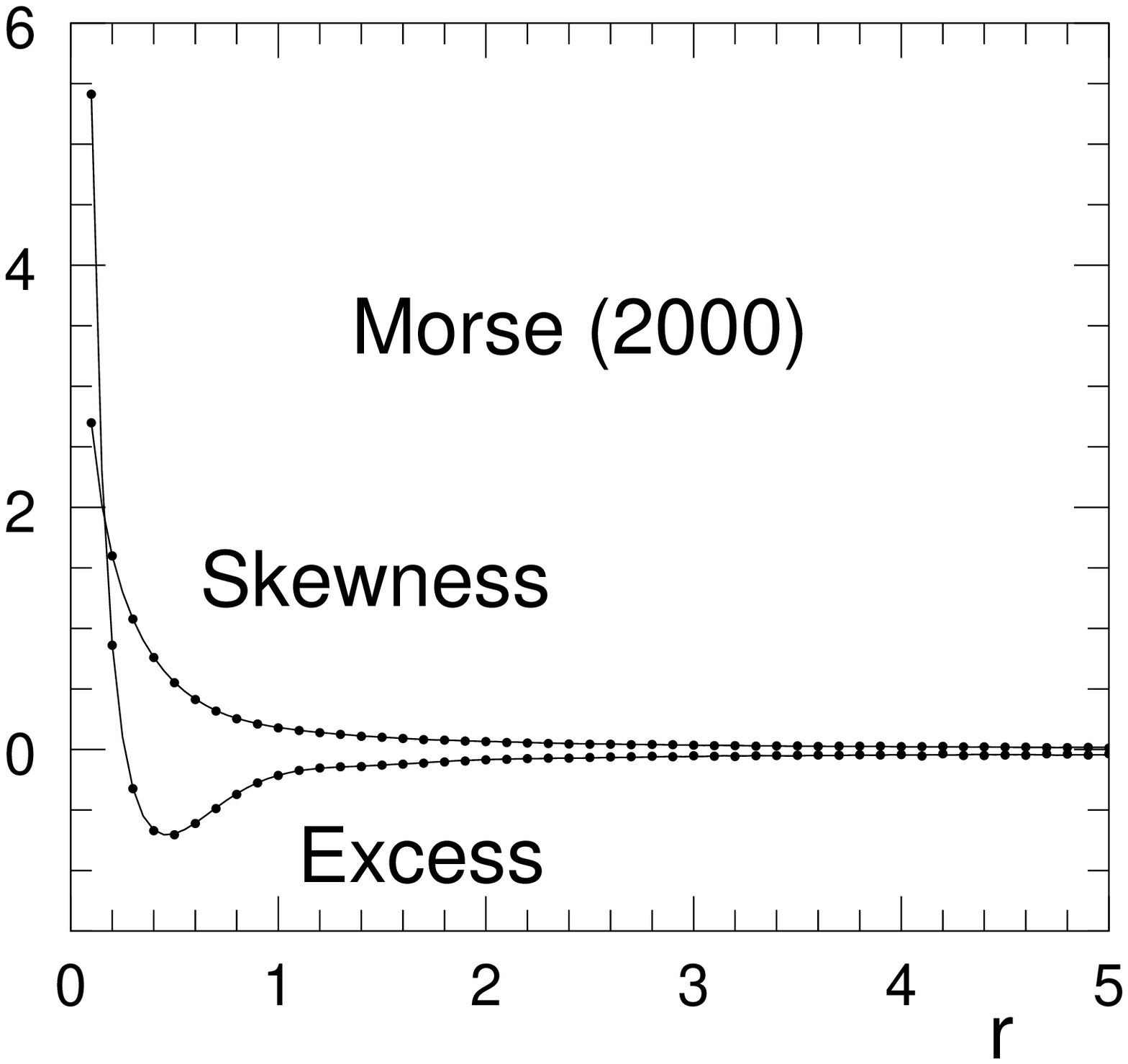}}
\caption{\sl Skewness and Excess parameters of the distribution of $n(r)$,
 the number of levels in interval of length $r$,
plotted as a function of $r$ for Morse potential with $N=2000$.
Continuous lines: Predictions for the GOE.
Filled Circles: Our data.}
\label{fig222}
\end{figure}

\begin{figure}[htp]
\vskip+0.5cm
\epsfxsize=3in
\centerline{\epsfbox{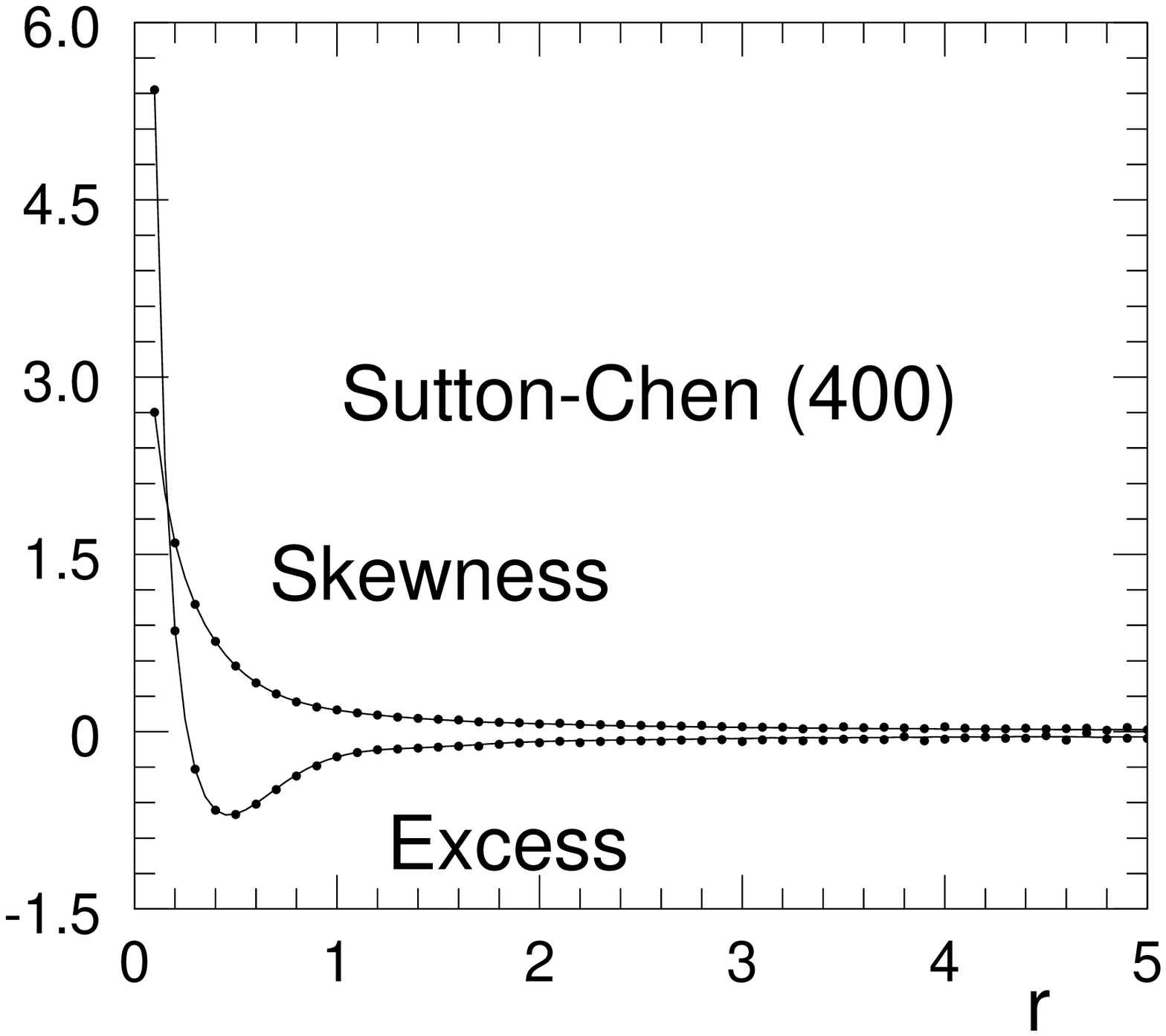}}
\caption{\sl Skewness and Excess parameters of the distribution of $n(r)$,
 the number of levels in interval of length $r$,
plotted as a function of $r$ for Sutton - Chen potential with $N=400$.
Continuous lines: Predictions for the GOE.
Filled Circles: Our data.}
\label{fig223}
\end{figure}

\begin{figure}[htp]
\vskip+0.5cm
\epsfxsize=3in
\centerline{\epsfbox{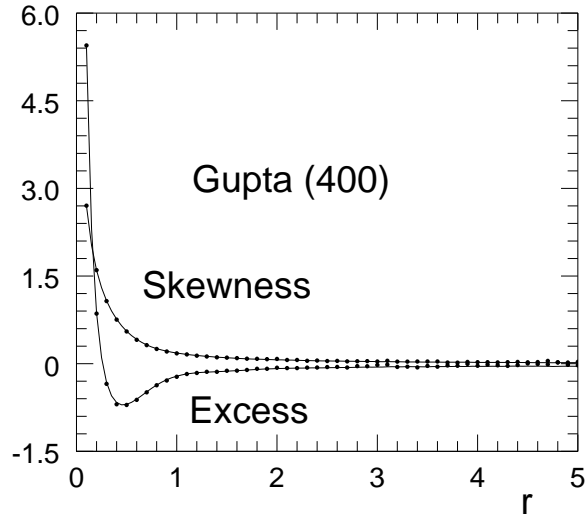}}
\caption{\sl Skewness and Excess parameters of the distribution of $n(r)$,
 the number of levels in interval of length $r$,
plotted as a function of $r$ for Gupta potential with $N=400$.
Continuous lines: Predictions for the GOE.
Filled Circles: Our data.}
\label{fig224}
\end{figure}
$N$. Common to all the
figures is the observation that the agreement between the predictions for
the GOE [69] and the computed data becomes closer as the system size increases.
Data for the skewness and excess parameters are presented in the figures
2.21, 2.22, 2.23 and 2.24 for the largest system size used with each of the
four potentials. In each case  the exact predictions for the GOE are also 
included. These ``exact predictions'' have been calculated on the basis of a
large ensemble of 500 X 500 matrices belonging to the Gaussian Orthogonal
Ensemble.

The reason why, for skewness and excess parameters, we show only the data
for the largest system size
in each case is that there is hardly any detectable dependence on the system
size.  Even for $N$ as small as 100 this is true - as can be seen in the
figures 2.25,
2.26, 2.27 and 2.28. Thus, along with the nearest neighbor spacing, skewness
and
excess parameters form ideal candidates for observing the GOE nature of the
fluctuations
in possible future experiments and ab - initio calculations. In all the data
on $\gamma_{1}(r)$ and $\gamma_{2}(r)$  we have included information up to
$r=5$ since both these functions become very small in absolute value beyond
\begin{figure}[htp]
\vskip+0.5cm
\epsfxsize=3in
\centerline{\epsfbox{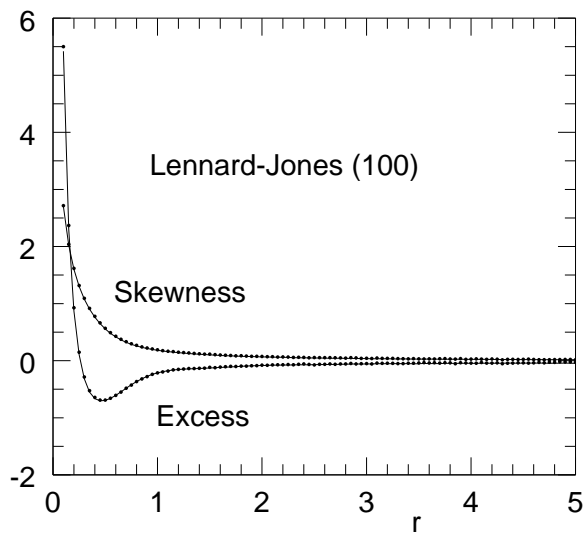}}
\caption{\sl Skewness and Excess parameters of the distribution of $n(r)$,
the number of levels in interval of length $r$,
plotted as a function of $r$ for Lennard-Jones potential with $N=100$.
Continuous lines: Predictions for the GOE.
Filled Circles: Our data.}
\label{fig225}
\end{figure}
       
\begin{figure}[htp]
\vskip+0.5cm
\epsfxsize=3in
\centerline{\epsfbox{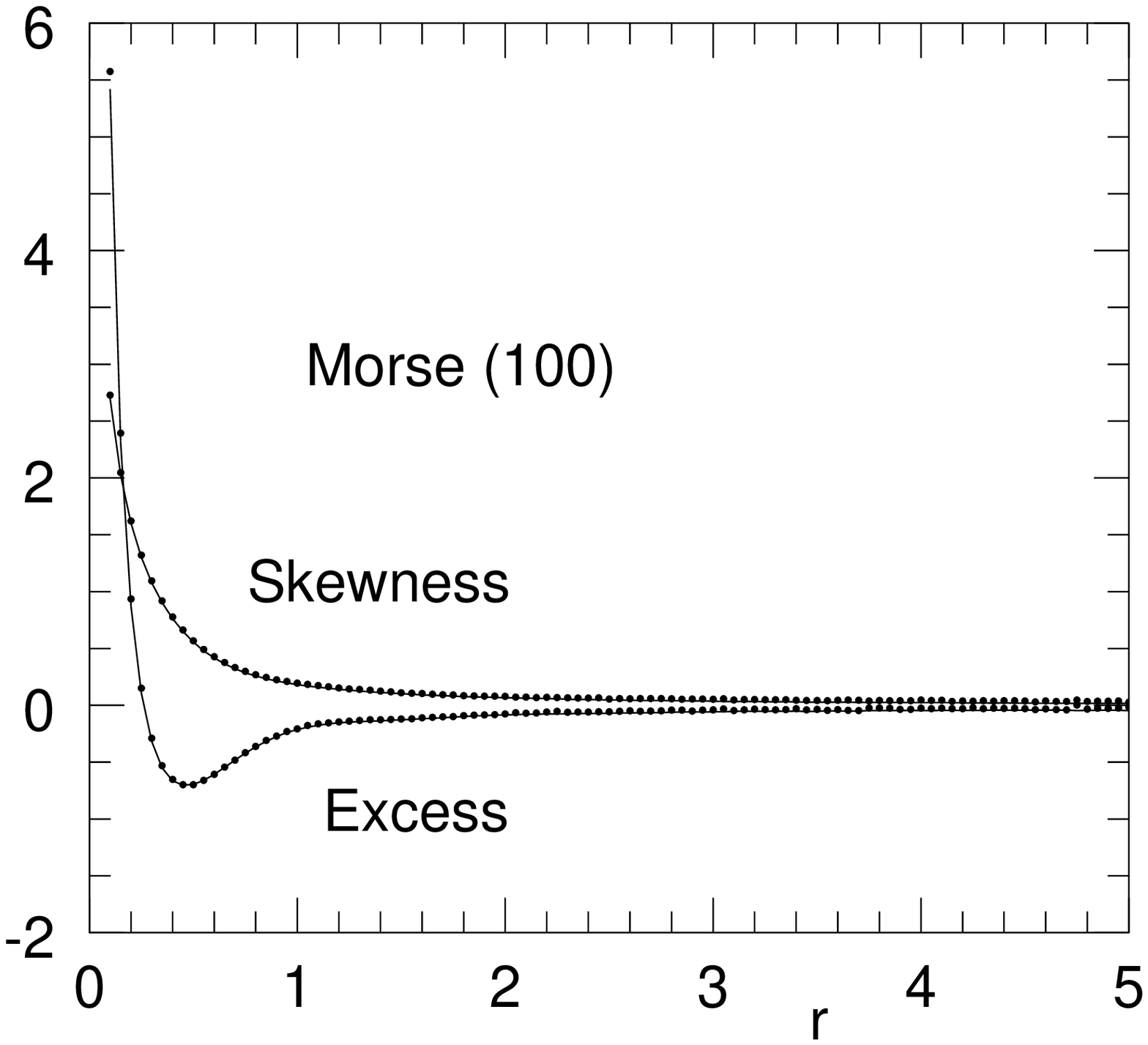}}
\caption{\sl Skewness and Excess parameters of the distribution of $n(r)$,
 the number of levels in interval of length $r$,
plotted as a function of $r$ for Morse potential with $N=100$.
Continuous lines: Predictions for the GOE.
Filled Circles: Our data.}
\label{fig226}
\end{figure}

\begin{figure}[htp]
\vskip+0.5cm
\epsfxsize=3in
\centerline{\epsfbox{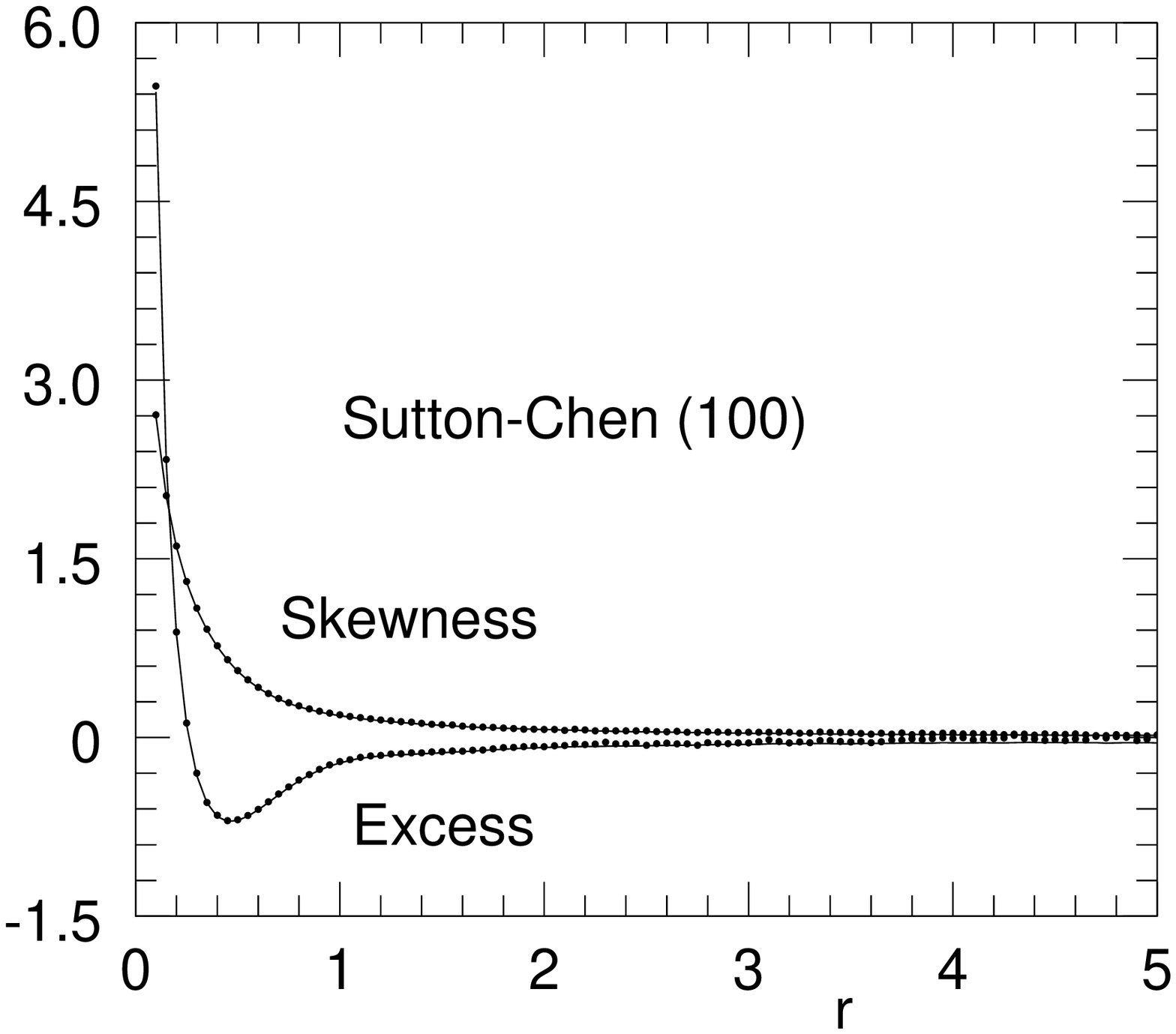}}
\caption{\sl Skewness and Excess parameters of the distribution of $n(r)$,
 the number of levels in interval of length $r$,
plotted as a function of $r$ for Sutton - Chen potential with $N=100$.
Continuous lines: Predictions for the GOE.
Filled Circles: Our data.}
\label{fig227}
\end{figure}

\begin{figure}[htp]
\vskip+0.5cm
\epsfxsize=3in
\centerline{\epsfbox{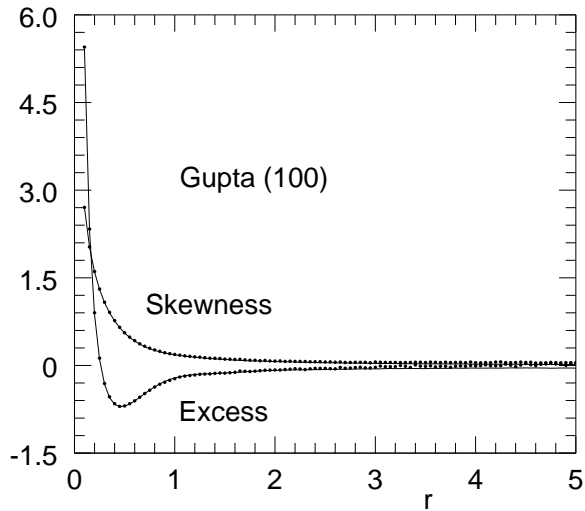}}
\caption{\sl Skewness and Excess parameters of the distribution of $n(r)$,
 the number of levels in interval of length $r$,
plotted as a function of $r$ for Gupta potential with $N=100$.
Continuous lines: Predictions for the GOE.
Filled Circles: Our data.}
\label{fig228}
\end{figure}
\newpage
this range. For $r \leq 5$, the agreement between the predictions for the GOE
and our calculations is very close - as can be seen in the figures 2.21 to
2.28.
As mentioned earlier, in the procedure adopted for the calculation of 
$\Sigma^{2}(r)$, $\gamma_{1}(r)$ and $\gamma_{2}(r)$, we have excluded
some regions
of spectra where the misfit function has irregular behavior. Although
the broad contours of the misfit function are more or less same for all
the spectra with a given potential and a given number of particles,
the exact locations of these irregular regions do vary somewhat from
spectrum to spectrum. However, since we are dealing with a very large
number of spectra (of the order of 1000 in some cases) and the exclusion of
the irregular regions is done only through visual inspection, we have actually
selected
the same groups of frequencies for all the spectra after inspecting only a few
of them. This process is somewhat subjective (and improper, strictly speaking)
and causes some degradation in the quality
of unfolding. All the data on $\Sigma^{2}(r)$ that have been presented
in figures 2.17, 2.18, 2.19 and 2.20 suffer from this limitation.
However, in the cases of Lennard-Jones potential and Morse potential 
with 2000 particles (the number of spectra being 262 and 49, respectively)
we have examined the misfit function for each spectrum separately
and excluded the irregular regions accordingly.
The results of this analysis for $\Sigma^{2}(r)$ are shown in 
figures 2.29 and 2.30.
\begin{figure}[htp]
\vskip+2cm
\epsfxsize=5.0in
\centerline{\epsfbox{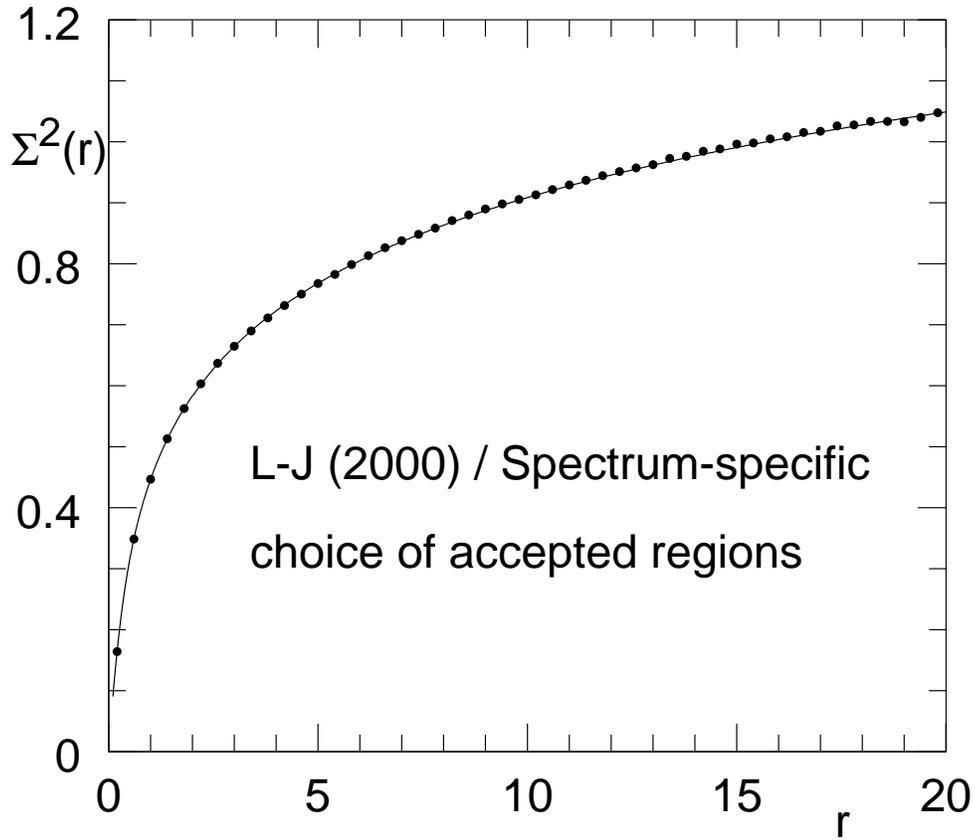}}
\caption{\sl Variance of the number of levels in interval of length
$r$ plotted as a function of $r$ for Lennard-Jones potential with $N$ = 2000
and with
spectrum-specific choice of accepted regions. Continuous line: Prediction for the
GOE. Filled Circles: Our data.}
\label{fig229}
\end{figure}

\begin{figure}[htp]
\vskip+2cm
\epsfxsize=6.0in
\centerline{\epsfbox{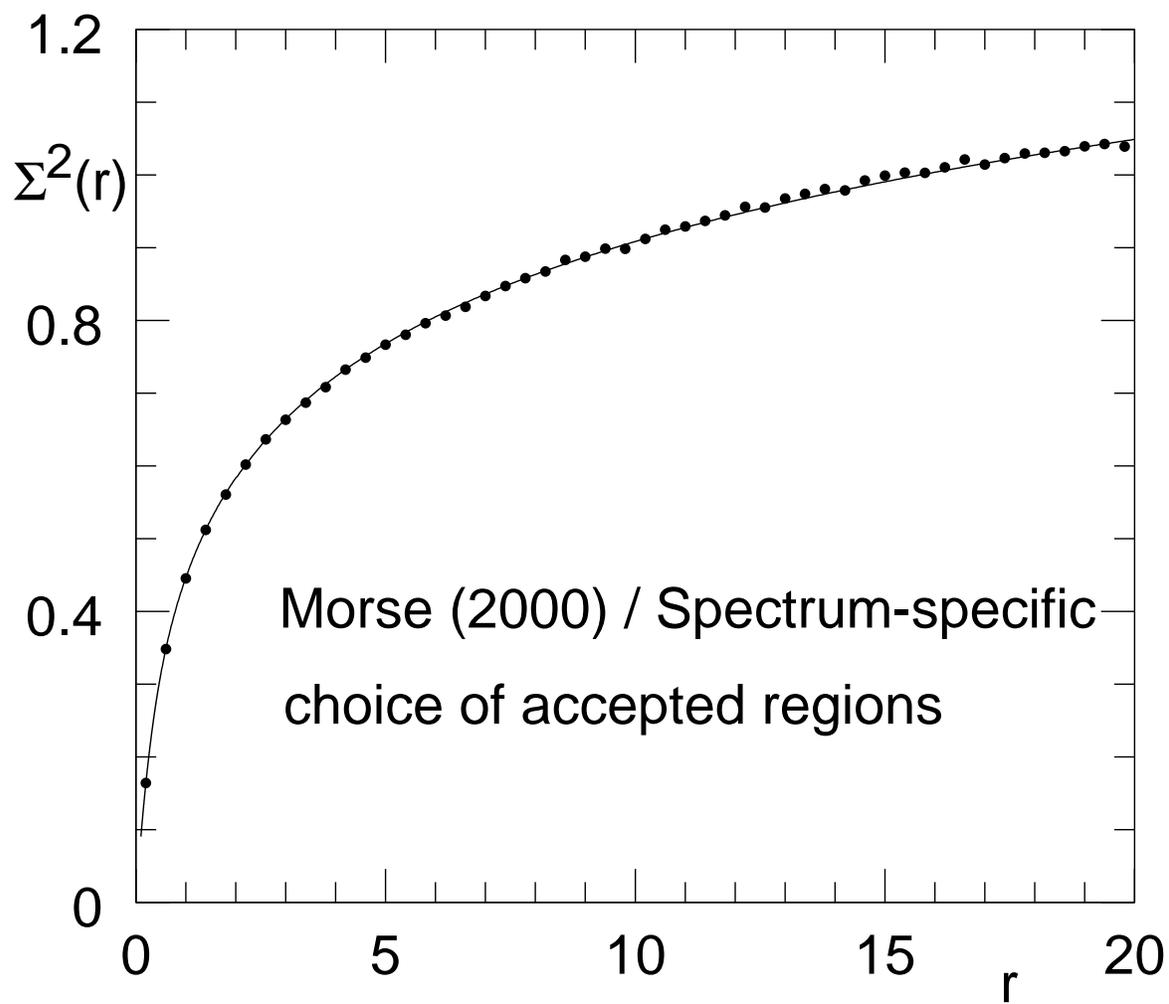}}
\caption{\sl Variance of the number of levels in interval of length
$r$ plotted as a function of $r$ for Morse potential with $N$ = 2000 and with
spectrum-specific choice of accepted regions. Continuous line: Prediction for the
GOE. Filled Circles: Our data.}
\label{fig230}
\end{figure}
One can see that now  there is essentially overlapping agreement with
the predictions of GOE all the way up to $r=20$. This is to be compared
with the data in figures 2.17 and 2.18.

\newpage
{ \huge \underline{Regions I and III}}

In this section  we present the results of the analysis of 
statistical fluctuations for the top 20\% (region III) and the 
bottom 10\% (region I) of the spectra. This is done only for
Lennard-Jones and Morse potentials. For the other two potentials
the number of eigenvalues in either of these two regions is too
small to permit any statistically significant analysis -- even for 
the largest system sizes used. For the Lennard-Jones and Morse cases
the number of levels in these two regions are reasonably high for
the largest system size of $N = 2000$. The reasons why we perform
the analysis separately for these two regions are two-fold: \newline

(1) As we have seen earlier, analysis of fluctuations is preceded
by the construction of an analytical unfolding function. However,
we are unable to construct such a function for the {\it entire} spectrum
in one go.

(2) We have calculated the participation ratios of the eigenmodes and 
the values indicate that the modes are extended in a large central part
of the spectrum overlapping region II. However, towards the top of the 
spectra, the eigenmodes are more and more localized and they are 
completely localized at the top of region III. 
In the region I  the eigenmodes seem to be a mixture of extended and
localized type. Since, from literature, we expect a connection between
the nature of spectral fluctuations and the localization characteristics
of the eigenmodes, it is better to analyze these regions separately so 
as not to mix up regions with possibly different types of spectral 
fluctuations [32,66]. 

As before, we have to make a proper choice of the unfolding function
that fits the cumulative density of states data sufficiently closely.
We choose this fitting function to be a quadratic function in region I
since the range of the fit is rather limited. For region III, the 
unfolding function has the same form as in region II but with different
values of parameters. We remove a further 5\% of the spectra from the
top of region III and 1\% from the bottom of region I in order to make
the fit sufficiently close. 

Figures 2.31, 2.32, 2.33 and 2.34 show the distribution of the normalized
nearest neighbor spacing for the regions I and III with the two
different potentials. Also shown in each figure are the two types of
theoretical
predictions. Inspection of these figures show that the closeness
of the numerically computed nearest neighbor spacing distribution
to the exact prediction for the GOE is essentially as good as it
is for region II. However, on account of the lower level of statistics
for regions I and III, scatter of the data around the prediction
is somewhat higher than it is for the region II. Thus the deviation
from the GOE statistics ,if any, is certainly weak even in those sections
of the eigenvalue spectra where localization is much more pronounced.
To summarize the results for the spectrum as a whole, any departure from
the GOE statistics much be limited to a small fraction of levels at the
two ends of the spectrum (five percent at the top and one percent at the
bottom). These regions, even for the maximum system sizes we have used,
contain such a small number of levels that it is not presently possible
to perform an independent study of the spectral fluctuations.
\begin{figure}[htp]
\epsfxsize=2.7in
\centerline{\epsfbox{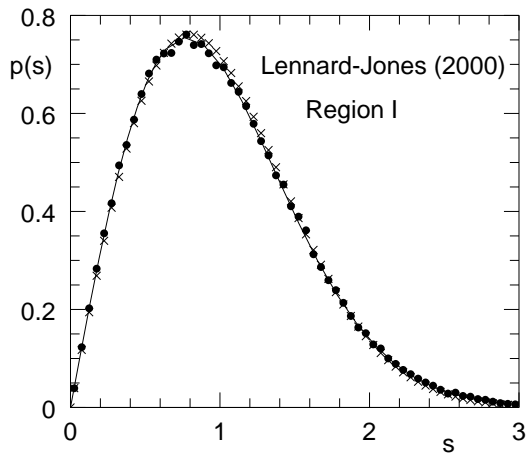}}
\caption{\sl Probability density [$p(s)$] for normalized nearest 
neighbor spacing $(s)$, obtained with Lennard-Jones potential
(N = 2000) in region I of the spectrum. Filled Circles: Our data.
Crosses: Wigner's surmise for GOE. Continuous line: Exact prediction
for the GOE.}
\label{fig231}
\end{figure}

\begin{figure}[htp]
\epsfxsize=2.7in
\centerline{\epsfbox{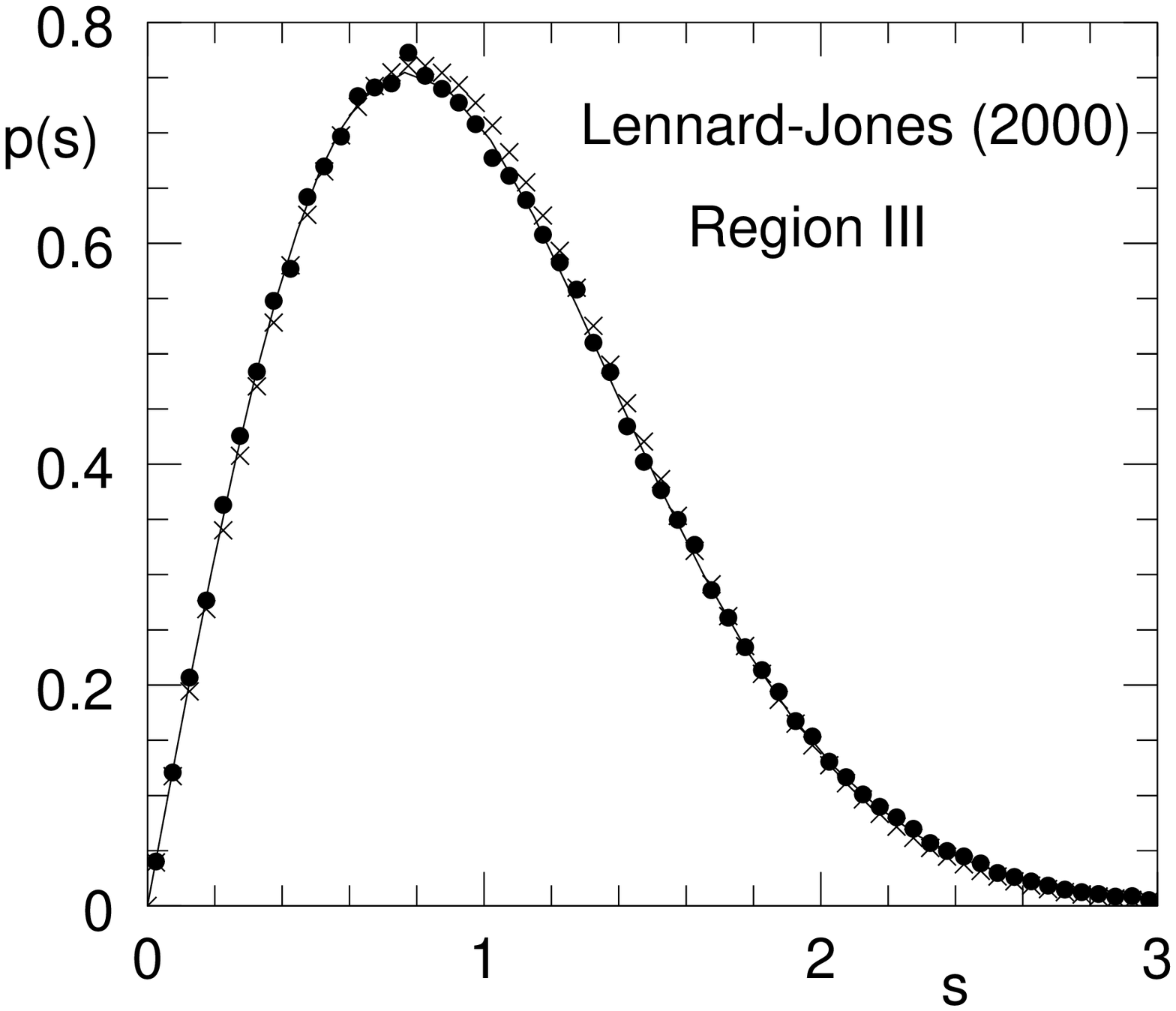}}
\caption{\sl Probability density [$p(s)$] for normalized nearest 
neighbor spacing $(s)$, obtained with Lennard-Jones potential
(N = 2000) in region III of the spectrum. Filled Circles: Our data.
Crosses: Wigner's surmise for GOE. Continuous line: Exact prediction
for the GOE.}
\label{fig232}
\end{figure}
\vspace{0.5in}
\begin{figure}[htp]
\epsfxsize=2.7in
\centerline{\epsfbox{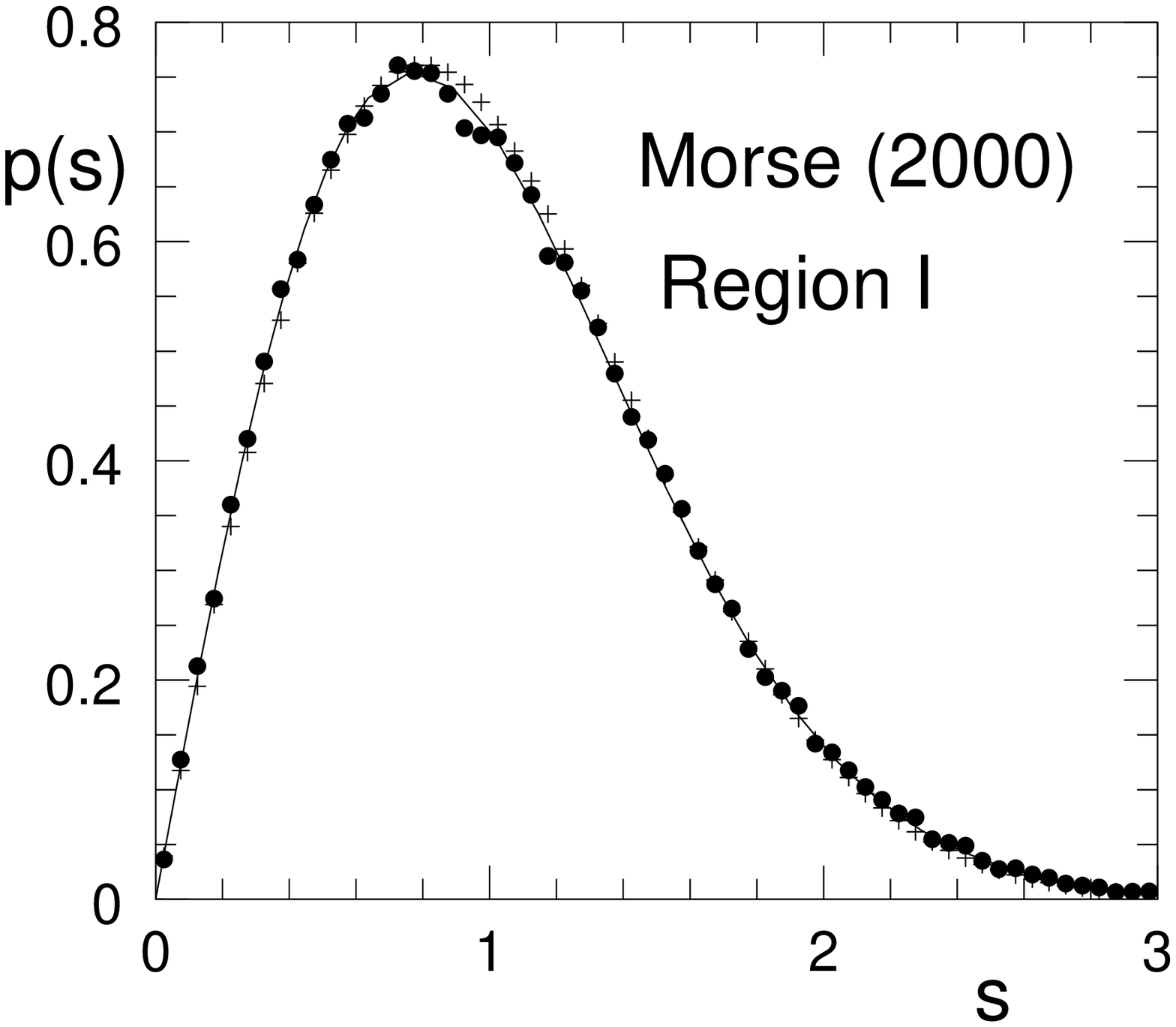}}
\caption{\sl Probability density [$p(s)$] for normalized nearest 
neighbor spacing $(s)$, obtained with Morse potential
(N = 2000) in region I of the spectrum. Filled Circles: Our data.
Crosses: Wigner's surmise for GOE. Continuous line: Exact prediction
for the GOE.}
\label{fig233}
\end{figure}

\begin{figure}[htp]
\vskip+0.5cm
\epsfxsize=3in
\centerline{\epsfbox{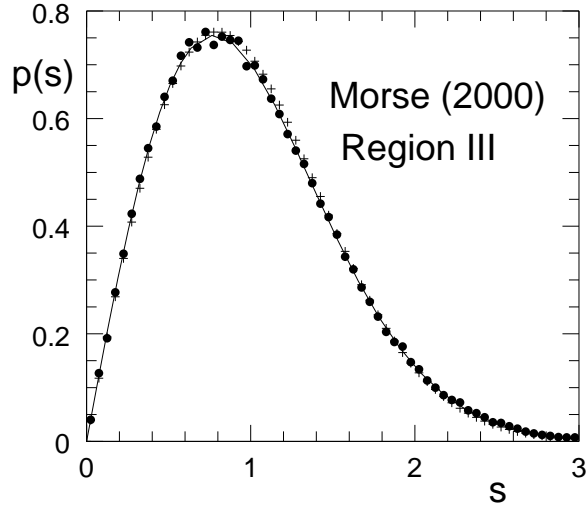}}
\caption{\sl Probability density [$p(s)$] for normalized nearest 
neighbor spacing $(s)$, obtained with Morse potential
(N = 2000) in region III of the spectrum. Filled Circles: Our data.
Crosses: Wigner's surmise for GOE. Continuous line: Exact prediction
for the GOE.}
\label{fig234}
\end{figure}

\vspace{2in}
Data for the variance $\Sigma^{2}(r)$ for the two regions and with 
the two potentials are presented in figures 2.35, 2.36, 2.37 and 2.38.
It is immediately obvious that the deviation of the numerically computed
values of the variance for the different window lengths from the 
exact predictions for the GOE is significantly higher now as compared 
to region II -- especially when higher values of $r$ are considered.
The probable reasons for this higher deviation are:\newline
(1) The quality of the fit of the unfolding function.\newline
(2) The possibility that there is a presence of Poissonian statistics
due to higher level of localization in regions I and III which are 
located at the two ends of the spectra.\newline
It is difficult to make a clear statement on the relative importance
of these two probable causes. However, since the agreement of the
nearest neighbor spacing distribution to the prediction for the GOE
is so close, it seems reasonable to conclude that it is the
quality of the unfolding that is the primary reason for the 
deviation of variance from the prediction for the GOE. In arriving at
this conclusion  we have kept in mind the fact that the nearest neighbor
spacing distribution is much less sensitive to the quality of the process of
unfolding.

\begin{figure}[htp]
\vskip+0.5cm
\epsfxsize=3in
\centerline{\epsfbox{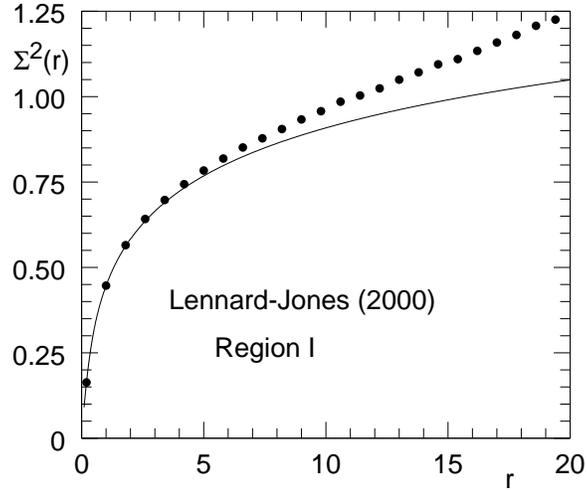}}
\caption{\sl Variance of the number of levels in intervals of
length $r$ plotted as a function of $r$ for region I of the spectrum
obtained with Lennard-Jones potential $(N = 2000)$. Continuous line:
Prediction for the GOE. Filled circles: Our data}
\label{fig235}
\end{figure}

\begin{figure}[htp]
\vskip+0.5cm
\epsfxsize=3in
\centerline{\epsfbox{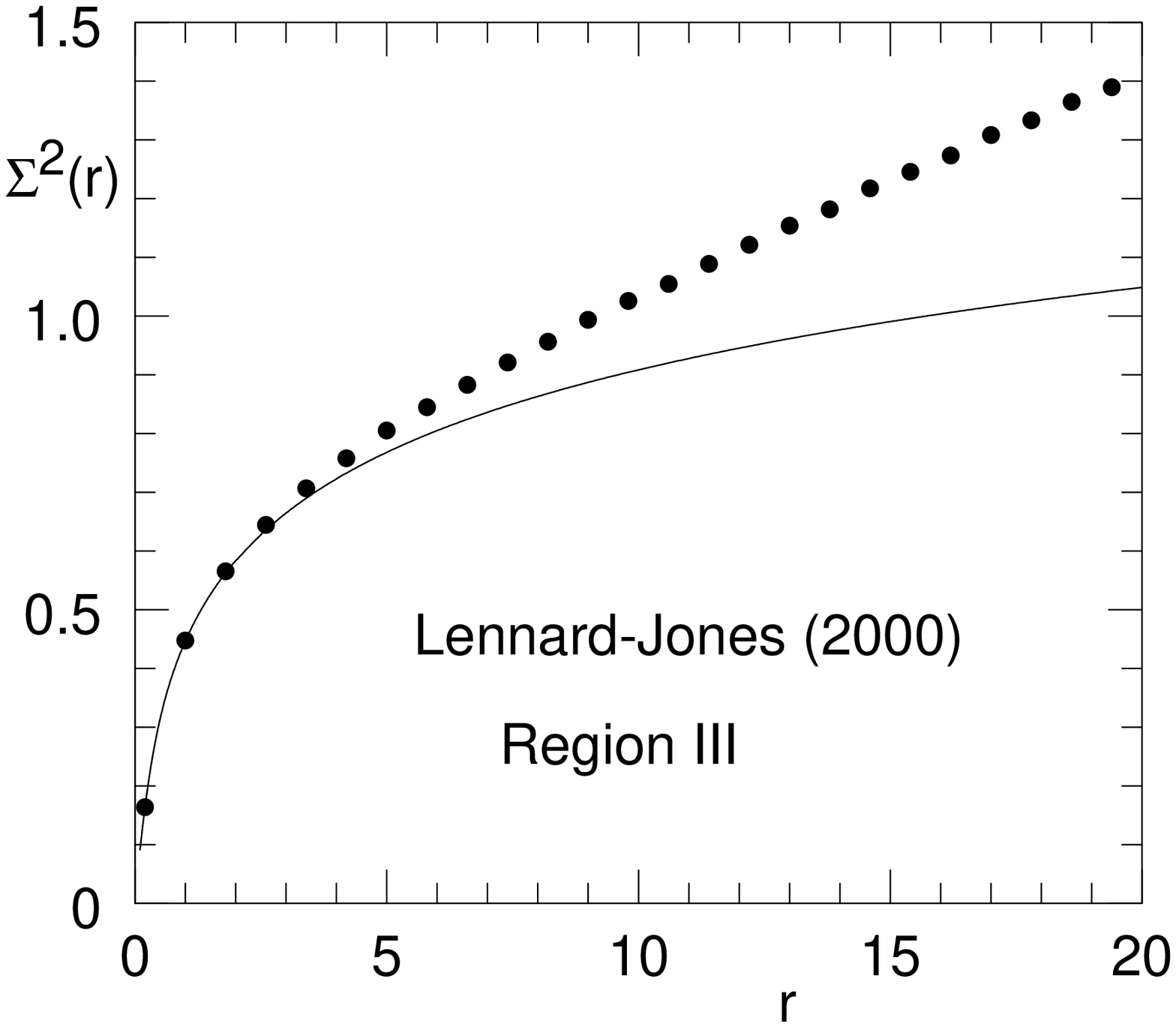}}
\caption{\sl Variance of the number of levels in intervals of
length $r$ plotted as a function of $r$ for region III of the spectrum
obtained with Lennard-Jones potential $(N = 2000)$. Continuous line:
Prediction for the GOE. Filled circles: Our data}
\label{fig236}
\end{figure}

\begin{figure}[htp]
\vskip+0.5cm
\epsfxsize=3in
\centerline{\epsfbox{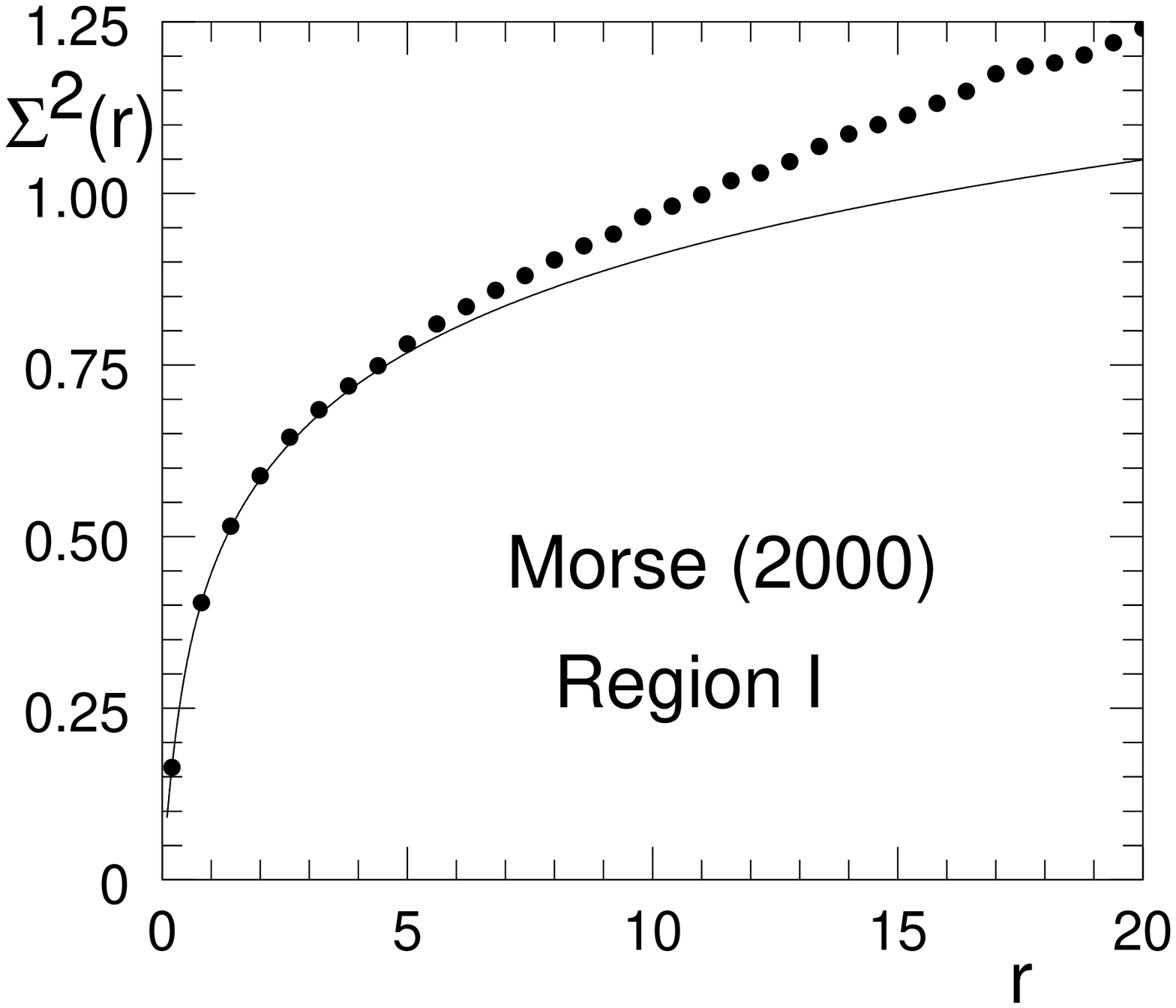}}
\caption{\sl Variance of the number of levels in intervals of
length $r$ plotted as a function of $r$ for region I of the spectrum
obtained with Morse potential $(N = 2000)$. Continuous line:
Prediction for the GOE. Filled circles: Our data}
\label{fig237}
\end{figure}

\begin{figure}[htp]
\vskip+0.5cm
\epsfxsize=3in
\centerline{\epsfbox{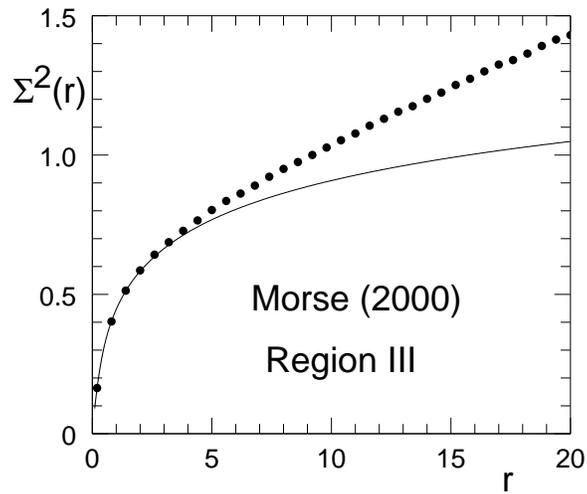}}
\caption{\sl Variance of the number of levels in intervals of
length $r$ plotted as a function of $r$ for region III of the spectrum
obtained with Morse potential $(N = 2000)$. Continuous line:
Prediction for the GOE. Filled circles: Our data}
\label{fig238}
\end{figure}

Finally, in figures 2.39, 2.40, 2.41 and 2.42  we display the plots of 
skewness and excess parameters as a function of window length for 
the two potentials . Also superimposed are the predictions for the 
GOE. The closeness of agreement are quite apparent in the figures.

\begin{figure}[htp]
\vskip+3cm
\epsfxsize=4.5in
\centerline{\epsfbox{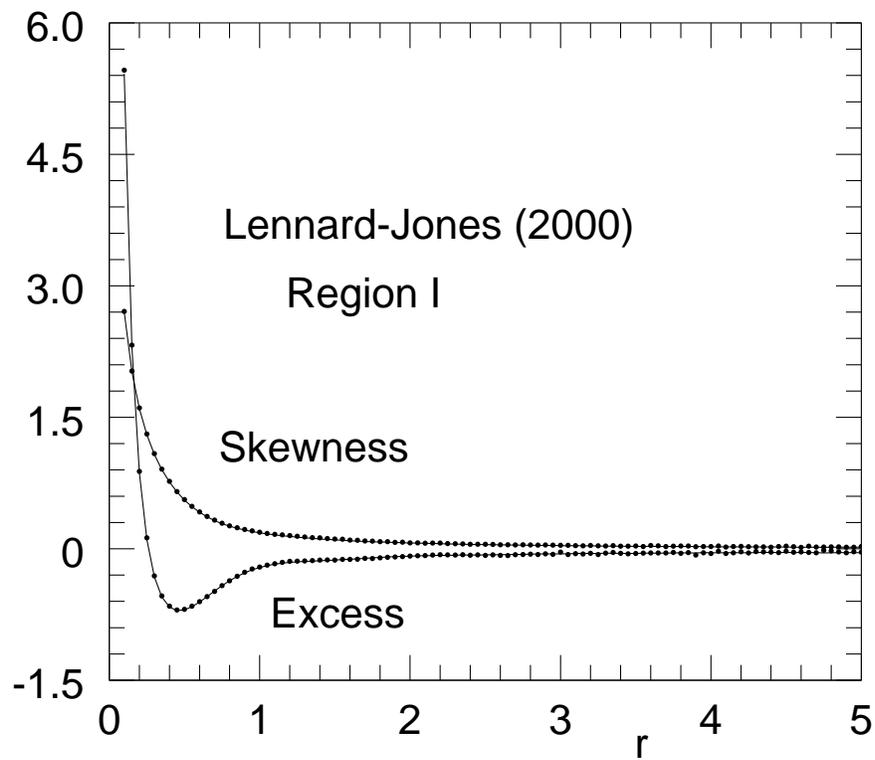}}
\caption{\sl Skewness and Excess parameters of the distribution of 
$n(r)$, the number of levels in interval of length $r$, plotted as 
a function of $r$ for region I of the spectra
obtained with Lennard-Jones potential $(N = 2000)$. Continuous lines:
Predictions for the GOE. Filled circles: Our data}
\label{fig239}
\end{figure}

\begin{figure}[htp]
\vskip+0.5cm
\epsfxsize=3in
\centerline{\epsfbox{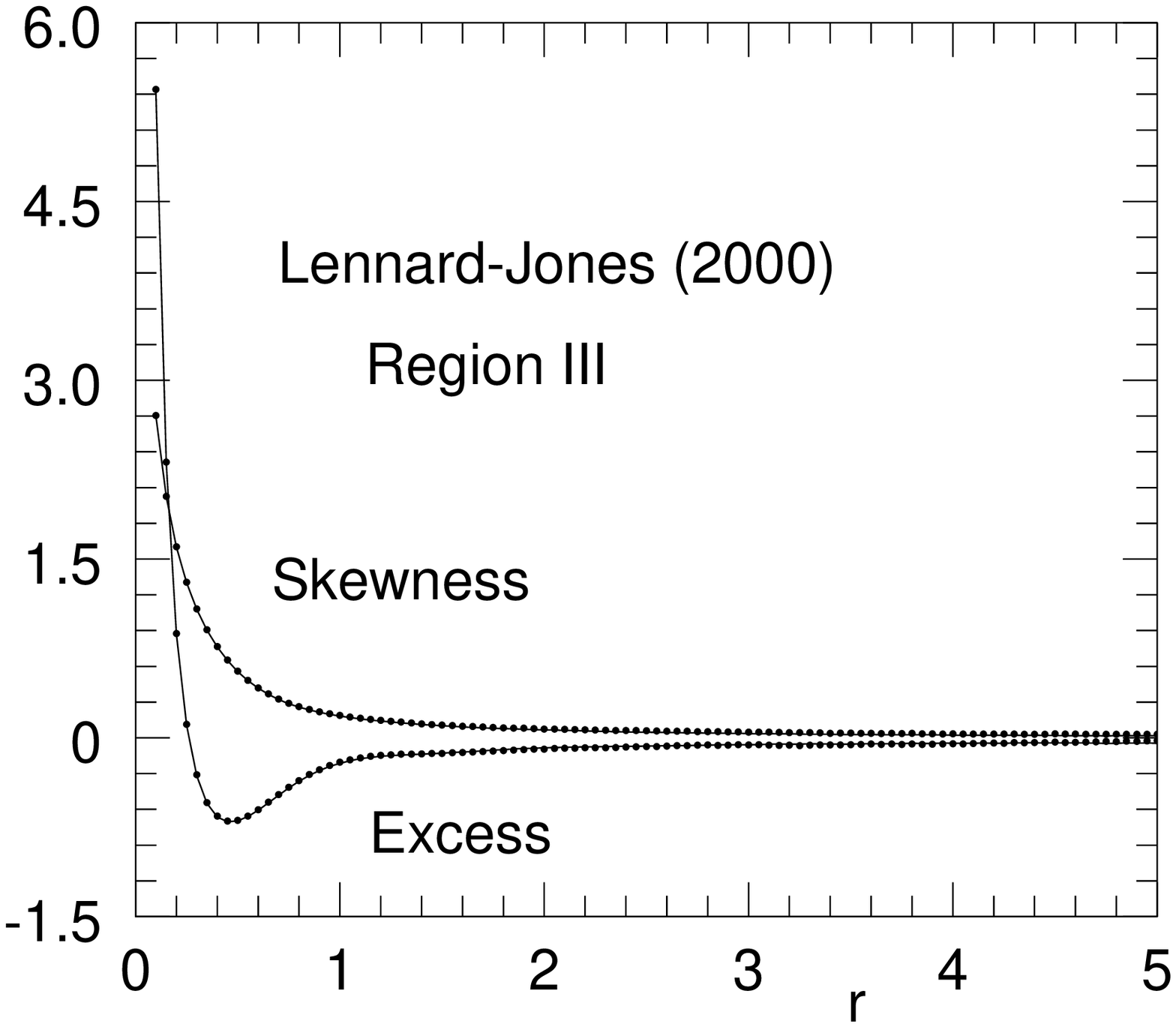}}
\caption{\sl Skewness and Excess parameters of the distribution of 
$n(r)$, the number of levels in interval of length $r$, plotted as 
a function of $r$ for region III of the spectra
obtained with Lennard-Jones potential $(N = 2000)$. Continuous lines:
Predictions for the GOE. Filled circles: Our data}
\label{fig240}
\end{figure}

\begin{figure}[htp]
\vskip+0.5cm
\epsfxsize=3in
\centerline{\epsfbox{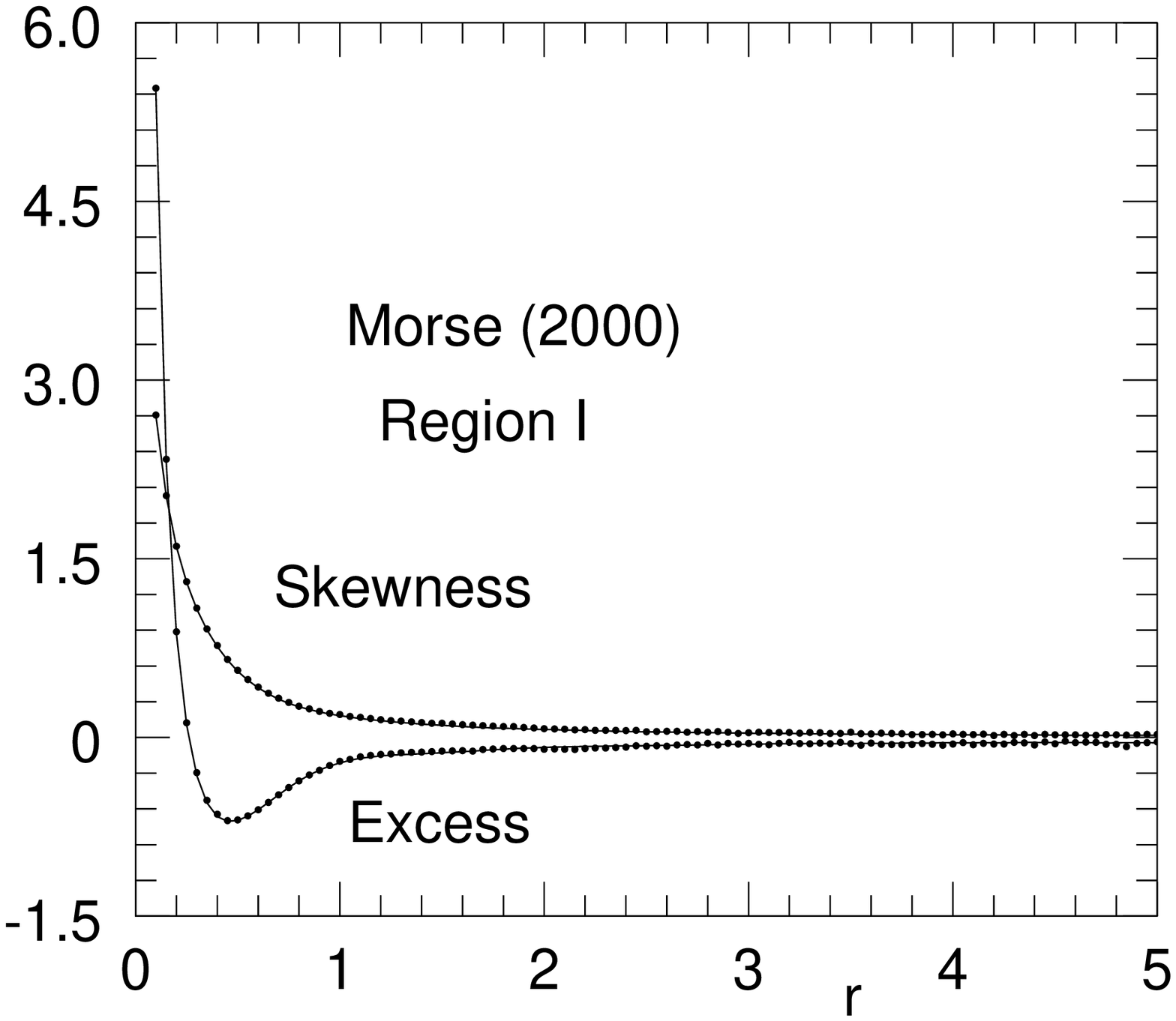}}
\caption{\sl Skewness and Excess parameters of the distribution of 
$n(r)$, the number of levels in interval of length $r$, plotted as 
a function of $r$ for region I of the spectra
obtained with Morse potential $(N = 2000)$. Continuous lines:
Predictions for the GOE. Filled circles: Our data}
\label{fig241}
\end{figure}

\newpage

\begin{figure}[htp]
\vskip+0.5cm
\epsfxsize=3in
\centerline{\epsfbox{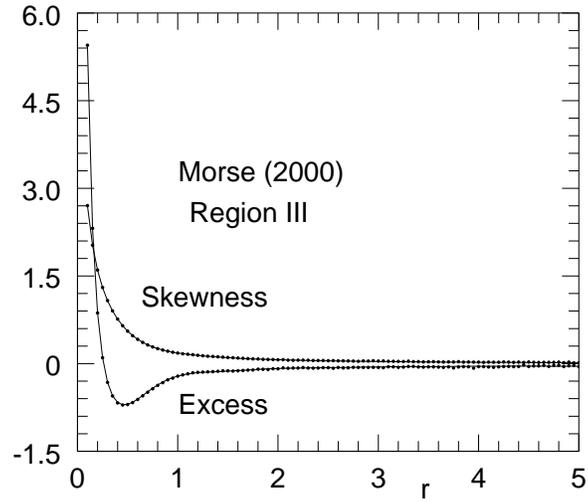}}
\caption{\sl Skewness and Excess parameters of the distribution of 
$n(r)$, the number of levels in interval of length $r$, plotted as 
a function of $r$ for region III of the spectra
obtained with Morse potential $(N = 2000)$. Continuous lines:
Predictions for the GOE. Filled circles: Our data}
\label{fig242}
\end{figure}

\chapter{Universality of Statistical Fluctuations : Binary Amorphous Clusters}

In this chapter  we continue the study of statistical fluctuations in the 
vibrational spectra of amorphous systems. But now  we consider systems that 
are made up of two different kinds of
constituent units. Studies in chapter 2 showed that for a large central
region of the vibrational spectra of single-component amorphous systems, the
integrated density of
states can be described very accurately by a functional form that does
not depend on the nature of the interaction. Also, the spectral fluctuations
turned out to be of the GOE type to a very high level of accuracy in
all cases. It is natural to ask whether these kinds of universal 
properties will hold even when the system is more complicated. The simplest 
situation to which one can extend these investigations is the one 
pertaining to binary systems. In situations where the potential
can be written as a sum over pairs, three functional forms are needed
for binary systems corresponding to the three distinct pairs of
constituent units that are possible. In our investigation  we take a
Lennard-Jones type of potential for every pair i.e. the potential
for the pair involving two units of type $P$ and $Q$ separated by a
distance $r$ is given by 4$\epsilon_{PQ}[(\sigma_{PQ}/r)^{12} - 
(\sigma_{PQ}/r)^{6}]$. Thus, the values of $\epsilon_{PQ}$ and
$\sigma_{PQ}$ for all the three pairs are needed to completely
specify the potential energy of the system for a given configuration.
We have studied situations with different ratios of the numbers of the
two different kinds of units, different system sizes and different 
rules for the construction of the Lennard-Jones interaction parameters
between the two species. Also the masses of the two types of units
are sometimes the same and sometimes different. Here we will present the
results for four different situations. Common to all of them are the
following: ( Denoting the two types of units as A and B) 
$\epsilon_{BB}/\epsilon_{AA} = 0.5$, $\sigma_{BB}/\sigma_{AA} = 0.88$
and $N_{A} + N_{B} = 2000$. Here  $N_{A}$ and $N_{B}$ denote the
numbers of units of type $A$ and $B$, respectively. The four cases
for which we present the results here are specified in {\bf Table 3.1}
in which $m_{A}$ and $m_{B}$ denote the masses of the two types of 
the units.
\begin{center}
{\bf Table 3.1}
\end{center}
\begin{center}
\begin{tabular}{|c|c|c|c|c|} \hline

&{$\bf N_{A}/N_{B}$}
&{$\bf \epsilon_{AB}/\epsilon_{AA}$} 
&{$\bf \sigma_{AB}/\sigma_{AA}$}
&{$\bf m_{A}/m_{B}$}\\ \hline

{Case I}
&{1}
&{$\sqrt{\epsilon_{BB}/\epsilon_{AA}}$}
&{$[1+(\sigma_{BB}/\sigma_{AA})]/2$}
&{2/3}\\ \hline
{Case II}
&{4}
&{1.5}
&{0.8}
&{2/3}\\ \hline
{Case III}
&{1}
&{$\sqrt{\epsilon_{BB}/\epsilon_{AA}}$}
&{$[1+(\sigma_{BB}/\sigma_{AA})]/2$}
&{1}\\ \hline
{Case IV}
&{4}
&{1.5}
&{0.8}
&{1}\\ \hline
\end{tabular}
\end{center}

It may be noted that in cases I and III, the parameters of the 
Lennard-Jones potential for the pair with nonidentical units are the same
and are constructed via the {\bf Lorentz - Berthelot} rule. However,
the ratios of the masses are different in the two cases. In cases II 
and IV, the parameters for the nonidentical pair correspond to a 
choice that has been used extensively in the literature [9-10,12-13]. Here
again, the only difference between the cases II and IV comes from the mass ratio.

While generating the random initial configuration for the process
of minimization, we assign a type ($A$ or $B$) to each particle 
and that completes the definition of the potential. After this
the process of generating the local minima is the same as that for a 
single-component cluster. The construction of the eigenvalue
problem is slightly more complicated when the masses of the two
units are not identical.

Once the spectra for the various local minima are generated, the
rest of the analysis is essentially the same as in the case of a
single-component cluster. For example, we find that the functional form
$D(\lambda) = a- b\exp(-c\lambda)$ fits the cumulative density of
states to a very high degree of accuracy in region II.
The maximum absolute value of the misfit function stays at the level
of 1.5\% or less of the range of the fit. Here again, we improve the 
process of unfolding by adding a quadratic correction to the $D(\lambda)$
function for various subdomains of each spectrum. Since our primary
interest here is to investigate whether the universal properties 
observed for single-component clusters are also present for the
binary systems, we present results only for the region II and ignore the
other parts of the spectrum (i.e. region I and region III).

Figures 3.1 to 3.12 show the data for the nearest neighbor spacing,
variance and skewness ( plus excess) for the four cases. 

\begin{figure}[htp]
\vskip+0.5cm
\epsfxsize=7.5in
\centerline{\epsfbox{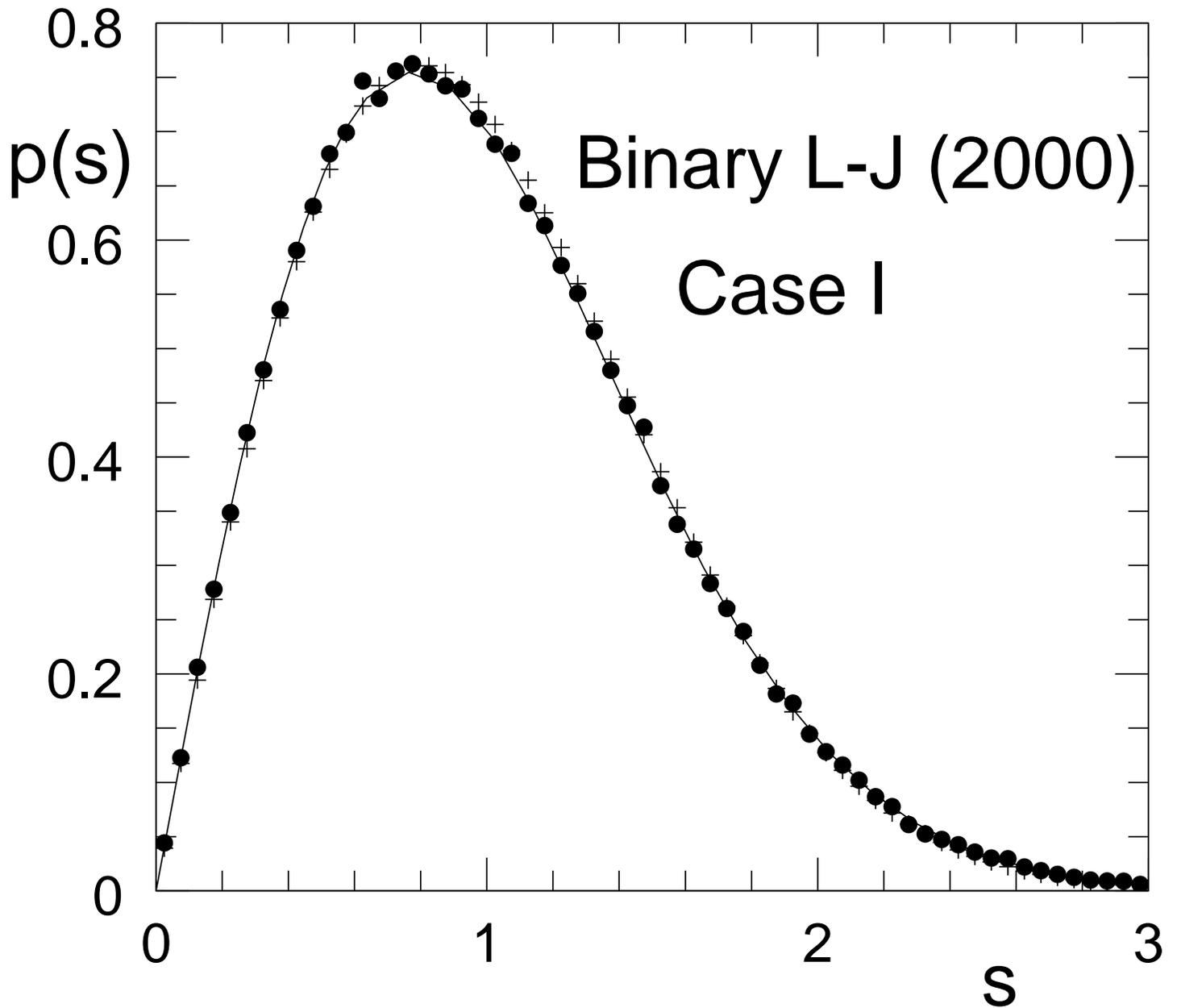}}
\caption{\sl Probability density [$p(s)$] for normalized nearest neighbor
spacing ($s$) for binary Lennard-Jones system ( Case I ).
Filled circles: Our data. 
Crosses: Wigner's surmise for GOE.
Continuous line: Exact prediction for the GOE.}
\label{fig301}
\end{figure}

\begin{figure}[htp]
\vskip+0.5cm
\epsfxsize=7.5in
\centerline{\epsfbox{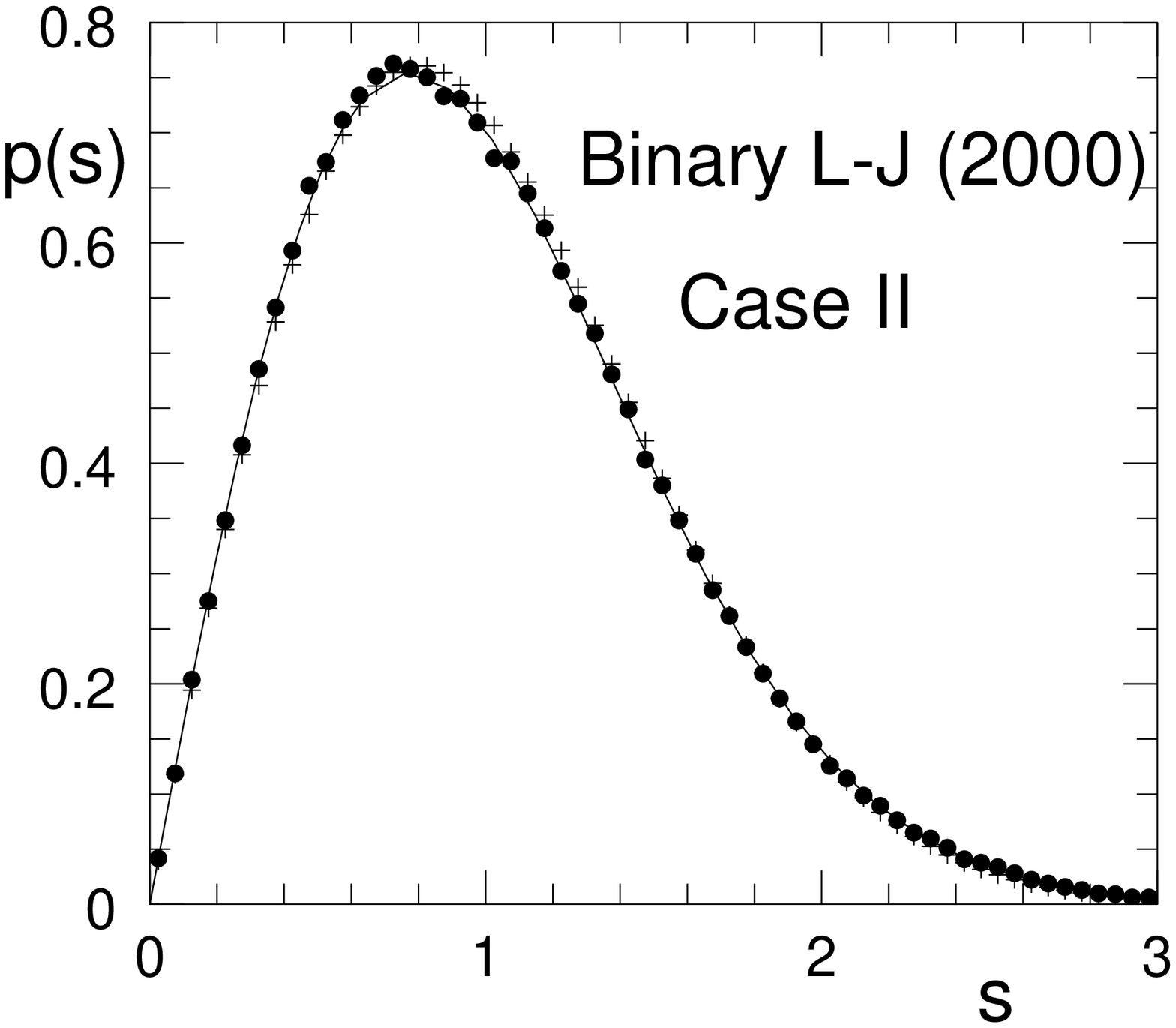}}
\caption{\sl Probability density [$p(s)$] for normalized nearest neighbor
spacing ($s$) for binary Lennard-Jones system ( Case II ).
Filled circles: Our data.
Crosses: Wigner's surmise for GOE.
Continuous line: Exact prediction for the GOE.}
\label{fig302}
\end{figure}

\begin{figure}[htp]
\vskip+0.5cm
\epsfxsize=7.5in
\centerline{\epsfbox{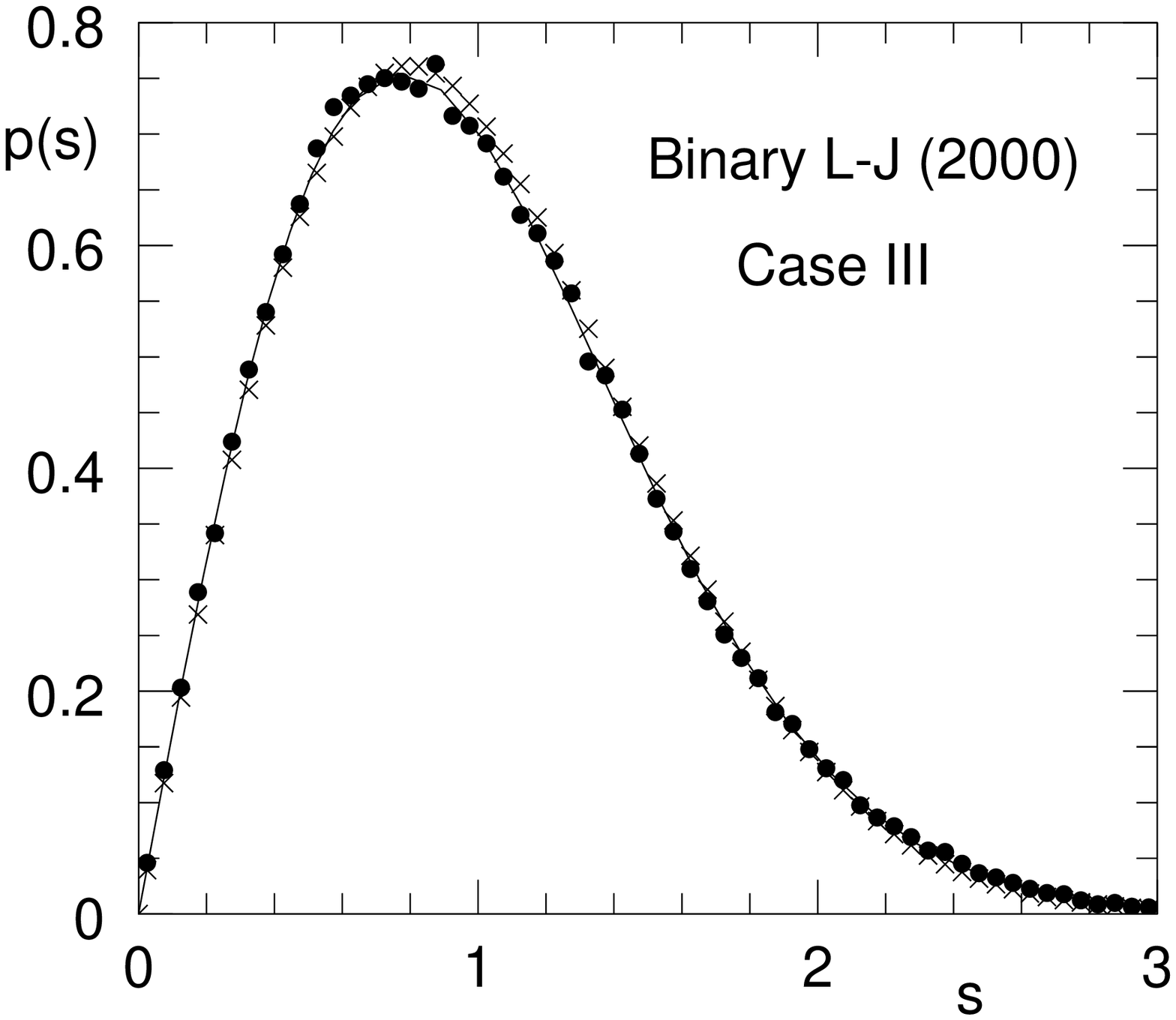}}
\caption{\sl Probability density [$p(s)$] for normalized nearest neighbor
spacing ($s$) for binary Lennard-Jones system ( Case III ).
Filled circles: Our data.
Crosses: Wigner's surmise for GOE.
Continuous line: Exact prediction for the GOE.}
\label{fig303}
\end{figure}

\begin{figure}[htp]
\vskip+0.5cm
\epsfxsize=7.5in
\centerline{\epsfbox{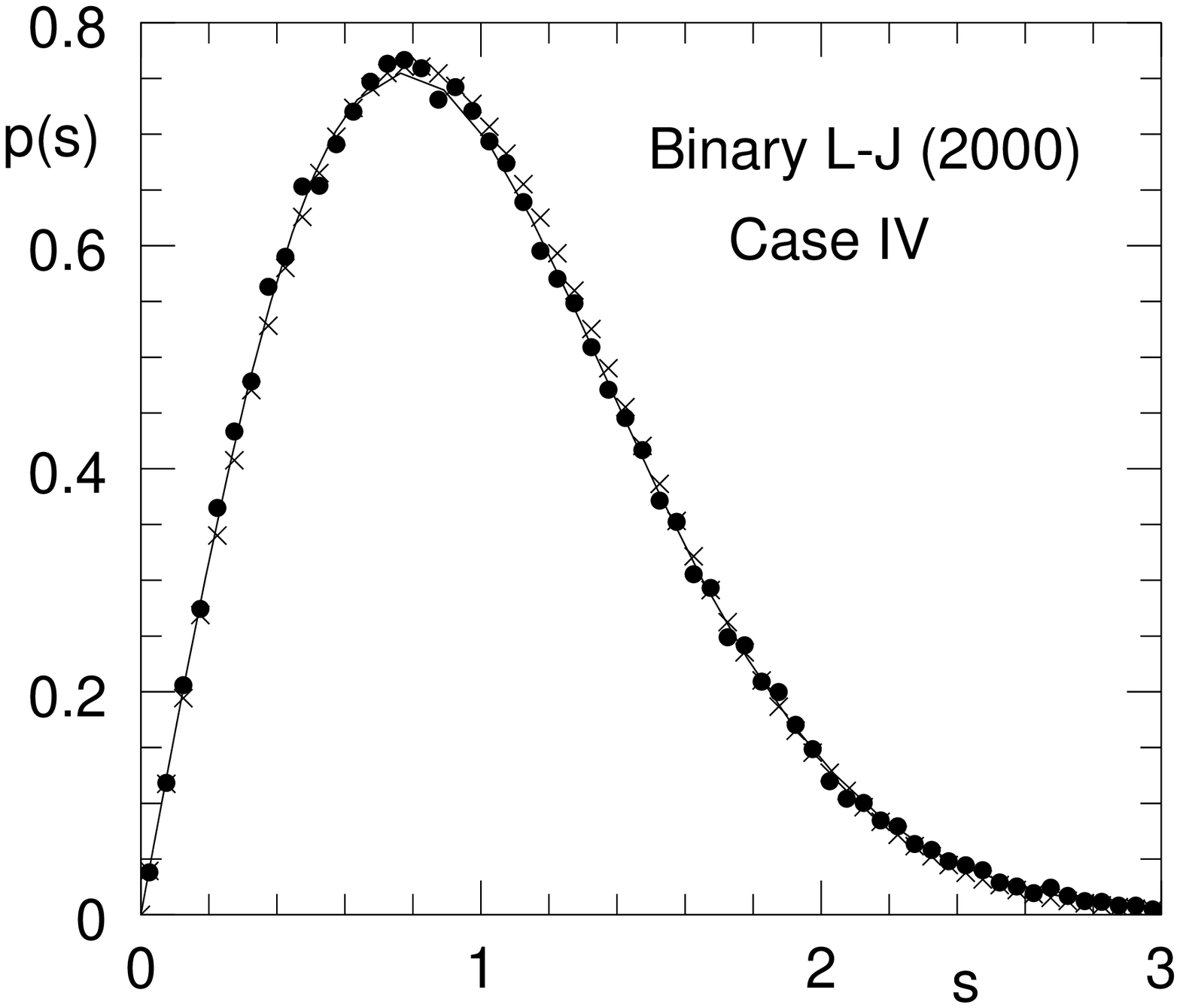}}
\caption{\sl Probability density [$p(s)$] for normalized nearest neighbor
spacing ($s$) for binary Lennard-Jones system ( Case IV ).
Filled circles: Our data.
Crosses: Wigner's surmise for GOE.
Continuous line: Exact prediction for the GOE.}
\label{fig304}
\end{figure}

\begin{figure}[htp]
\vskip+0.5cm
\epsfxsize=3in
\centerline{\epsfbox{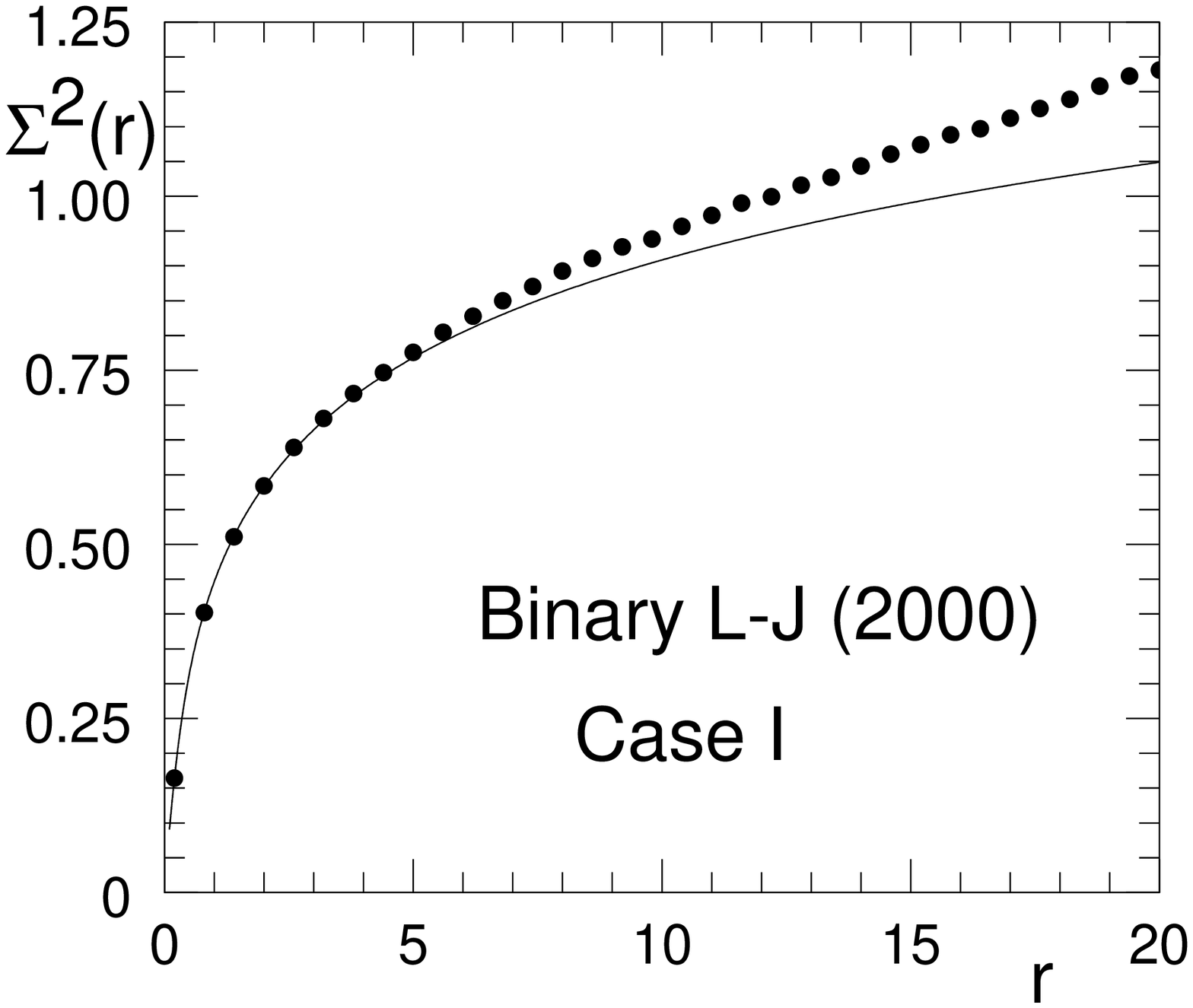}}
\caption{\sl Variance of the number of levels in intervals of
length $r$ plotted as a function of $r$ for the spectra
of binary Lennard-Jones system ( Case I ). Continuous line:
Prediction for the GOE. Filled circles: Our data}
\label{fig305}
\end{figure}

\begin{figure}[htp]
\vskip+0.5cm
\epsfxsize=3in
\centerline{\epsfbox{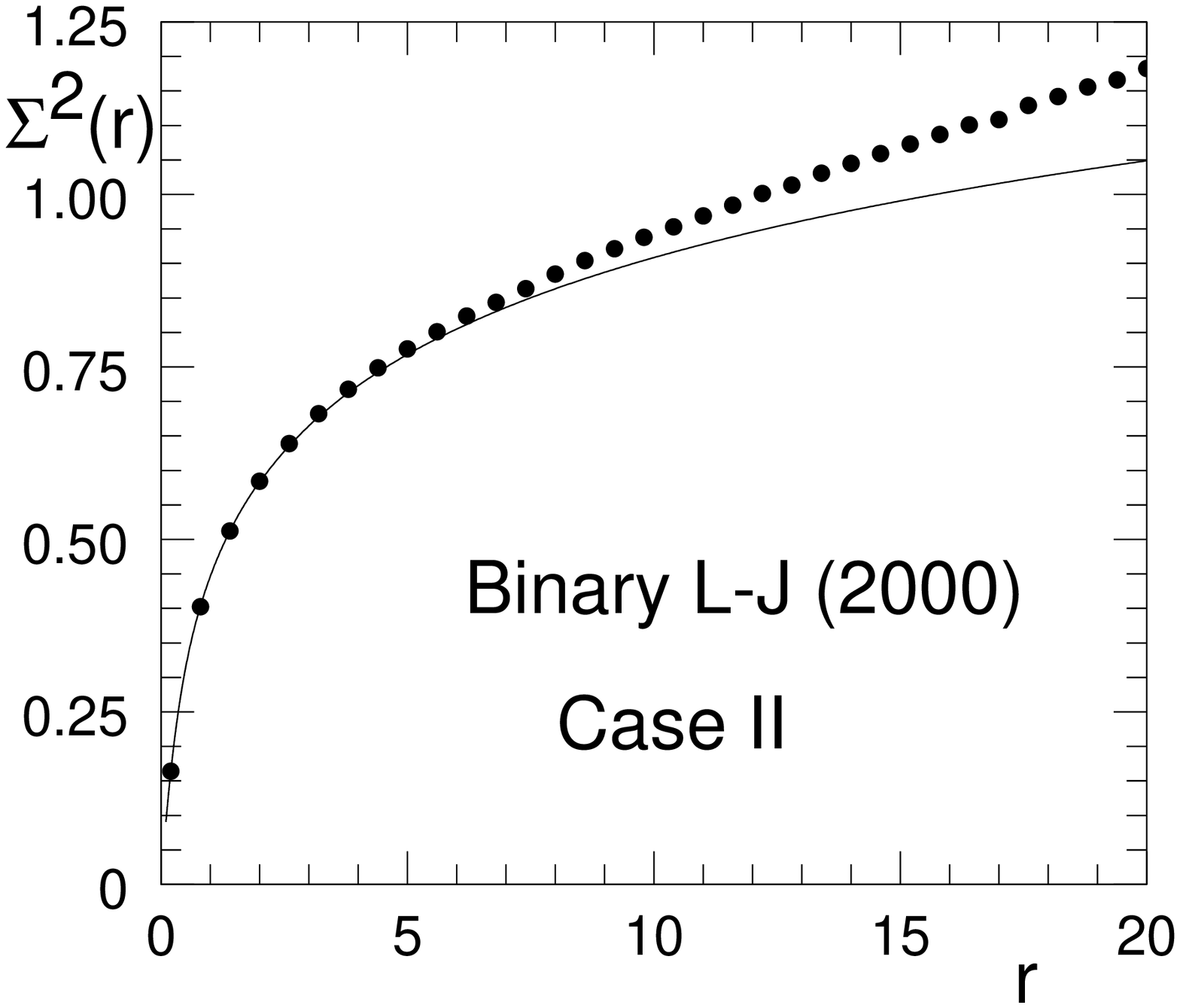}}
\caption{\sl Variance of the number of levels in intervals of
length $r$ plotted as a function of $r$ for the spectra
of binary Lennard-Jones system ( Case II ). Continuous line:
Prediction for the GOE. Filled circles: Our data}
\label{fig306}
\end{figure}

\begin{figure}[htp]
\vskip+0.5cm
\epsfxsize=3in
\centerline{\epsfbox{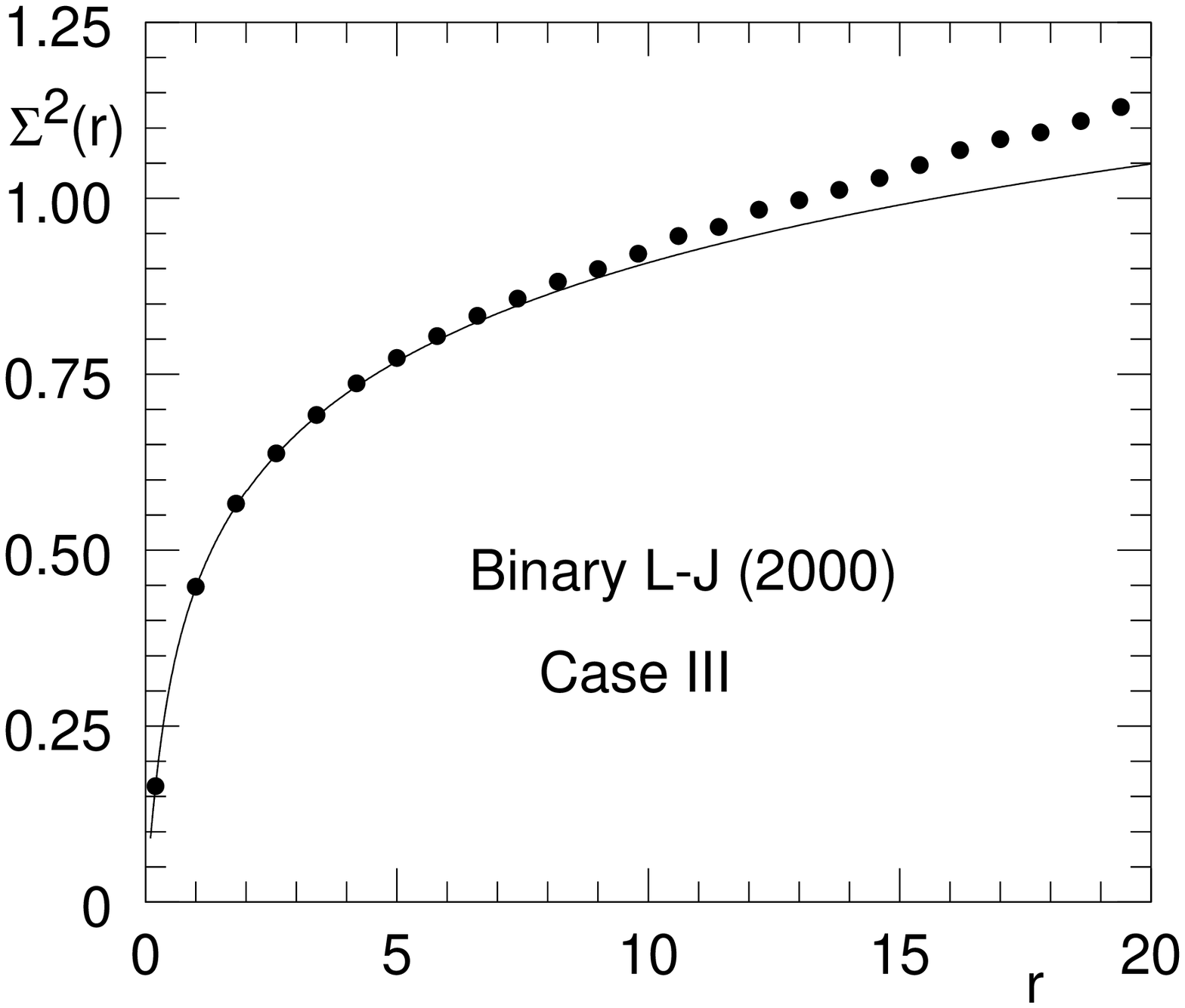}}
\caption{\sl Variance of the number of levels in intervals of
length $r$ plotted as a function of $r$ for the spectra
of binary Lennard-Jones system ( Case III ). Continuous line:
Prediction for the GOE. Filled circles: Our data}
\label{fig307}
\end{figure}

\begin{figure}[htp]
\vskip+0.5cm
\epsfxsize=3in
\centerline{\epsfbox{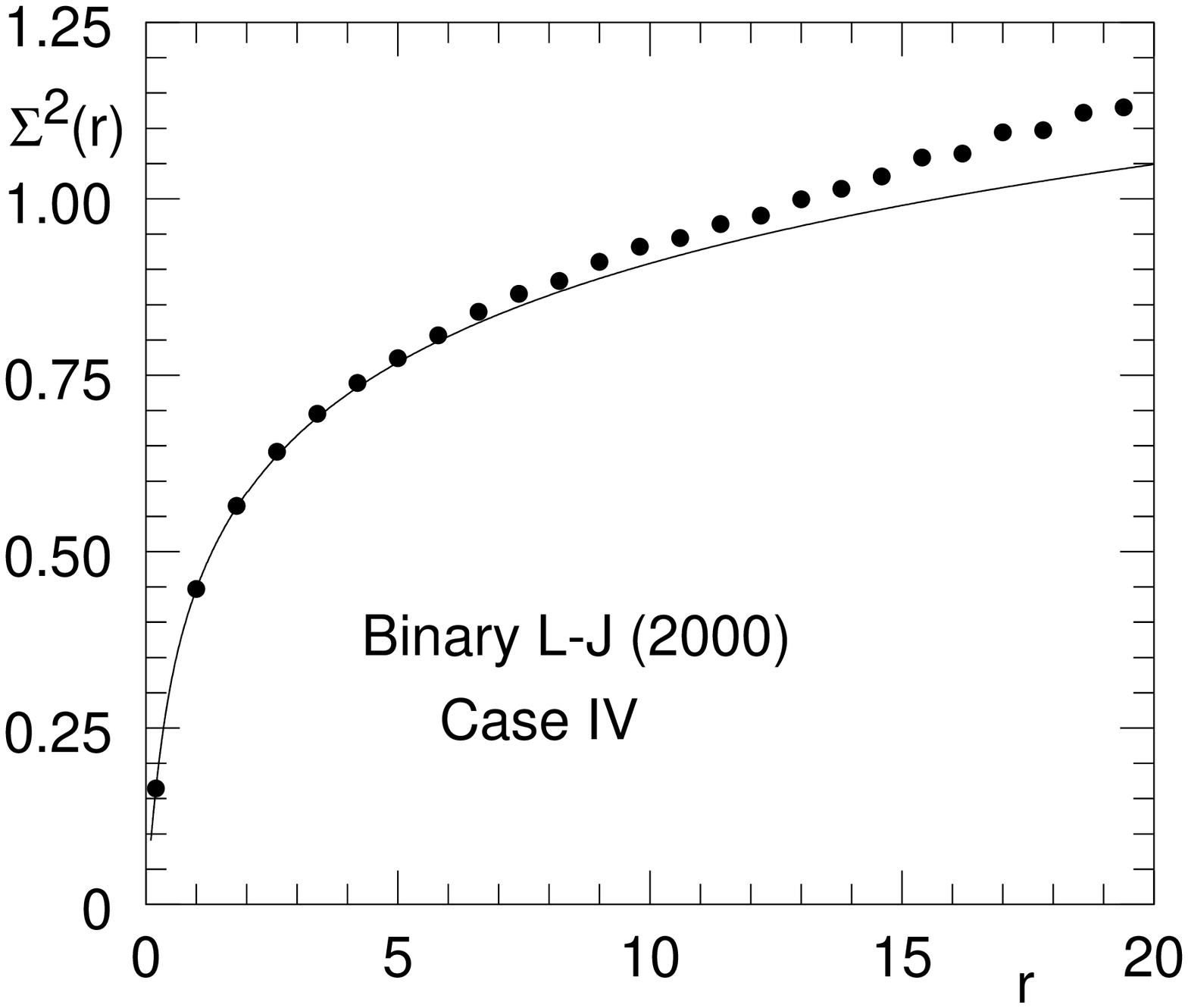}}
\caption{\sl Variance of the number of levels in intervals of
length $r$ plotted as a function of $r$ for the spectra
of binary Lennard-Jones system ( Case IV ). Continuous line:
Prediction for the GOE. Filled circles: Our data}
\label{fig308}
\end{figure}

\begin{figure}[htp]
\vskip+0.5cm
\epsfxsize=3.0in
\centerline{\epsfbox{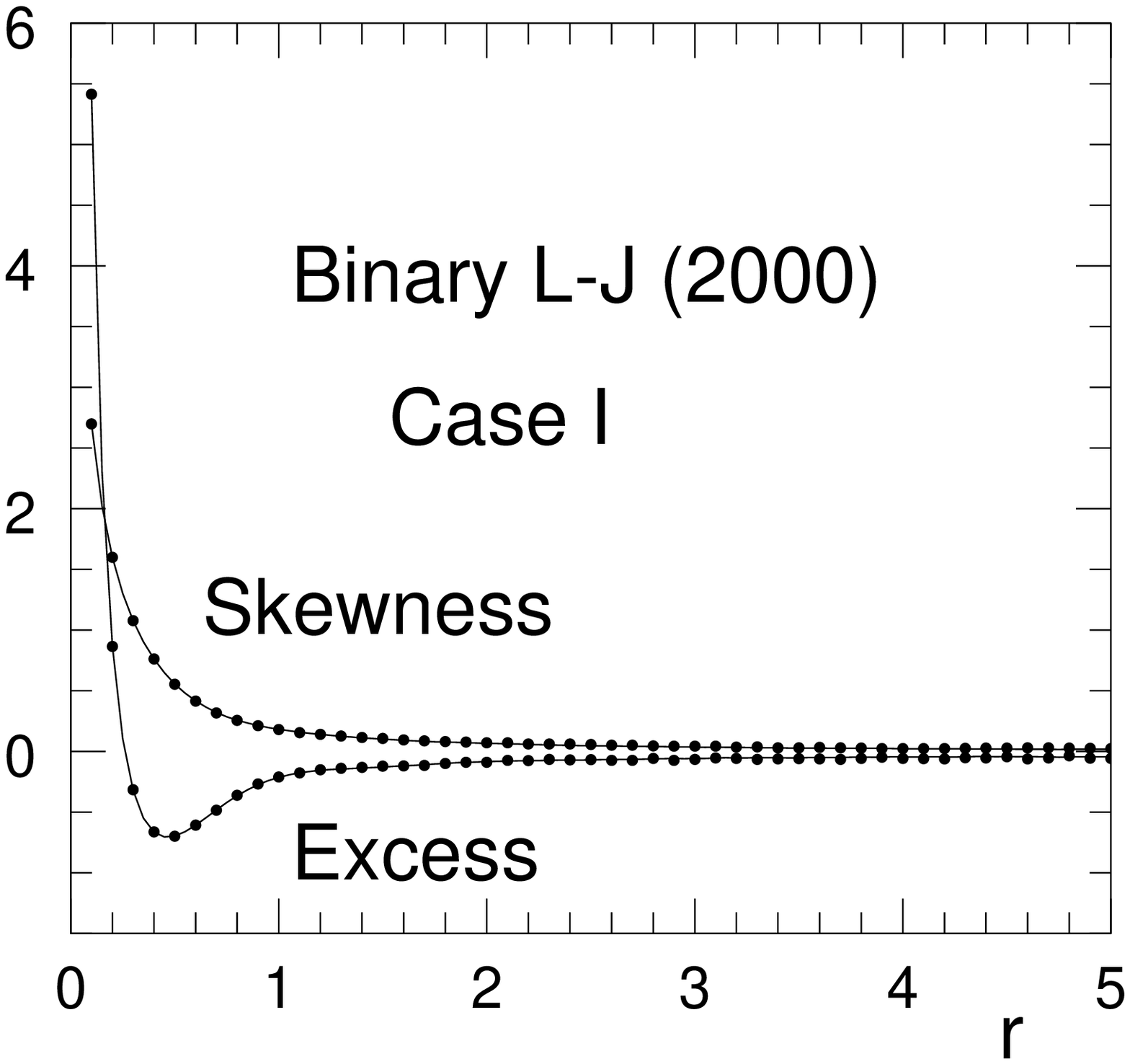}}
\caption{\sl Skewness and Excess parameters of the distribution of 
$n(r)$, the number of levels in interval of length $r$, plotted as 
a function of $r$ for the spectra
of binary Lennard-Jones system ( Case I ). Continuous lines:
Predictions for the GOE. Filled circles: Our data}
\label{fig309}
\end{figure}

\begin{figure}[htp]
\vskip+0.5cm
\epsfxsize=3in
\centerline{\epsfbox{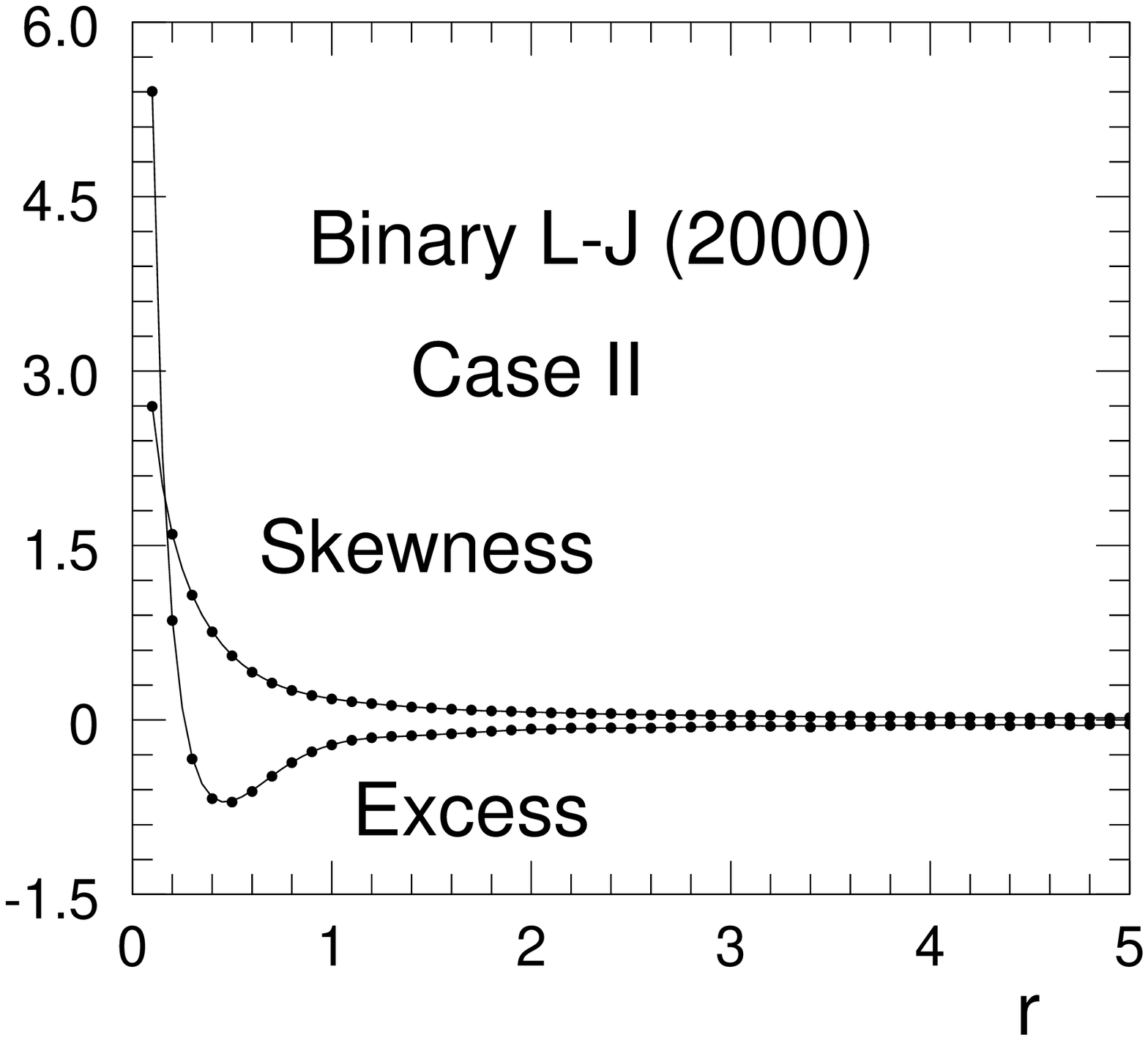}}
\caption{\sl Skewness and Excess parameters of the distribution of 
$n(r)$, the number of levels in interval of length $r$, plotted as 
a function of $r$ for the spectra
of binary Lennard-Jones system ( Case II ). Continuous lines:
Predictions for the GOE. Filled circles: Our data}
\label{fig310}
\end{figure}

\begin{figure}[htp]
\vskip+0.5cm
\epsfxsize=3in
\centerline{\epsfbox{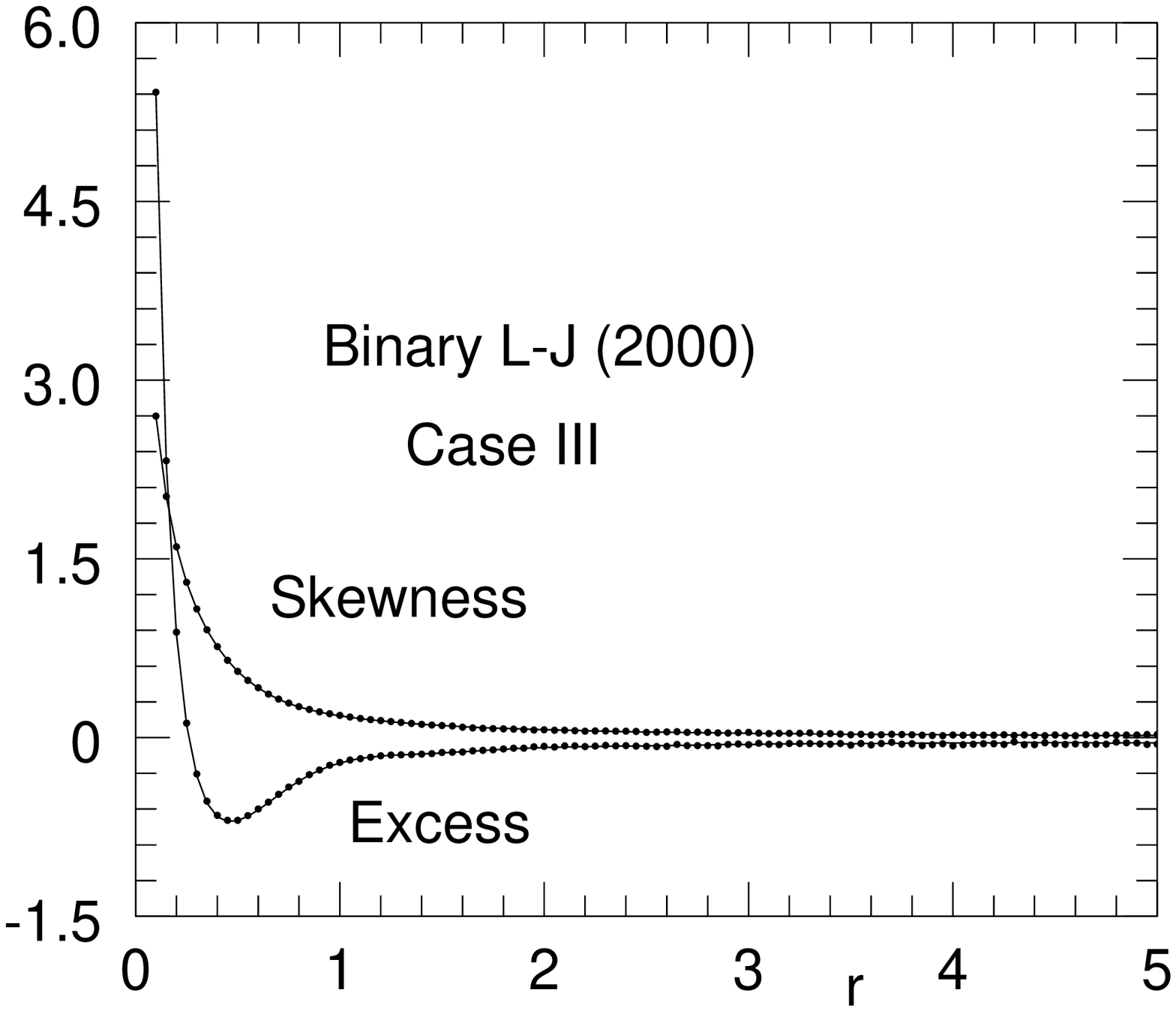}}
\caption{\sl Skewness and Excess parameters of the distribution of 
$n(r)$, the number of levels in interval of length $r$, plotted as 
a function of $r$ for the spectra
of binary Lennard-Jones system ( Case III ). Continuous lines:
Predictions for the GOE. Filled circles: Our data}
\label{fig311}
\end{figure}

\begin{figure}[htp]
\vskip+0.5cm
\epsfxsize=3in
\centerline{\epsfbox{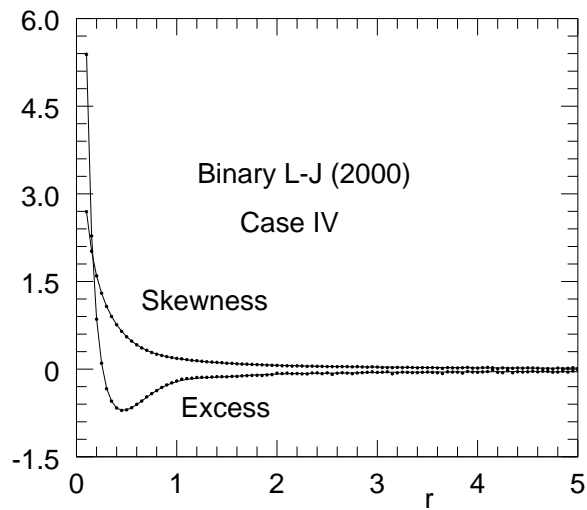}}
\caption{\sl Skewness and Excess parameters of the distribution of 
$n(r)$, the number of levels in interval of length $r$, plotted as 
a function of $r$ for the spectra
of binary Lennard-Jones system ( Case IV ). Continuous lines:
Predictions for the GOE. Filled circles: Our data}
\label{fig312}
\end{figure}
\newpage
It can be seen that the agreement with the predictions for the GOE
is at the same level as that for single component systems as far
as the nearest neighbor spacing and skewness ( plus excess ) are
concerned. However, in the case of variance ($\Sigma^2(r)$) the
disagreement with the predictions for the GOE seems to be somewhat
higher for larger values of $r$. It is quite likely that this is 
merely a consequence of not choosing the accepted regions of the spectra
separately for each local minimum.

\chapter{Universality of the Density of States for Amorphous Clusters}

The two previous chapters on the vibrational properties of amorphous
clusters have dealt mostly with an examination of the nature
of statistical fluctuations in the language of Random Matrix 
Theories. The first step in such an analysis is the unfolding of the 
spectrum. Let us recall that we needed a smooth and highly accurate 
fitting function for the plot of eigenvalue versus eigenvalue number.
The analytical form of this fitting function $D(\lambda)$ is given by
[$a - b \exp(-c\lambda)$]. This function fitted the cumulative 
density of states quite closely over the large central region of the
spectrum which was labelled as Region II. This was true for all the 
potentials and all the system sizes that we studied. Note that the 
function $D(\lambda)$ has only one scale ($\lambda_{0}$) for 
$\lambda$, namely $1/c$. This means that if $\lambda_{0}$ is chosen
as the unit of $\lambda$, all the density of states curves can be 
mapped into a single master curve in the Region II to a rather good
approximation - which is what we mean by universality in the density
of states. However, since the fit is close only in Region II it is
not clear whether the universality extends over the whole spectrum.
It is certainly possible, in principle, that there exists a universal
fitting function over the {\it entire} spectrum but we do not know 
its functional form and the functional form $D(\lambda)$ ( = $a -
b \exp(-c\lambda)$)
happens to provide a very good approximation to the true underlying
universal function in the large central region provided a, b and c
are chosen properly. In this chapter we examine this possibility 
through a numerical analysis of the vibrational spectra. In this 
connection  it is appropriate to remember that for a given potential
and a given number of particles, the local minima that correspond 
to amorphous clusters will be distributed over a range of energies.
Of all these possible energies, we obtain the ones that have the 
largest basin of attraction with respect to our method of generation
of local minima. Thus, our procedure permits us to examine the theme
of universality over the entire spectrum only for these minima.

Since different potentials will have different scales of frequency
$\omega$, ($\omega = \sqrt{\lambda}$) it is necessary  first of all
to decide on the appropriate unit of frequency in each case before
the density of states functions can be compared. We choose, for any 
given spectrum, the unit of frequency to be the average frequency
of {\it that} spectrum. Once the frequencies are normalized in all
the spectra corresponding to a given potential and a given number of 
particles, we can combine the histograms for the normalized frequencies
to generate its distribution. The normalized form of the distribution
of the normalized frequency ($\nu$) is denoted here by $n(\nu)$.
In figure 4.1 we present the data for the normalized density of states 
function for  six different cases for which the cluster sizes are 
comparable (400 or 500). This grouping of data according to system 
size is done keeping in mind the presence of finite size effects.
The cases which are shown in figure 4.1 correspond to (with the size
of the cluster shown in parenthesis): (1) Single-component
Lennard-Jones (500), (2)
Morse (500), (3) Sutton - Chen (400), (4) Gupta (400) for nickel,
(5) Gupta (400) for vanadium and  (6) Binary Lennard-Jones mixture
with parameters of Case I of chapter 3 (500). Before normalization
of frequency, the maximum values of $\omega$ for these six cases are
approximately 16, 33, 150, 18, 5 and 32 respectively.
\begin{figure}[htp]
\vskip+0.5cm
\epsfxsize=7.5in
\centerline{\epsfbox{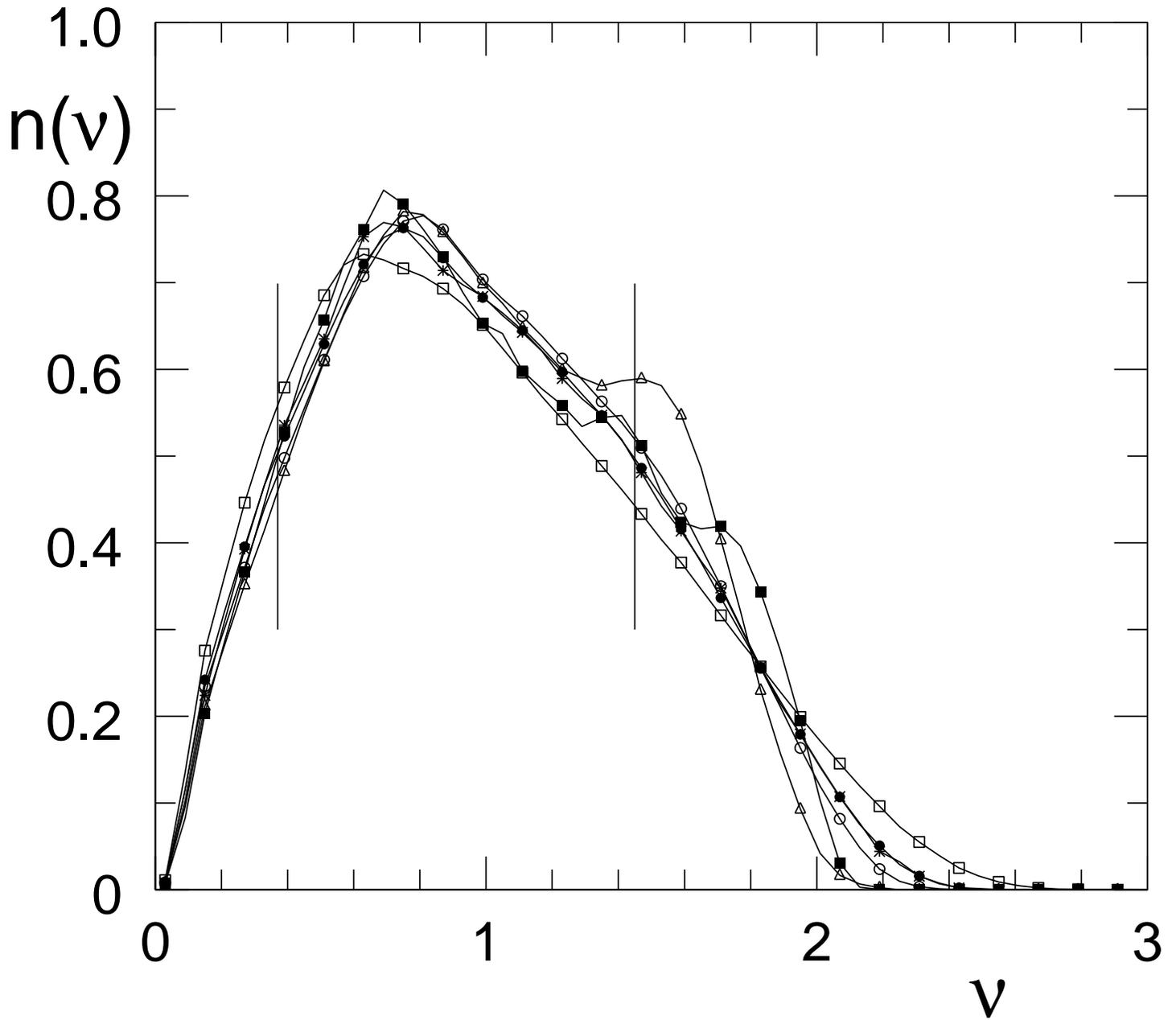}}
\caption{\sl Normalized density of states $[n(\nu)]$  for rescaled 
frequency $(\nu)$ with rescaling done by the average frequency of
the corresponding spectrum. Vertical bars denote approximately the
limits of region II. Filled circles: Lennard-Jones (500). Open
circles: Morse (500). Open triangles: Sutton - Chen (400). Stars:
Gupta for nickel (400). Filled squares: Gupta for vanadium (400).
Open squares: Binary Lennard-Jones (500), Case I.}
\label{fig401}
\end{figure}

\begin{figure}[htp]
\vskip+0.5cm
\epsfxsize=7.5in
\centerline{\epsfbox{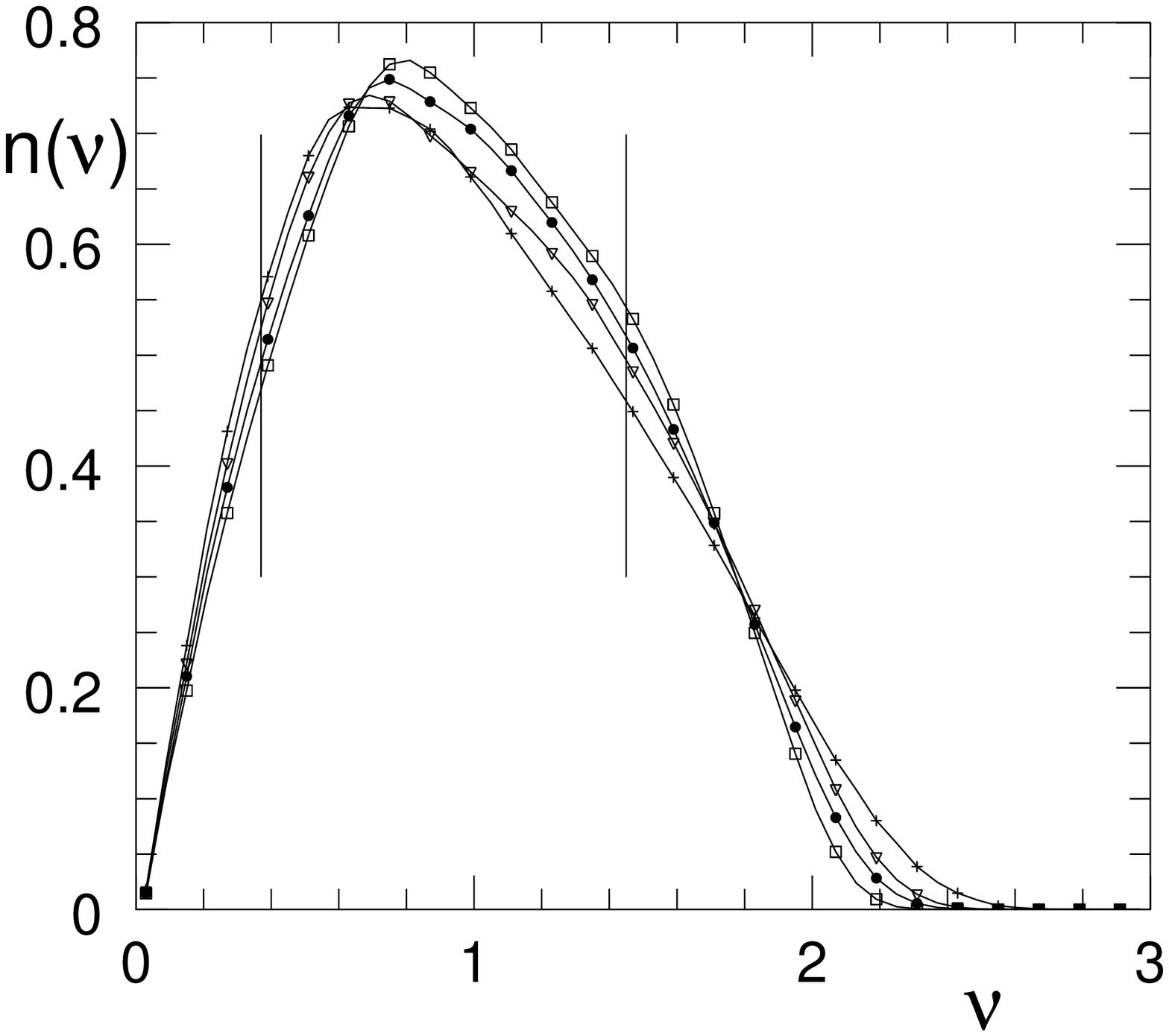}}
\caption{\sl Normalized density of states $[n(\nu)]$  for rescaled 
frequency $(\nu)$ with rescaling done by the average frequency of
the corresponding spectrum. Vertical bars denote approximately the
limits of region II. Filled circles: Lennard-Jones (2000). Open
squares: Morse (2000). Crosses: Binary Lennard-Jones , Case I (2000).
Open inverted triangles: Binary Lennard-Jones, Case II (2000).}
\label{fig402}
\end{figure}

Figure 4.2 shows similar data with the number of particles 
$N = 2000$ for: (1) Single - Component Lennard-Jones,
(2) Morse, (3) Binary Lennard-Jones, Case I and (4)Binary
Lennard-Jones, Case II. The scales of intrinsic frequencies
for the six cases in figure 4.1 vary by a factor of almost 30.
Given such variations of the intrinsic scales, the extent of 
overlap of the normalized density of states curves in figures 4.1
and 4.2 is certainly strongly suggestive. But in the light of
the observation that the function $D(\lambda) = a - b\exp(-c\lambda)$
describes the cumulative density of states to within 1\% in Region II,
which includes approximately 70\% of the spectrum, this apparent overlap
of the
different normalized density of states functions is somewhat misleading.
To demonstrate this, we replot the data of figures 4.1 and 4.2 in figures
4.3 and 4.4 by choosing the unit of frequency to be the inverse of
the square root of the best fit value of c -- which is the natural scale
of frequency suggested by the functional form of $D(\lambda)$. It is 
obvious from figures 4.3 and 4.4 that the quality of overlap in
Region II is much better now.

\begin{figure}[htp]
\vskip+0.5cm
\epsfxsize=7.5in
\centerline{\epsfbox{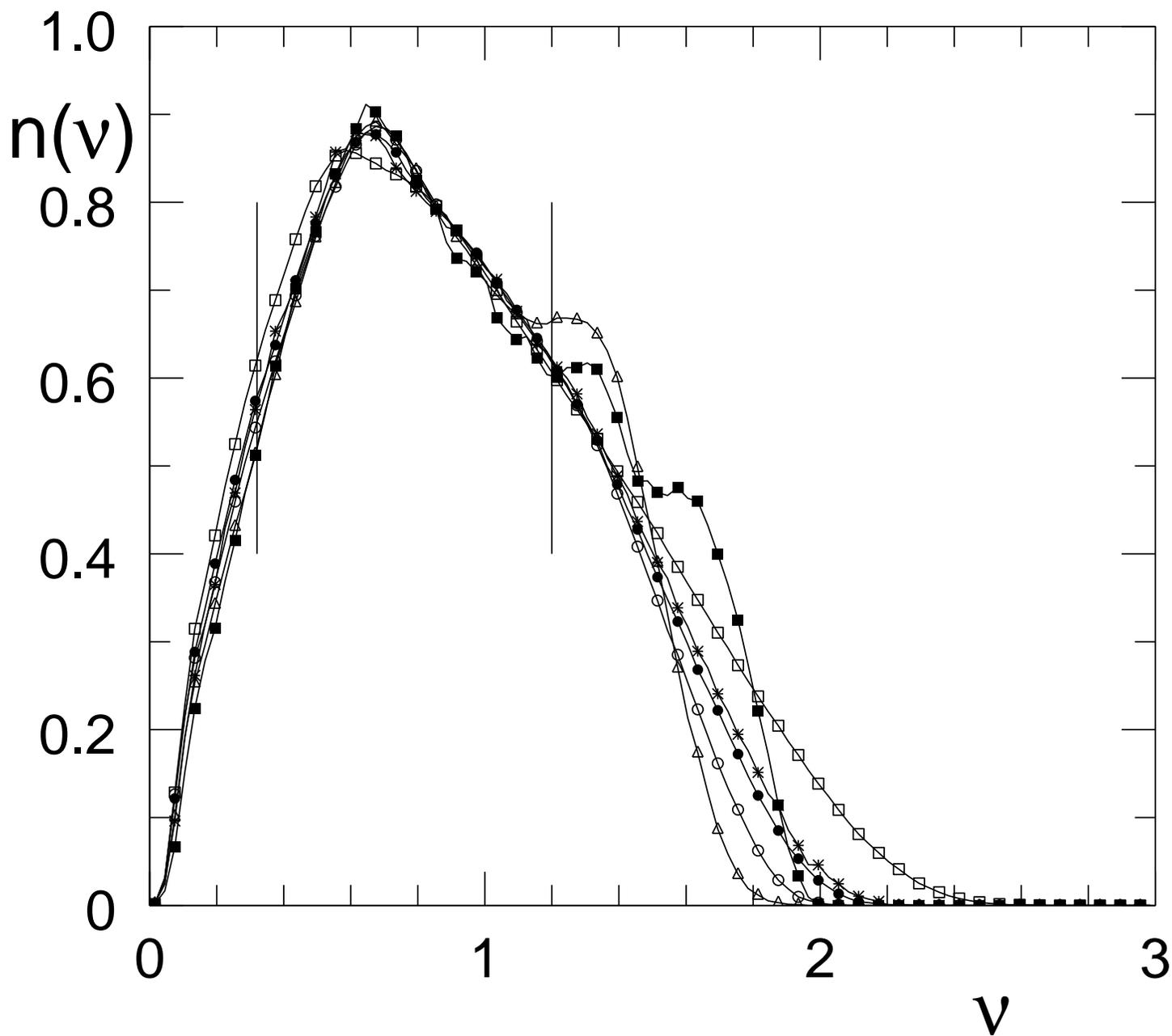}}
\caption{\sl Normalized density of states $[n(\nu)]$ for rescaled 
frequency $(\nu)$ with rescaling done by the best-fit frequency
parameter of the region II of the corresponding spectrum. Vertical 
bars denote approximately the limits of region II.
Filled circles: Lennard-Jones (500). Open circles: Morse (500).
Open triangles: Sutton - Chen (400). Stars: Gupta for nickel (400).
Filled squares: Gupta for vanadium (400). Open squares: Binary
Lennard-Jones (500), Case I.}
\label{fig403}
\end{figure}

\begin{figure}[htp]
\vskip+0.5cm
\epsfxsize=7.5in
\centerline{\epsfbox{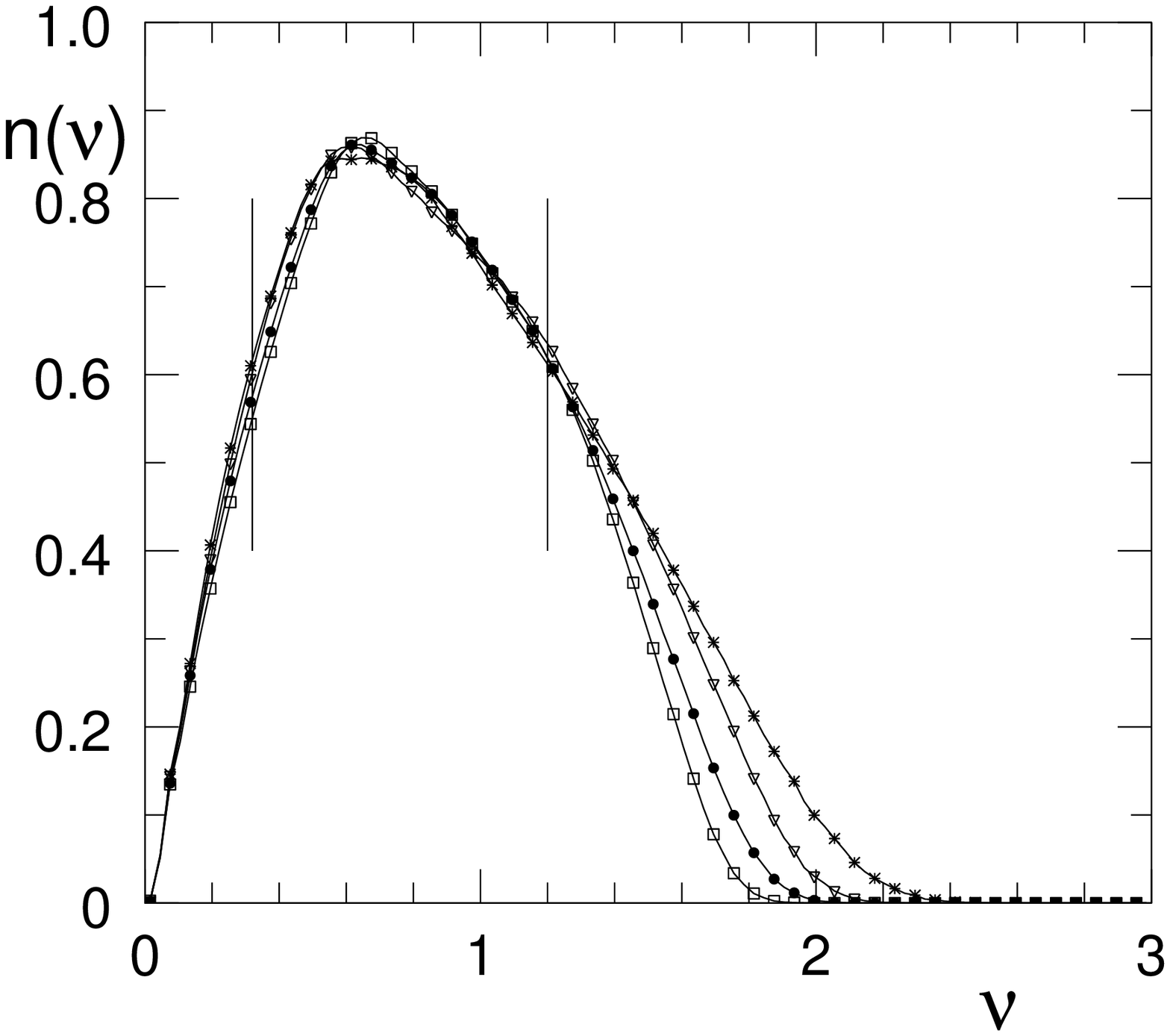}}
\caption{\sl Normalized density of states $[n(\nu)]$ for rescaled 
frequency $(\nu)$ with rescaling done by the best-fit frequency
parameter of the region II of the corresponding spectrum. Vertical 
bars denote approximately the limits of region II.
Filled circles: Lennard-Jones (2000). Open squares: Morse (2000).
Crosses: Binary Lennard-Jones, Case I (2000).
Open inverted triangles: Binary Lennard-Jones, Case II(2000).}
\label{fig404}
\end{figure}

Data presented in figures 4.1, 4.2, 4.3 and 4.4 suggest that the
universality of the density of states holds fairly accurately over
the large central region of the spectrum but not over the rest of
the spectrum. The inclusion of regions I and III, where there is a
violation of the universality, while computing the scale of frequency
in figures 4.1 and 4.2 causes the significant deviation around any
master curve in these two figures. It can also be seen in figures 4.1
and 4.3 that the Sutton - Chen spectrum for nickel and the Gupta 
spectrum for vanadium contain small but visible peaks in addition 
to the broad central peak.

To understand these features we need to use a physical insight
that was first provided by Rehr and Alben [38] regarding the mechanism
of how the sharp peaks of a crystalline spectrum transform into
the rather broad peaks of an amorphous solid as more and more 
disorder is introduced in the system. According to this line of
reasoning, the computation of the vibrational spectrum of a 
disordered system can be divided into two steps. In the first step
it is necessary to construct a geometry for the configuration around
which the vibration takes place (For us  any configuration 
corresponding to a local minimum of the potential energy 
function is a candidate). The next step is to construct a model of vibration.
This is
done by connecting all pairs of elements within an appropriate cut-off
distance via linear springs. The spring constants for the pairs within the
cut-off distance have a well defined functional dependence on
the corresponding pair separation. Thus, the vibrational modes of a 
solid are computed as the normal modes of this interconnected spring-mass
system. Now  even if the spring constants for all the pairs
included are taken to be the same, the sharp peaks of the crystalline
spectrum will still transform to the broad peaks for disordered systems.
This is due to the change in the topology of the spring-mass system 
induced by disorder which changes the connectivity pattern. Thus, this
`topological disorder'[38] will destroy the van-Hove singularities of the 
crystalline spectrum but the broad features will still survive -- since,
even for amorphous states, the local geometry is still substantially
like that present in a crystal. Now if we consider a situation in 
which the spring constants do change from pair to pair according to
the pair separation, disorder will introduce another source of 
broadening the time scales of vibration via the random variability
of the pair separation distances. The precise extent of this effect is
controlled by the sharpness of the variation of the spring constant
with pair separation and is quantified by the third derivative of 
the potential. This has been referred to as `quantitative disorder'[38].
If we keep on increasing the strength of the dispersion of the
spring constant, it will ultimately cause the disappearance of even 
the broad features that resulted from the imposition of only 
topological disorder. Hence, it is possible to have a spectrum with
just one peak in the presence of disorder if the nature of the 
inter-particle potential is such that there is strong enough 
`quantitative disorder'. This also suggests that if the dispersion of the
spring constant is not strong enough, even for disordered systems
vibrational spectra may show more than one peak. We believe this to
be the explanation for the existence of weak extra peaks in case of
Sutton - Chen potential for nickel and Gupta potential for vanadium.
Please also note that
we are dealing with clusters here and the presence of the surface can
be considered to be a source of `topological disorder' -- since the elements 
on the surface will have a different connectivity pattern compared
to those in the interior. However, the effect of this `topological
disorder' due to finite size will decrease as the cluster size increases
since the fraction of elements on the surface will keep decreasing.
In that limit  more crystalline features will be recovered. For example,
the extra peaks in the two cases that we have just discussed are expected
to grow in strength as the cluster size increases and this will cause
an even stronger violation of universality.

The idea that there may be universality in atleast parts of the 
vibrational spectra of disordered systems has also been put forward
recently on the basis of laboratory experiments with a variety of
glasses [65]. These experiments are on bulk systems and the constituent
units are too complex to be realistically represented by only the
positional degree of freedom. These extra complications will 
introduce additional features in the vibrational spectra. Nevertheless,
if we confine our attention to collective vibrations with coherence
length greater than the size of a constituent unit, it will still be 
permissible to compare the experimental results with theoretical
calculations of the type we are considering. The fact that we are 
dealing with clusters rather than bulk systems will certainly make a 
quantitative difference. But the main point here is the very existence
of universality itself in the vibrational spectra of  amorphous 
systems. To that extent  these new experiments put forward the same 
concept of universality that we have discussed here in the context of
clusters. The functional form that has been proposed for the universal
density of states function for the bulk glasses is different from what
is implied by  our $D(\lambda)$ function but  it also has only one scale
of frequency and thus satisfies the necessary condition of shape 
universality. Infact  we have used this functional form also to construct
an alternative cumulative density of states function for the purpose
of analyzing statistical fluctuations for clusters. To 
do this  please note that the density of states function for $\omega$
in ref.65 is given by $G(\omega) = 
\alpha\omega^{2}\exp(\beta\omega)$. This implies that the cumulative 
density of states has the form 
$I(\omega) = const - (2\alpha/\beta^{3})[1 + \beta\omega + (1/2){(\beta
\omega})^{2}] \exp(-\beta\omega)$.
Now  we can repeat the analysis of fluctuations by replacing $\lambda$
and $D(\lambda)$ [as used in chapters 2 and 3] by $\omega$ and 
$I(\omega)$, respectively. The misfit function now has amplitude that 
is comparable to or somewhat smaller than what is obtained with $\lambda$
and $D(\lambda)$. However, this has no quantitative consequence for the
analysis of statistical fluctuations since any difference in the quality
of unfolding caused by the difference of functional forms at the first
level is made up by the process of correction with the quadratic fit to
the residue.

It is also possible to study the issue of universality of the density 
of states function by examining the various moments of the frequency.
Let us remember that the information content of the density of states 
function can be expressed equivalently through the discrete but infinite
collection of all the moments. Although we do not presently have a theory
for the universality that is observed empirically, it is likely that the
preliminary effort in that direction will be in terms of a random matrix
type theory which typically concentrates on these moments. We can define 
these moments both in terms of $\lambda$ and $\omega$ but since $\lambda$
appears in a somewhat more immediate way in theoretical calculations,
we have chosen to calculate the moments of only $\lambda$ here. In the
present context  universality is signalled by the presence of only one
scale of $\lambda$ in its density of states function. This implies that
if we define $R(n) \equiv  <\lambda^{n}>/{<\lambda>^{n}}$ for every positive 
integral value of $n$ and if universality holds over the entire spectrum,
$R(n)$ should be universal. We show the data for this ratio of moments
in {\bf Table 4.1} for the various cases we have studied.
\newpage
\begin{center}
{\bf Table 4.1}
\end{center}
\begin{center}
\begin{tabular}{|c|c|c|c|c|c|} \hline

&{\bf Number}&&&& \\
{\bf Potential}&{\bf of}&{\bf $\bar{\lambda}$}&{\bf $\bar{\lambda^2}$/$\bar{\lambda}^2$}&
{\bf $\bar{\lambda^3}$/$\bar{\lambda}^3$}&{\bf $\bar{\lambda^4}$/$\bar{\lambda}^4$} \\
&{\bf Particles}&&&& \\ \hline 
    
&100
&5644.5&1.6676&3.4641&8.1032\\ \cline{2-6}
\bf Sutton - Chen &200
&5657.8&1.6513&3.3821&7.7913\\ \cline{2-6}
&300
&5677.3&1.6368&3.3139&7.5395\\ \cline{2-6}
&400
&5693.4&1.6286&3.2775&7.4116\\ \hline
\bf Gupta&100
&56.9&1.8360&4.5041&13.0656\\ \cline{2-6}
\bf (Nickel)&200
&61.7&1.8087&4.3018&11.9598\\ \cline{2-6}
&400
&66.3&1.7621&4.0328&10.7267\\ \hline
\bf Gupta(Vanadium)&400
&7.9&1.7034&3.6336&8.7015\\ \hline
&200
&261.5&1.7452&3.9473&10.3628\\ \cline{2-6}
\bf Morse&500
&280.4&1.7063&3.7216&9.3429\\ \cline{2-6}
&1000
&291.6&1.6825&3.5934&8.7990\\ \cline{2-6}
&2000
&300.8&1.6634&3.4933&8.3869\\ \hline
&200
&62.8&1.7968&4.2480&11.7732\\ \cline{2-6}
\bf Lennard-Jones&500
&69.0&1.7621&4.0360&10.7592\\ \cline{2-6}
\bf (Monatomic)&1000
&72.7&1.7375&3.8950&10.1187\\ \cline{2-6}
&2000
&75.8&1.7148&3.7687&9.5627\\ \hline
\bf Lennard-Jones&500
&195.8&1.9088&4.9043&14.9385\\ \cline{2-6}
\bf (Binary, Case I)&1000
&206.2&1.8798&4.7182&13.9731\\ \cline{2-6}
&2000
&214.7&1.8558&4.5728&13.2627\\ \hline
\bf Lennard-Jones&1000
&376.6&1.7662&4.0473&10.7777\\ \cline{2-6}
\bf (Binary, Case II)&2000
&364.2&1.7684&4.0430&10.7030\\ \hline

\bf (L-J Binary, Case III)&2000
&250.8&1.7539&3.9856&10.5159\\ \hline

\bf (L-J Binary, Case IV)&2000
&389.8&1.7463&3.9386&10.301\\ \hline

\end{tabular}
\end{center}

Please note that 
the scale of frequency in various situations (as measured by the 
corresponding average frequency) varies very widely -- infact over 
almost three decades. Second point to be noted is that with increase in the
order of moment the high frequency domain of the spectrum plays 
progressively more dominant role in determining the ratio. Thus, 
the effect of any deviation from universality in the high frequency
domain will be visible more and more in the moment ratio for higher 
order moments. In {\bf Table 4.1} $R(n)$ decreases gradually with
increase
in the system size while keeping the composition, the potential and the
value of $N$ fixed. However, the signature of convergence is not strong
enough . If we focus on the pattern of convergence for the various cases,
we cannot draw definite conclusions but the data is also not consistent
with the conjecture of universality at this level.

\chapter{Universality in the Vibrational Spectra of Bulk Amorphous
Systems}

In this chapter  we address the issue of universality in the vibrational
spectra of {\it bulk} amorphous systems rather than clusters. In chapter 4 we
observed the presence of shape universality over a large central region of
the vibrational spectra of clusters with various types of interparticle
interactions. The questions we want
to address here are: (1) What happens when the system size approaches the 
bulk limit and (2) Whether there are any situations in which shape universality
extends over the {\it entire} spectrum rather than over only a large
central region.

The  methodology that is used to produce the bulk amorphous systems is also 
used here to address another question which is not a central topic of
this thesis but is a rather important issue in its own right. This 
concerns the presence of `boson peaks' in the vibrational spectra of
disordered systems [21-22, 45-65]. This results from the existence of extra
modes in the low frequency region of the spectrum. Here  by `extra' one
means that the density of states is higher than what would be predicted
by a Debye type theory based on the speed of sound in the zero 
frequency limit. Although there are many theories regarding the 
origin of these extra modes, one aspect to which enough attention
has not been paid is to have a detailed and realistic picture of 
how the build up of the modes in the low frequency region (as well
as in the high frequency region) takes place as one goes from the
crystalline ground state to the amorphous states. It may be remembered
that crystalline vibrational spectra have sharp peaks and 
van-Hove singularities  whereas amorphous states don't have such
features. Here  we generate a detailed description of this 
morphological change in the spectra through a numerical study in which
we can create stable solid structures with variable amounts of disorder
i.e. any solid structure ranging from perfect crystal to completely
amorphous solids. We check our results regarding the evolution of the
vibrational
spectrum with the predictions of some existing theoretical
works in which disorder itself is modelled in a specific fashion [85-90]. In 
contrast, we model only the interaction between the particles. 
Disorder follows naturally from it and does not have to be modelled
separately.

To generate stable solid structures with variable extent of disorder
the first step is to select a model for the interaction among the
constituent units (atoms or molecules). Here we assume that every 
unit has only a position vector as its degree of freedom and the 
potential energy of the system is of the form of sum over pairs. The
route to generating amorphous structures which, strictly speaking,
have no periodicity on any length scale is actually via crystalline
structures. Let us denote the number of constituent units per primitive
cell by $N$. Ideally, $N$ should be infinity to produce systems with
finite disorder. In practice, however, one takes as large a value as
possible within the available computational resources. Now let us
denote the three edges of the unit cell by $\vec{a}$, $\vec{b}$ and
$\vec{c}$. The positions of the particles within the unit cell can be
defined by $\vec{r_{i}} = \theta_{1}(i) \vec{a} + \theta_{2}(i) \vec{b}
+ \theta_{3}(i) \vec{c}$ with the values of $\theta_{1}(i)$, 
$\theta_{2}(i)$ and $\theta_{3}(i)$ being all between 0 and 1 for
every value of $i=1,...,N$. For economy of notation, we denote
the triad $(\theta_{1}(i), \theta_{2}(i), \theta_{3}(i))$ by 
$\vec{\theta}(i)$. 

In order to generate states with variable amount of disorder  we begin
with a fcc lattice for which the lattice constant is adjusted to
minimize the potential energy. Now  this system is subjected to a
$NPT$ type Monte-Carlo simulation. Here  $N$, $P$ and $T$ represent
the number of particles per unit cell, pressure and temperature, respectively,
and are held constant
during the simulation in which the variables are $\vec{a}$, $\vec{b}$
and $\vec{c}$ (which define the geometry of the unit cell) and the
$\vec{\theta}$s for all the particles. In the present calculation
we have taken $P = 0$ and thus the potential energy per unit cell 
($U$) constitutes the Hamiltonian of the system. At the end of every m 
(typically 2) cycles of the $NPT$ simulation, we take the instantaneous
configuration and subject it to a conjugate gradient minimization for $U$
with respect to $\vec{a}$, $\vec{b}$, $\vec{c}$ and all the 
$\vec{\theta}$s [84]. We repeat the process of
generating the local minima with various values of the seed of the
random number generator in the Monte-Carlo simulation. Finally, the
information regarding this collection of solid structures is displayed
in the form of a volume per particle ($\Omega$) versus energy per
particle ($\epsilon$) diagram.
\begin{figure}[htp]
\vskip+0.5cm
\epsfxsize=5.0in
\centerline{\epsfbox{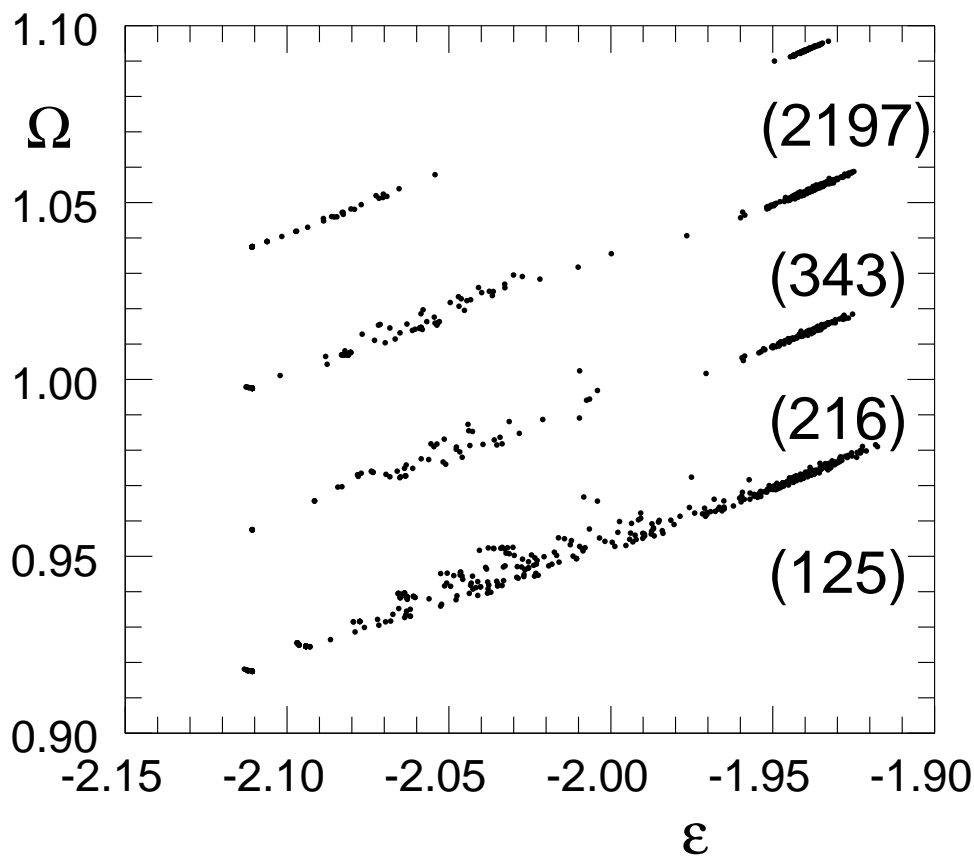}}
\caption{\sl Volume per particle $\left(\Omega\right)$ vs. energy per
particle $\left(\epsilon\right)$ diagram for Lennard-Jones potential.
Data for $N$ = 216, 343 and 2197 are shifted upwards with respect to the
data for $N$ = 125 by 0.04, 0.08 and 0.12, respectively to avoid overlap.}
\label{fig501}
\end{figure}
Figure 5.1 provides an example of this 
kind of volume versus energy diagram for the standard Lennard-Jones
potential [ i.e. $V(r) = {(1/r)^{12}} - {(1/r)^{6}}$ ] which has been 
properly cut off and smoothed so that the function as well as all
derivatives upto second order are continuous for all values of r. In
figure 5.1 the completely amorphous states correspond to the group at
the highest energies. Isolation of this group from the rest of the minima
is visible more and more
clearly as the number of particles in the unit cell increases. This is
due to the progressively decreasing probability of generating states in
the domain of energies just below the amorphous group as the number of
particles in the unit cell increases. For the purpose of demonstrating
the evolution of the
vibrational spectrum with disorder, these minima are of critical 
importance. Any calculation such as ours will naturally generate only
a small subset of all the local minima that exist. In particular, the
data shown in Figure 5.1 certainly do not correctly represent the relative
densities of the local minima except in the completely amorphous band.

For the amorphous band  the functional dependence of the relative density
of the local minima on the value of energy per particle ($\epsilon$) is
given by a Gaussian function whose width is proportional to $1/\sqrt{N}$.
This follows from the extensivity property of the configurational entropy
and has been observed earlier also in calculations with constant density.
It should be noted that our calculations are done under the condition
of constant `inherent' pressure. For a given value of energy per
particle, dispersion in the value of volume per particle is expected to
go to zero in the thermodynamic limit. For the amorphous band we have
verified the validity of
this observation in our case in the following manner: Compute the best
fit quadratic function  $S(\epsilon)$ through the data points in the amorphous
band of the
$\Omega - \epsilon$ diagram. Then calculate the root mean square
deviation of $\Omega$ from the best fit curve. We find that this
measure of scatter around a smooth curve indeed goes to zero in the
completely amorphous band as $1/\sqrt{N}$.
\newpage
\begin{figure}[htp]
\vskip+0.5cm
\epsfxsize=5.0in
\centerline{\epsfbox{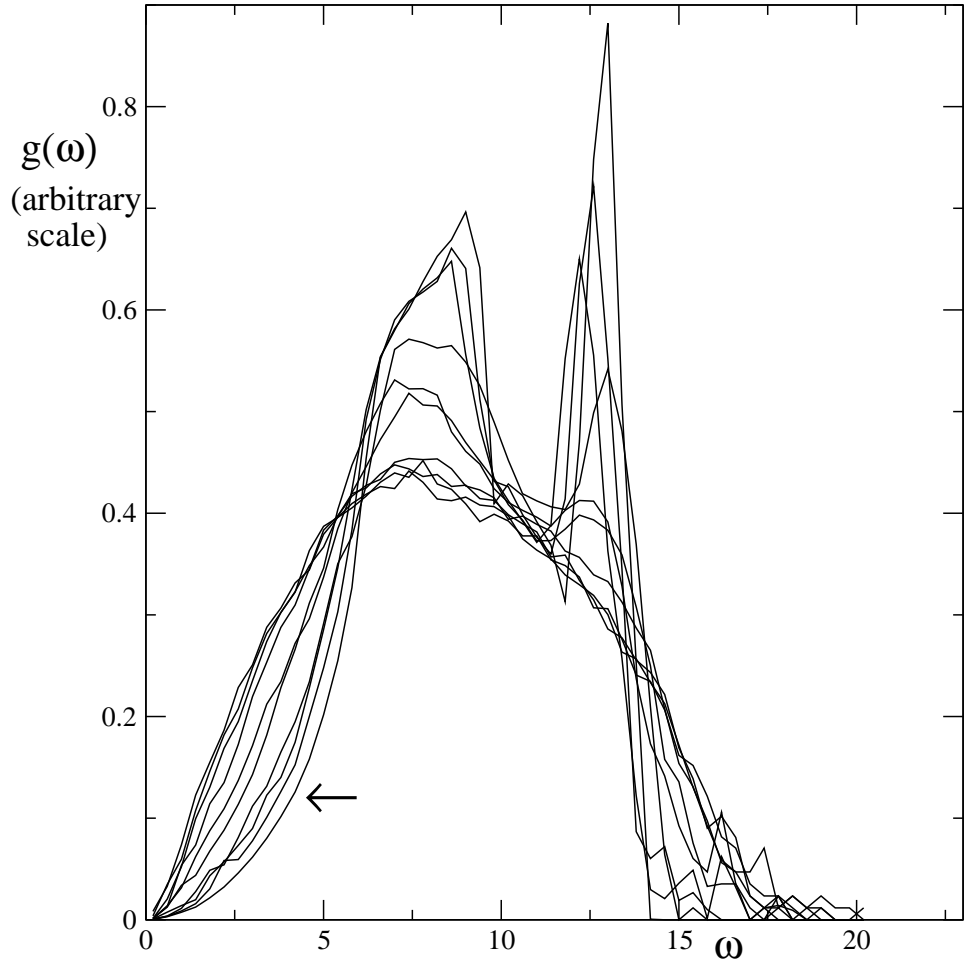}}
\caption{\sl Density of states $\left(g\left(\omega\right)\right)$ plotted
against frequency $\left(\omega\right)$ for Lennard-Jones potential with
$N$=343. Following the arrow in the figure the plots correspond to energy
per unit cell equal to -724, -713, -706, -700, -689, -685, -677, -672, -666
and -662, respectively.}
\label{fig502}
\end{figure}

For any local minimum in the $\Omega - \epsilon$ plane  the associated
vibrational spectrum can be computed through standard methods. As energy
per particle increases the system becomes more and more amorphous and the
density decreases i.e. there is swelling in the system. At the same time
there is a continuous change in the nature of the associated vibrational
spectrum. Figure 5.2 illustrates this for the case of Lennard-Jones
potential  with $N = 343$. Here  we have picked, from figure 5.1, a few
local minima which are spaced roughly equally in energy. For the
selected minima  we calculate the vibrational spectra and the density
of states for each case is plotted in figure 5.2. One can clearly see
a gradual transformation from a crystal-like spectrum to one which is
characteristic of amorphous states.
\begin{figure}[htp]
\vskip+2cm
\epsfxsize=3.7in
\centerline{\epsfbox{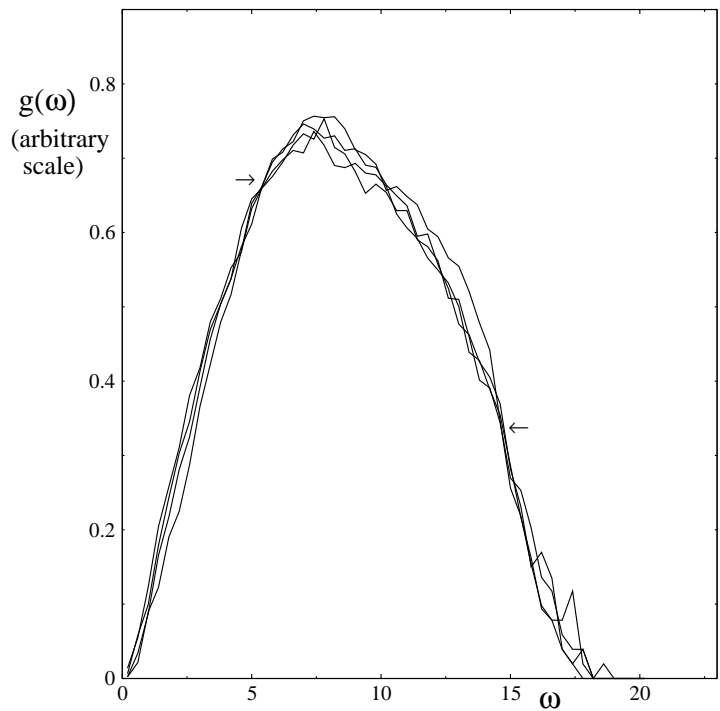}}
\caption{\sl Density of states $\left(g\left(\omega\right)\right)$ plotted
against frequency $\left(\omega\right)$ for Lennard-Jones potential (with
$N$=343) for the four highest energy values out of the set
for which the density of states plots are shown in figure 5.2.
The two arrows point to the two approximate fixed points of $g(\omega)$.}
\label{fig503}
\end{figure}
We can compare the data available
in this figure with the results presented by Mattis and
coworkers for some model calculations on the evolution of the density
states as the strength of disorder changes [85-90]. These calculations are for
a quantum mechanical model of vibration in which a specific model term
is included in the Hamiltonian to describe the effect of 
disorder. The amplitude of this term quantifies the strength of disorder.
It so happens that the spectrum for the model becomes unstable when the
strength of the disorder exceeds a critical value. To avoid this
it is technically necessary to include an anharmonic term in the
Hamiltonian -- although, at the end, the strength of anharmonicity can
be made arbitrarily small. On the other hand, we use harmonic approximation
to compute the normal modes of a classical disordered solid and no
explicit model of disorder is needed. To compare the results of our
calculation regarding the evolution of the vibrational spectrum
with disorder with previous theoretical calculations , it is useful
to summarize the main results of these latter calculations.

For a simple cubic lattice, they are the following [90]: (1) For any finite
value of the disorder parameter  there are no van-Hove singularities.
(2) As the strength of the disorder parameter increases, more and more
modes are transferred to the domains with the lowest and highest
frequencies. (3) There is a critical value of the disorder parameter
which marks the boundary between finite long range order [disordered
crystal] and vanishing long range order [glass]. When the disorder
parameter exceeds this value the density of states function develops
two fixed points i.e. there are two frequencies $\omega_{1}$ and
$\omega_{2}$ where the value of the density of states function does not change
with change in the value of the disorder parameter. The value of $\omega_{2}$,
the higher of the two frequencies, is approximately 3.5 times the
value of $\omega_{1}$. (4) When the disorder parameter exceeds the
critical value, the value of the density of states  function at zero
frequency becomes finite and increases with disorder.

While comparing the predictions listed in the previous paragraph
with our results, two limitations must be kept in mind. (1) Our
starting crystal has a face centered cubic lattice rather than a
simple cubic lattice. (2) As we have noted earlier, modelling
any finite extent of disorder requires the size of the unit cell
to go to infinity, strictly speaking. In our case, the value of
$N$ is only 343. Despite these limitations, we find that our data
is consistent with the first two predictions. Also, if we examine
only those spectra in figure 5.2 which correspond to completely
disordered states (shown separately in figure 5.3), we notice the
presence of two approximate fixed points. However, the ratio
$\omega_{2}/\omega_{1}$ is close to 2.4. The fact that it is
substantially different from the value of 3.5 obtained in the analytical
calculation may be due to the difference between the starting crystal
structures of the two calculations. However, our data certainly does not
corroborate the fourth prediction of finite value of the density of
states function in the $\omega \rightarrow 0$ limit. We always find
that the density of states goes to zero as $\omega \rightarrow 0$.

Now we address the central problem of this chapter. The question here
is: Are there any situations where, for bulk amorphous systems, the
vibrational density of states function assumes a {\it shape} that is
universal over the {\it entire} spectrum? This question is motivated partially
by our observations regarding the universality of the shape of the
spectrum ({\it but only in the large central region}) for the cases of
clusters interacting via several different kinds of potentials -- as
reported in the previous chapters [67-68]. The other motivation behind these
investigations is to understand the aspect of universality observed
for the density of states function in recent experiments with bulk
molecular glasses [65] -- as noted in chapter 4. In these
experiments  the constituent units are too complicated to be adequately
modelled as single particles. However, for vibrational modes with
coherence lengths much larger than nearest neighbor spacing, replacing
a somewhat complicated constituent unit by a point particle may not be
a bad approximation. Our approach to discovering the possible situations
in which there will be universality in the shape of the entire vibrational
spectrum is influenced by the discussion ( see chapter 4 ) of the effect 
of the dispersion of the second derivative of the potential energy function
around its minimum. Let us recall that, according to the insight provided 
by the work of Rehr and Alben [38], in the limit of the dispersion of the 
spring constant becoming very large, we expect only a single broad peak
in the density of states function. In our calculations  the geometries
around which vibration takes place correspond to the local minima 
of the potential energy function. For such a configuration the significant
pair distances will largely be distributed around the minimum of the 
pair potential. Thus, a quantitative dimensionless measure of the 
dispersion of the spring constant (DSC) is given by
$\phi =\left [\left| \partial^3 V/\partial r^3\right |_{r = r_{0}}/
      \left| \partial^2 V/\partial r^2\right |_{r = r_{0}}\right] r_{0}$
where $r_{0}$ is the distance at which the pair potential $V(r)$ assumes its 
minimum value. Thus, $\phi \rightarrow \infty$ is the limit where 
we expect a featureless spectrum with a single broad maximum. The
numerical data reported later in this chapter is indeed consistent 
with this expectation. But what is more important and interesting
is the following: If the manner in which $\phi$ approaches infinity
satisfies an
additional condition (to be stated shortly), the vibrational spectrum
approaches a shape that is universal to within the uncertainities of
our numerical calculation. Here the property of universality implies
that the
asymptotic shape of the vibrational spectrum does not depend on the
analytical form of the potential energy function whose parameters are varied 
appropriately to realize the $\phi \rightarrow \infty$ limit.

It is possible, in principle, to construct many parametric functional 
forms which, with suitable variation of the parameters, can be made to
approach
the $\phi \rightarrow \infty$ limit. We have used only exponential and
power law functions in our parametric functional forms since interatomic
or intermolecular forces will typically have rapid spatial variation
and the only elementary functions that are capable of describing such
variations are the exponential and power law functions. The two types
of functional forms that we have actually used are as follows:\newline
(1) The Generalized Lennard-Jones (GLJ) family where the pair potential
has the form $V(m,n; r) = (1/r^{m} - 1/r^{n})$ where $m$ and $n$ are 
positive integers with $m > n$.\newline
(2) The Morse family in which the potential has only one parameter
$\alpha$ and the expression is given by $V(\alpha; r) = \exp(-2\alpha(r-1))
- 2\exp(-\alpha(r-1))$.
The overall shape of the pair potential is largely the same for both the 
types. They always have a single minimum where the potential function
assumes a negative value. The function rises rapidly to zero at
larger distances. At shorter distances also the potential rises fast but
to a finite value for the Morse case and to infinity for the GLJ case. 
In the Morse case  the rapidity with which the potential increases away
from the minimum is controlled by the single parameter $\alpha$. 
However, for the the GLJ type  the rise of the attractive and repulsive 
sides are controlled separately by $n$ and $m$, respectively. It is easy
to verify that the DSC parameter ($\phi$) is given by ($m + n + 3$) and
$3\alpha$ for the GLJ and Morse families, respectively.
\begin{figure}[htp]
\vskip+0.5cm
\epsfxsize=6.5in
\centerline{\epsfbox{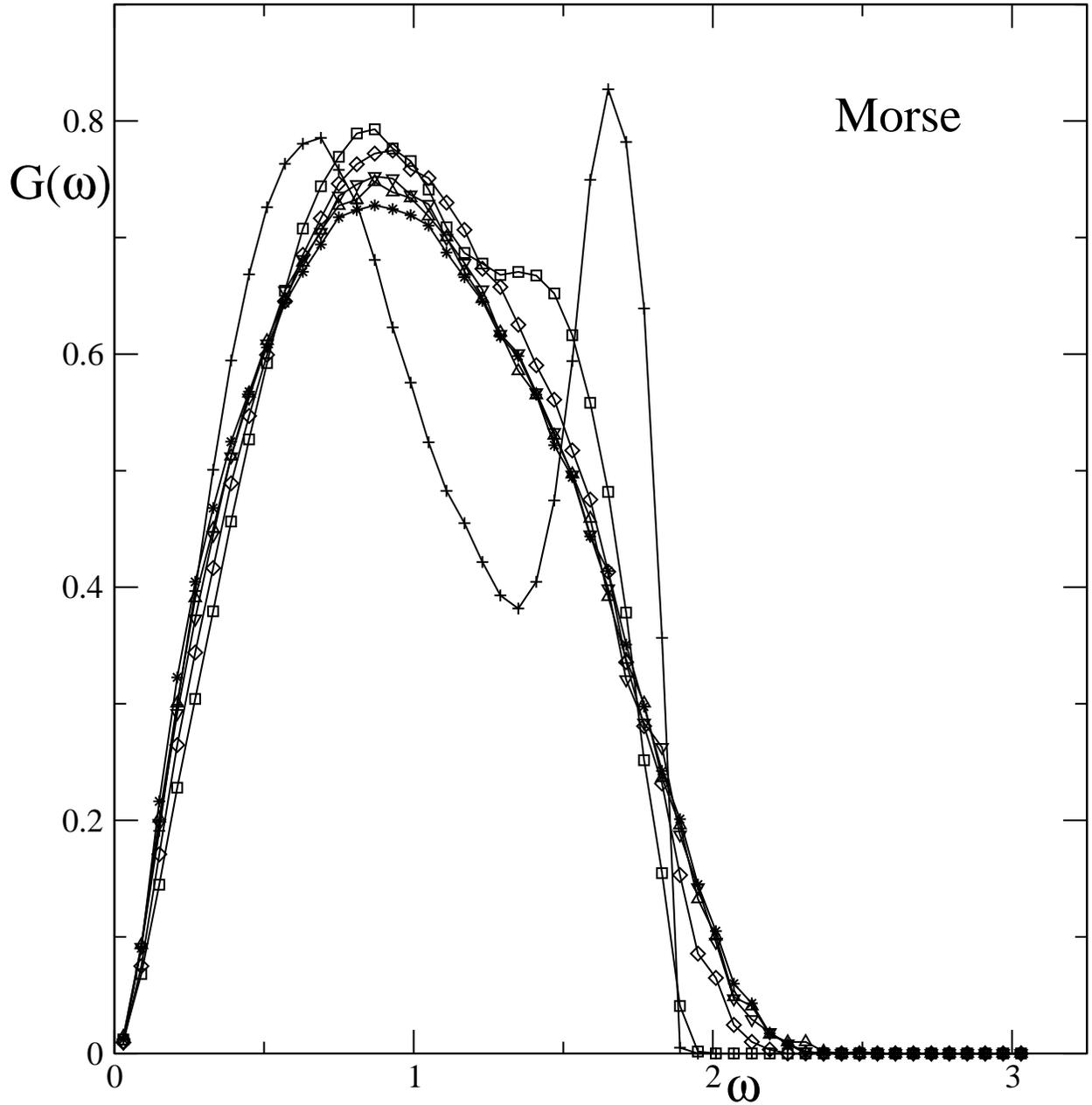}}
\caption{\sl Normalized density of states $\left(G\left(\omega\right)\right)$
vs. normalized frequency $\left(\omega\right)$ in the fully disordered region
for some selected cases of Morse potential. The values of $\alpha$ are:
3.0 (cross), 4.5 (square), 7.5 (diamond), 10.5 (inverted triangle), 13.5 (
triangle) and 16.5 (star).}
\label{fig504}
\end{figure}

\begin{figure}[htp]
\vskip+0.5cm
\epsfxsize=6.5in
\centerline{\epsfbox{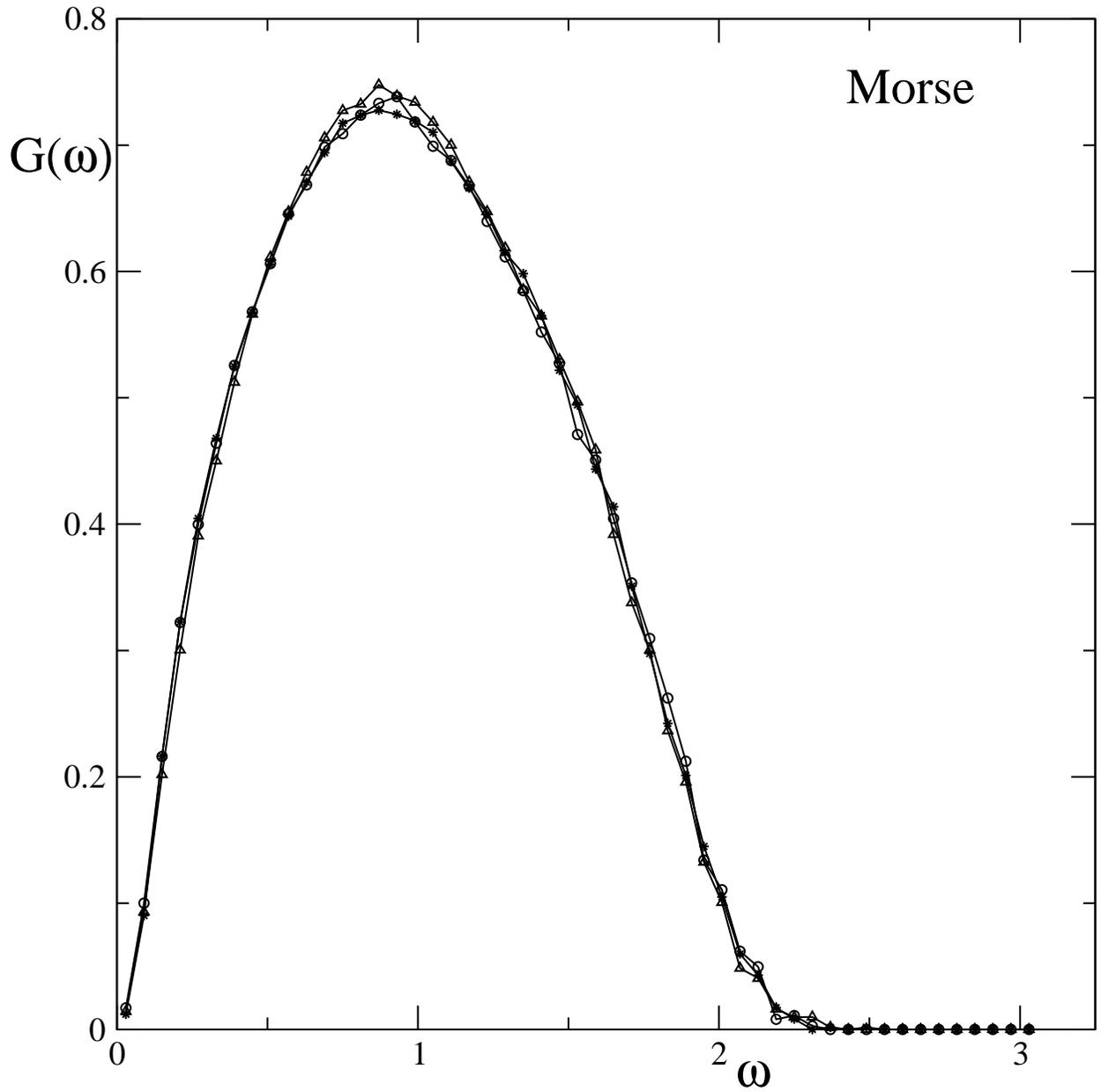}}
\caption{\sl Normalized density of states $\left(G\left(\omega\right)\right)$
vs. normalized frequency $\left(\omega\right)$ in the fully disordered region
for some selected cases of Morse potential. The values of $\alpha$ are:
13.5 (triangle), 15.0 (circle) and 16.5 (star).}
\label{fig505}
\end{figure}
\begin{figure}[htp]
\vskip+0.5cm
\epsfxsize=6.5in
\centerline{\epsfbox{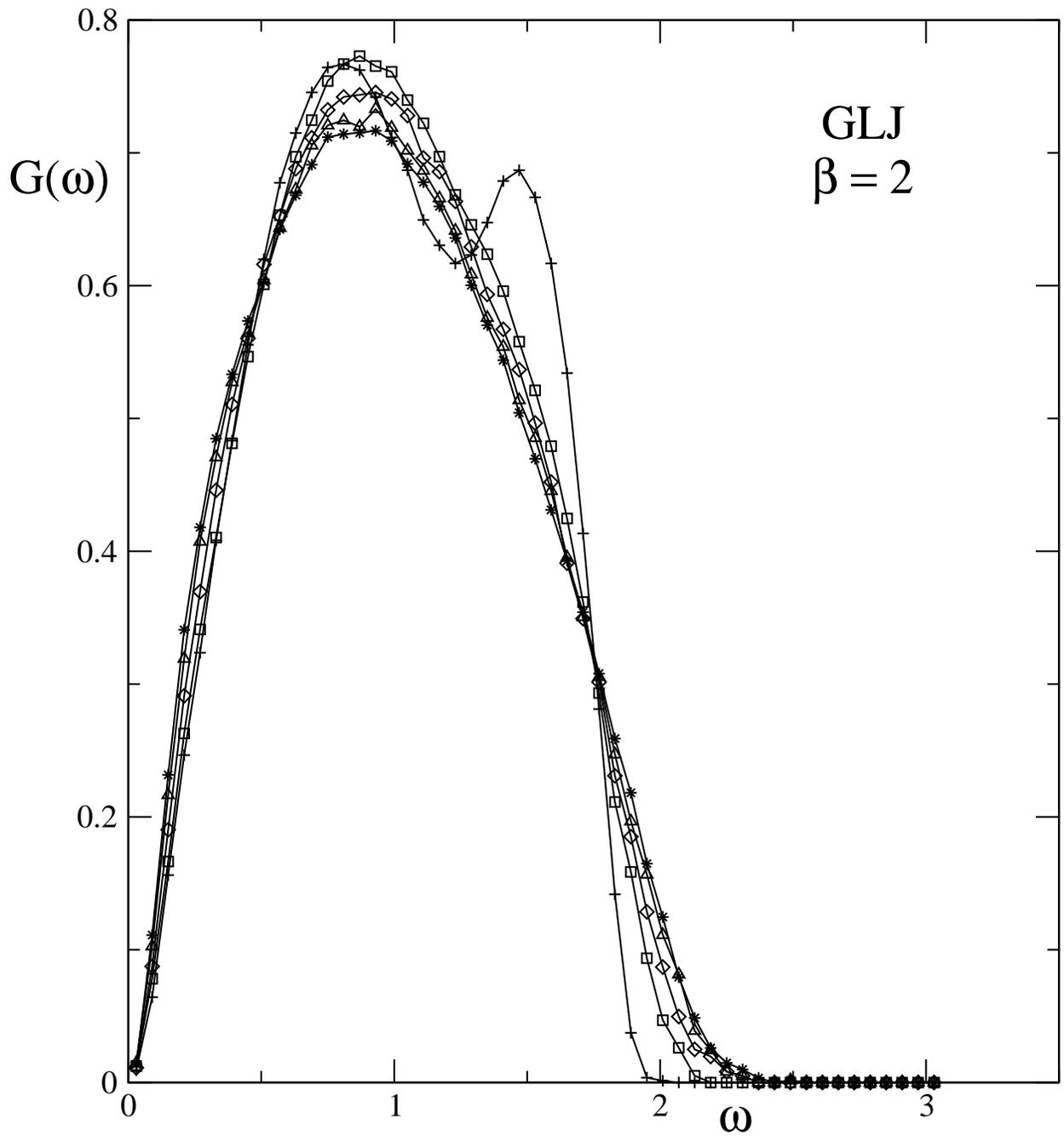}}
\caption{\sl Normalized density of states $\left(G\left(\omega\right)\right)$
vs. normalized frequency $\left(\omega\right)$ in the fully disordered region
for some selected cases of GLJ potential. The values of [m,n] are
[6,4] (plus), [10,8] (square), [14,12] (diamond), [18,16] (triangle) and
[22,20] (star).}
\label{fig506}
\end{figure}

\begin{figure}[htp]
\vskip+0.5cm
\epsfxsize=6.5in
\centerline{\epsfbox{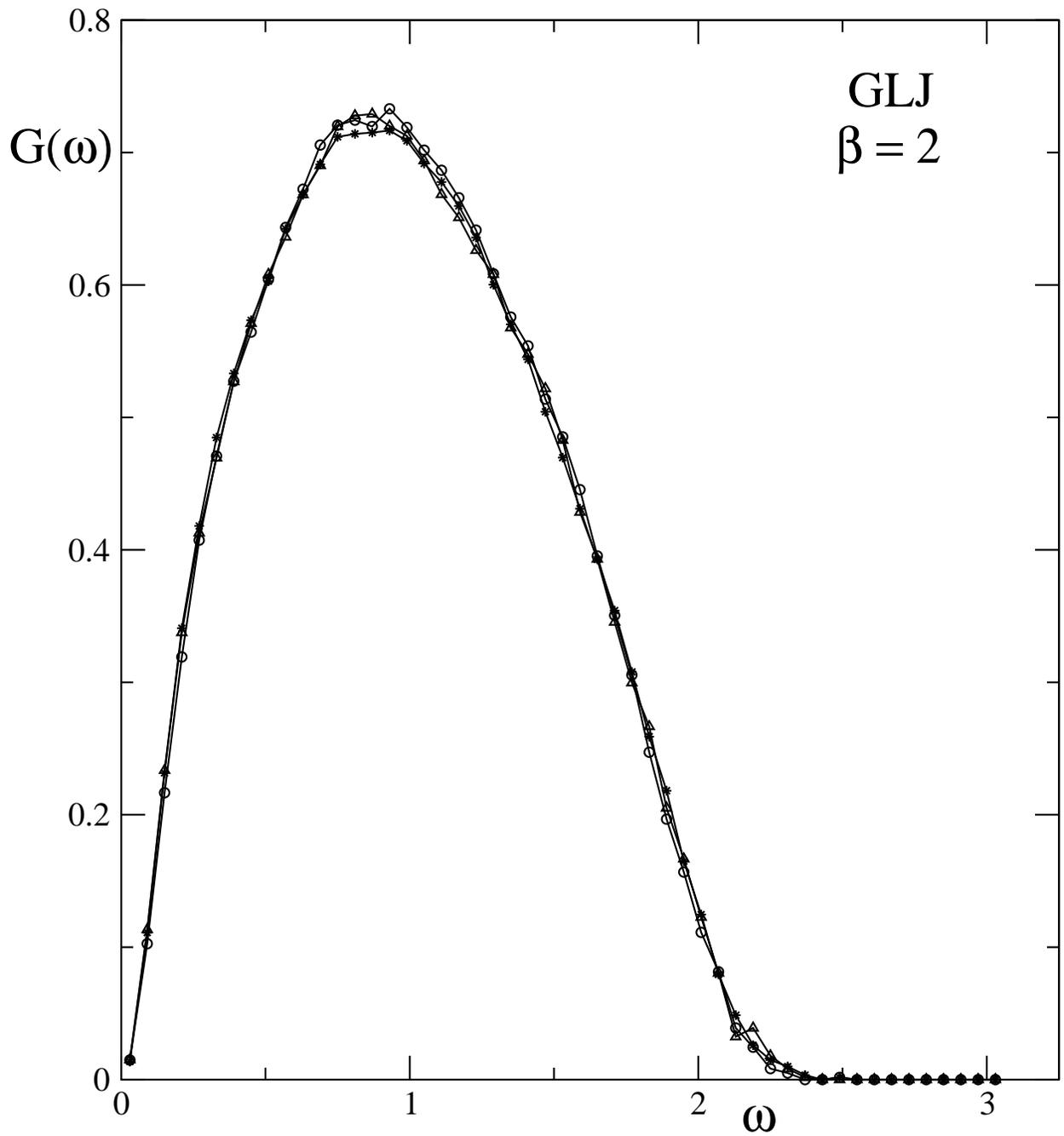}}
\caption{\sl Normalized density of states $\left(G\left(\omega\right)\right)$
vs. normalized frequency $\left(\omega\right)$ in the fully disordered region
for some selected cases of GLJ potential. The values of [m,n] are
[18,16] (circle), [20,18] (triangle) and [22,20] (star).}

\label{fig507}
\end{figure}

Obviously, for
the GLJ case there is no unique sequence for approaching the 
$\phi \rightarrow \infty $ limit even under the restriction of $m > n$. 
Here, either only $m$ can go to infinity or both $m$ and $n$ can go to 
infinity while always satisfying $m > n$. Our observation is that only 
when both $m$ and $n$ go to infinity, universal shape of the vibrational 
spectrum is realized i.e. while the dispersion of spring constant goes
to infinity, contribution must come both from the attractive and the 
repulsive sides of the minimum.

We have already described how to generate the completely disordered 
states for any given potential. Ofcourse they span a range of 
energies and there is variation of the vibrational spectrum with energy.
Thus, a pertinent question is: which vibrational spectra should be 
considered representative of these band of states? We use the criterion
of maximum likelihood of physical relevance. Thus, for each potential,
we calculate the mean and the standard deviation of energy in the
completely disordered region. Next, we choose a set of 10 local minima
which are approximately equally spaced in energy in the band contained
within one standard deviation of the mean. For each local minimum, the
spectrum is computed and the frequencies are rescaled so that the average
frequency is unity. We then calculate the distribution of these normalized
frequencies and average the distribution over the 10 local minima. 
Finally, this averaged histogram is rescaled to compute the normalized
density of states ($G(\omega)$) for normalized frequencies. Figure 5.4
shows the result of such calculation for a series of potentials 
belonging to the Morse family and figure 5.5 displays the data with the
three highest values of $\alpha$ so as to show the pattern of
convergence more clearly. For the GLJ family, the route to 
$\phi \rightarrow \infty $ has been taken to be of two different types
characterized by the value of $m - n ( =\beta)$ for each potential. Here,
ofcourse $m$ keeps increasing but the value of $\beta$ is taken to be
either 2 or 4. The spectra for $\beta = 2$ are presented in figure 5.6
with figure 5.7 showing the cases corresponding to the three largest
values of $m$ -- the idea being again to show the pattern of convergence.
\begin{figure}[htp]
\vskip+0.5cm
\epsfxsize=6.5in
\centerline{\epsfbox{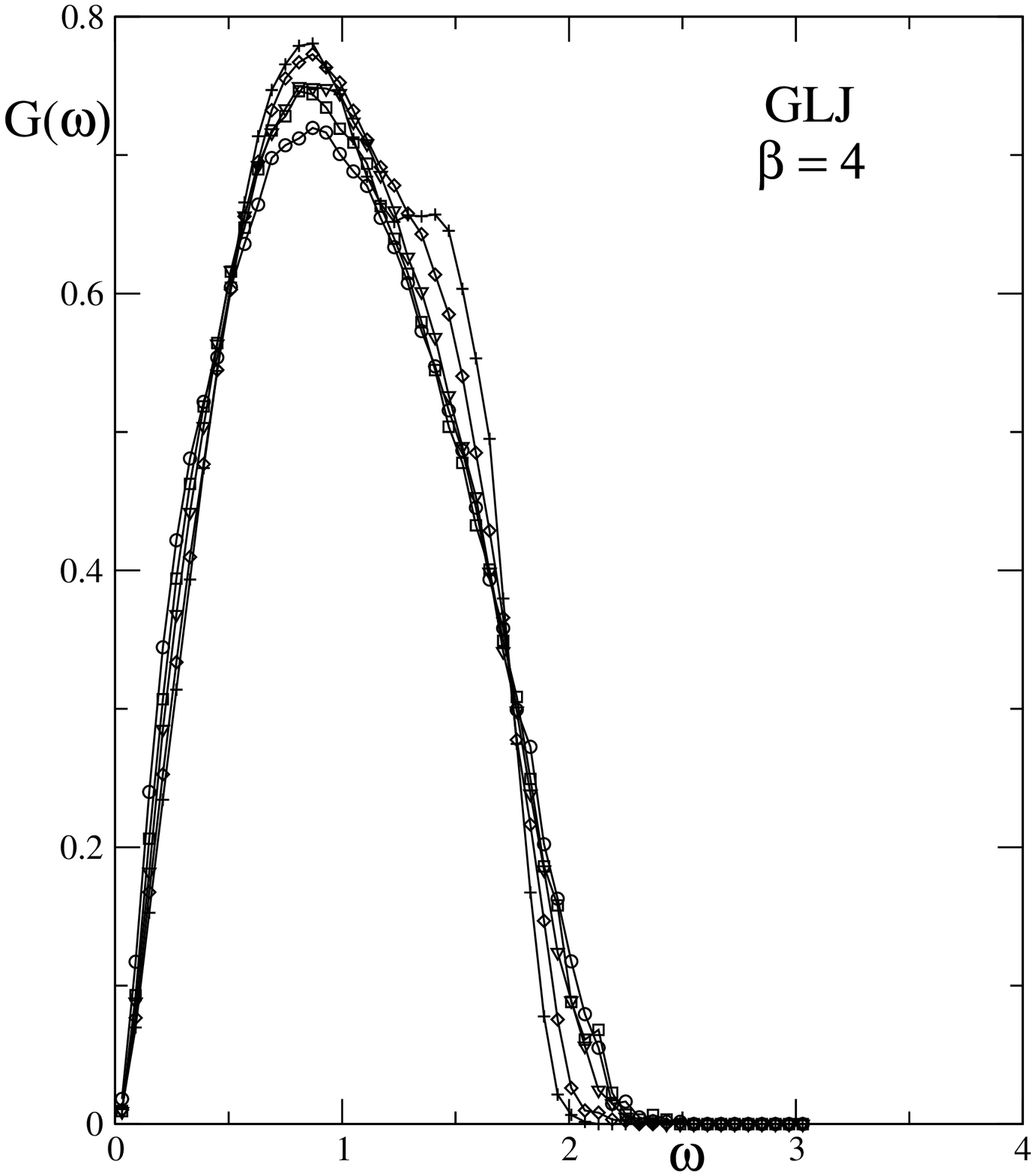}}
\caption{\sl Normalized density of states $\left(G\left(\omega\right)\right)$
vs. normalized frequency $\left(\omega\right)$ in the fully disordered region
for some selected cases of GLJ potential. The values of [m,n] are
[8,4] (plus), [10,6] (diamond), [14,10] (triangle), [18,14] (square) and
[22,18] (circle).}
\label{fig508}
\end{figure}

\begin{figure}[htp]
\vskip+0.5cm
\epsfxsize=6.5in
\centerline{\epsfbox{fig509.eps}}
\caption{\sl Normalized density of states $\left(G\left(\omega\right)\right)$
vs. normalized frequency $\left(\omega\right)$ in the fully disordered region
for some selected cases of GLJ potential. The values of [m,n] are
[18,14] (triangle), [20,16] (square) and [22,18] (circle).}
\label{fig509}
\end{figure}

\begin{figure}[htp]
\vskip+0.5cm
\epsfxsize=6.5in
\centerline{\epsfbox{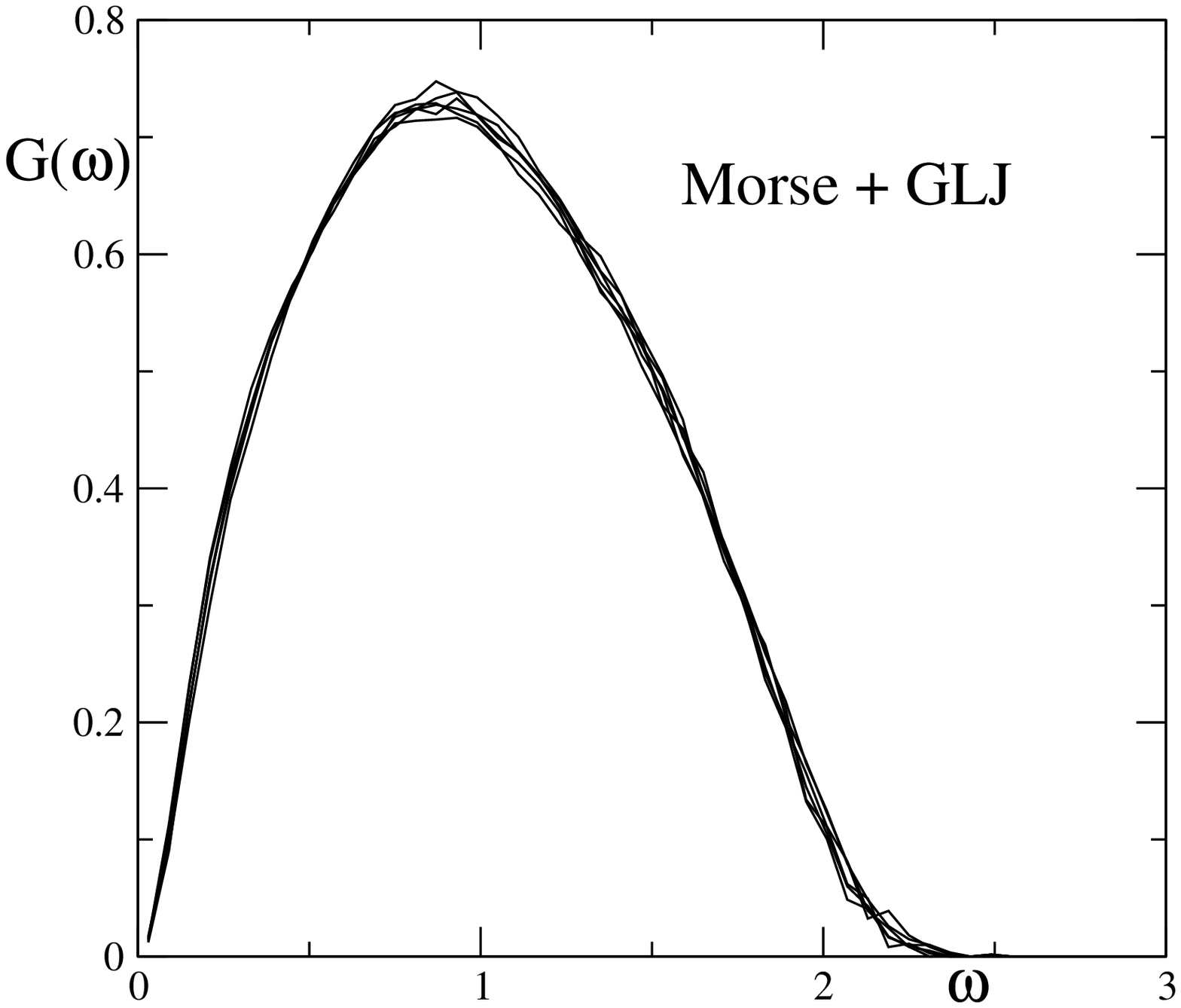}}
\caption{\sl Superposition of the data in figures 5.5 and 5.7 to compare
the convergence pattern of the GLJ and Morse families.}
\label{fig510}
\end{figure}
For the sake of completeness, we also present the data for $\beta = 4$
in figures 5.8 and 5.9. Please note that the highest value of $m$ that is
used is 22
for both $\beta = 2$ or 4. But since $\beta = 2$ corresponds to the
highest value of n, in figure 5.10 we superimpose the data for the three
highest values of $\alpha$ in the Morse family and the three highest
values of $m$ in the GLJ family with $\beta = 2$ to show the
relationship between the asymptotic normalized spectra for the two 
families. Within the error bars of our calculations, estimated by the
standard deviation of normalized density of states in each bin with
respect to the 10 local minima in each case, the asymptotic normalized
density of states function seem to be identical over the entire spectrum.
This is the most important result of this chapter. To restate it,
asymptotic $\left ( \phi \rightarrow \infty \right )$ density of states
is independent of the explicit form of the parametric potential function
which is used to achieve the asymptotic limit provided both the attractive
and repulsive sides of the minimum of the pair potential contribute to
the approach of $\phi$ to infinity.

If we assume the validity of the observation at the end of the previous
\begin{figure}[htp]
\vskip+0.5cm
\epsfxsize=4.0in
\centerline{\epsfbox{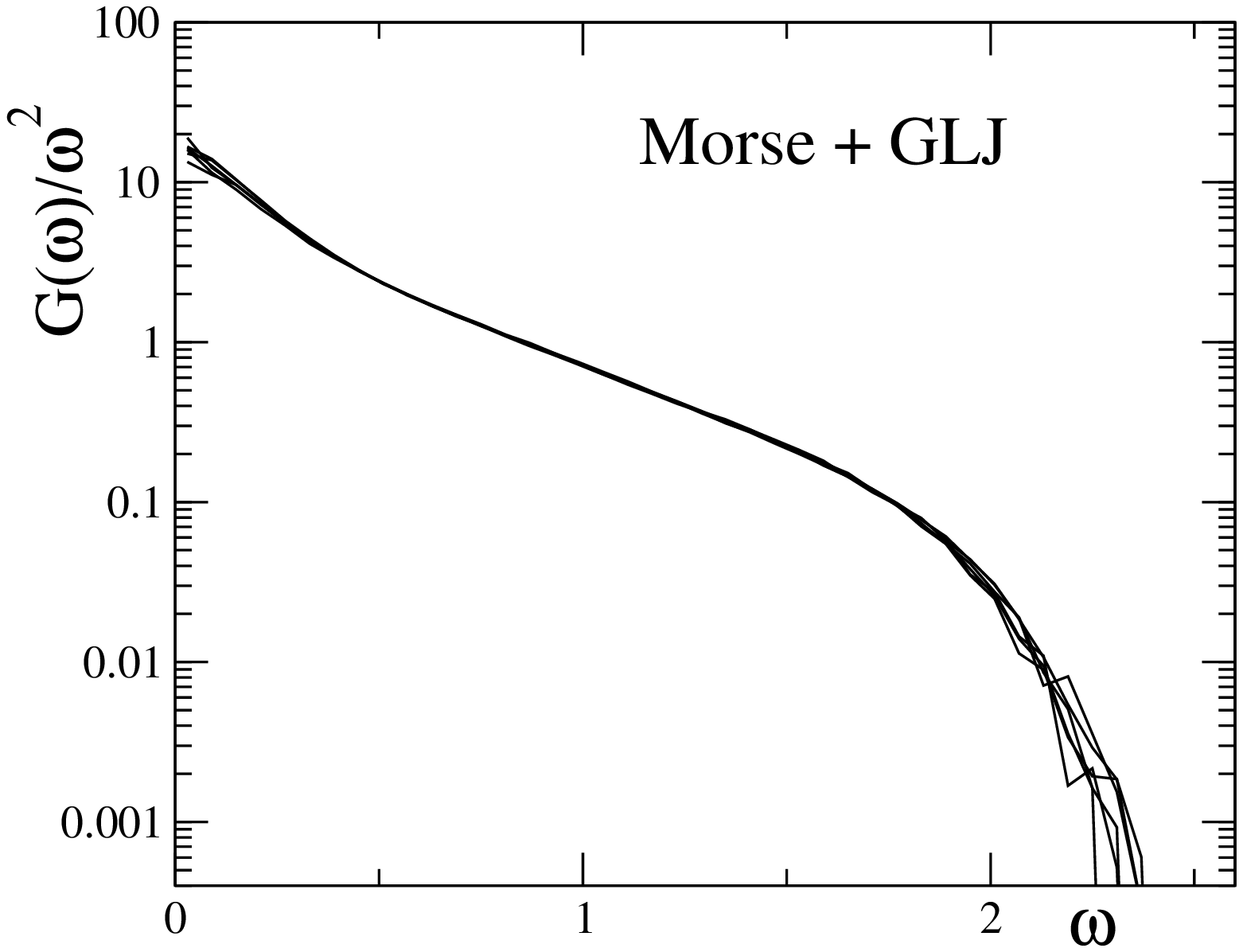}}
\caption{\sl Data on $G(\omega)$ shown in figure 5.10 is divided by
$\omega^2$ and shown in a semi-logarithmic plot for the purpose of
comparison with experimental data.}
\label{fig511}
\end{figure}
paragraph, an empirical explanation of the observation of universality 
in the trans-boson peak region of the vibrational spectra of several
molecular glasses that has been observed recently emerges. In these
experiments [65] it was observed that the semilogarithmic plot of
$g(\omega)/\omega^2$ as a function of $\omega$  is a straight line in
the trans-boson peak region for several different kinds of glasses.
This implies the following functional form of $g(\omega)$: $g(\omega)
= \alpha \omega^2\exp(-\omega/\omega_{0})$ where $\omega_{0}$ is the
scale of frequency specific to the material. 
Our empirical explanation
of this observation has to do with two facts: (1) The existence (or presumption)
of the asymptotic universal density of states and (2) Change in the vibrational
spectrum with the variation of parameter(s) becomes rather weak significantly
before asymptotic conditions are reached.
Thus, a large family of potentials will have vibrational spectra (in the
amorphous region) that is close to, but not quite, universal. Since
we do not know the exact nature of the potential function for the
specific materials, it is not possible to make a precise comparison with
our calculations. The next best course of action is to check whether our
asymptotic and (presumably) universal density of states has the
functional form seen in the experiments. In figure 5.11  we show a 
semi-logarithmic plot of the asymptotic normalized density of states
(divided by $\omega^2$) against $\omega$. There is a substantial linear 
segment in the middle in common with the experimental observations [65]. From
our discussion and also from the data presented for both the Morse
family and the GLJ family  for smaller values of $\phi$, it should
be clear that if the material at hand is described by a potential
which is far from satisfying the asymptotic conditions, the vibrational
spectrum will not be universal and can even show more than one peak.

The asymptotic universality which is suggested by our numerical results
poses a major theoretical challenge. At this moment, we have no 
understanding of this empirical observation. 


\newpage
{\Huge \bf Bibliography}
\begin{enumerate}

\item S. -K. Ma, {\it Modern Theory of Critical Phenomena} (Benjamin / Cummings,
Reading, 1976).

\item M. Goldstein, J. Chem. Phys. {\bf 51}, 3728 (1969).

\item T. B. Schr$\mbox\o$der, S. Sastry, J. C. Dyre, and S. Glotzer, J. Chem. Phys.
{\bf 112}, 9834 (2000).

\item A. Cavagna, Europhys. Lett. {\bf 53}, 490 (2001).

\item G. Adams and J. H. Gibbs, J. Chem. Phys. {\bf 43} , 139 (1965).

\item F. H. Stillinger and T. A. Weber, Phys. Rev. A {\bf 25} , 978 (1982).

\item F. H. Stillinger and T. A. Weber, Science {\bf 225} , 983 (1984).

\item F. H. Stillinger, Science {\bf 267} ,1935 (1995).

\item S. Sastry, P. G. Debenedetti, and F. H. Stillinger, Nature (London)
{\bf 393} , 554 (1998).

\item S. Sastry, Nature (London) {\bf 409} , 164 (2001).

\item C. A. Angell, J. Non-Cryst. Solids {\bf 131-133} ,13 (1991).

\item W. Kob and H. C. Andersen, Phys. Rev. Lett. {\bf 73} , 1376 (1994).

\item F. Sciortino, W. Kob, and P. Tartaglia, Phys. Rev. Lett. {\bf 83} ,3214
(1999).

\item K. K. Bhattacharya, K. Broderix, R. Kree, and A. Zippelius, Europhys.
Lett. {\bf 47} , 449 (1999).

\item S. B\"{u}chner and A. Heuer, Phys. Rev. Lett. {\bf 84} , 2168 (2000).

\item R. A. Denny, D. R. Reichman, and J. -P. Bouchaud, Phys. Rev. Lett. {\bf 90} ,025503
(2003).

\item W. Kob, F. Sciortino, and P. Tartaglia, Europhys. Lett. {\bf 49} ,590
(2000).

\item K. Broderix, K. K. Bhattacharya, A. Cavagna, A. Zippelius, and I. Giardina,
Phys. Rev. Lett. {\bf 85} , 5360 (2000).

\item L. Angelani, R. Di Leonardo, G. Ruocco, A. Scala, and F. Sciortino,
Phys. Rev. Lett. {\bf 85} , 5356 (2000).

\item P. Shah and C. Chakravarty, Phys. Rev. Lett. {\bf 88} , 255501 (2002).

\item Special issue of Physica (Amsterdam) {\bf 107D}, Issue 2-4 (1997).

\item {\it Proceedings of 7th International Workshop on Disordered Systems},
Molveno, Italy, edited by A. Fontana and G. Viliani [ Philos. Mag. {\bf 79}
(1999)].

\item R. J. Bell, Reports on Progress in Physics {\bf 35}, 1315 (1972).

\item S. R. Elliott, {\it Physics of Amorphous Materials} (Longmans, New York,1990),
2nd ed.

\item M. Sampoli, P. Benassi, R. Eramo, L. Angelani, and G. Ruocco, J. Phys. :
Cond. Matt. {\bf 15} , S1227 (2003).

\item G. Ruocco, F. Sette, R. Di Leonardo, G. Monaco, M. Sampoli, T. Scopigno,
and G. Viliani, Phys. Rev. Lett. {\bf 84}, 5788 (2000).

\item A. Cavagna, I. Giardina, and G. Parisi, Phys. Rev. Lett. {\bf 83},108
(1999).

\item T. S. Grigera, A. Cavagna, I. Giardina, and
G. Parisi, Phys. Rev. Lett. {\bf 88}, 055502 (2002).

\item E. La Nave, S. Mossa, and F. Sciortino, Phys. Rev. Lett. {\bf 88} , 225701
(2002).

\item S. Mossa, E. La Nave, H. E. Stanley, C. Donati, F. Sciortino, and
P. Tartaglia, Phys. Rev. E {\bf 65} , 041205 (2002).

\item S. Mossa, E. La Nave, P. Tartaglia, and F. Sciortino, J. Phys. : Cond.
Matt. {\bf 15} , S351 (2003).

\item P. Carpena and P. Bernaola-Galvan, Phys. Rev. B {\bf 60}, 201 (1999).

\item G. Parisi, cond-mat/0301282.

\item M. M\'{e}zard, G. Parisi, and A. Zee, cond-mat/9906135.

\item A. Fielicke, A. Kirilyuk, C. Ratsch, J. Behler, M. Scheffler, G. von Helden,
and G. Meijer, Phys. Rev. Lett. {\bf 93} , 023401 (2004).

\item L. V. Heimendahl and M. F. Thorpe, J. Phys. F: Metal Physics. {\bf 5},
L87 (1975).

\item A. Rahman, M. J. Mandell, and J. P. McTague, J. Chem. Phys.
{\bf 64}, 1564 (1976).

\item J. J. Rehr and R. Alben, Phys. Rev. B {\bf 16}, 2400 (1977).

\item J. -B. Suck, H. Rudin, H. -J. G\"{u}ntherodt, and H. Beck, J. Phys. C: Solid
State Physics, {\bf 14}, 2305 (1981).

\item S. R. Nagel, G. S. Grest, S. Feng and L. M. Schwartz, Phys. Rev. 
B {\bf 34}, 8667 (1986).

\item K. Vollmayr, W. Kob, and K. Binder, J. Chem. Phys.
{\bf 105} , 4714 (1996).

\item S. N. Taraskin and S. R. Elliott, Phil. Mag. B {\bf 79}, 1747 (1999).

\item A. F. Ioffe and A. R. Regel, Prog. Semicond. {\bf 4}, 237 (1960).

\item P. B. Allen, J. L. Feldman, J. Fabian, and F. Wooten, Phil. Mag. B {\bf 79},
1715 (1999).

\item {\it Amorphous Solids: Low Temperature Properties}, edited by W. A. Philips
(Springer-Verlag, Berlin, 1981).

\item U. Buchenau, N. N\"{u}cker and A. J. Dianoux, Phys. Rev. Lett. {\bf 53},
2316 (1984).

\item F. J. Bermejo, J. Alonso, A. Criado, F. J. Mompe\'{a}n, J. L. Mart\'{i}nez,
M. Garc\'{i}a-Hern\'{a}ndez, and A. Chahid, Phys. Rev. B {\bf 46} , 6173 (1992).

\item V. A. Luchnikov, N. N. Medvedev, Y. I. Naberukhin, and V. N. Novikov,
Phys. Rev. B {\bf 51} , 15569 (1995).

\item B. Fultz, C. C. Ahn, E. E. Alp, W. Sturhahn, and T. S. Toellner, Phys.
Rev. Lett. {\bf 79} , 937 (1997).

\item C. A. Tulk, D. D. Klug, E. C. Svensson, V. F. Sears, and
J. Katsaras, Appl. Phys. A [Suppl.] {\bf 74} , S1185 (2002).

\item M. A. Parshin, C. Laermans, D. A. Parshin, and V. G. Melehin,
Physica B {\bf 316-317} , 549 (2002).

\item M. A. Ramos, C. Tal\'{o}n, R. J. Jim\'{e}nez-Riob\'{o}o,
and S. Vieira, J. Phys.: Cond. Matt. {\bf 15} , S1007 (2003).

\item N. V. Surovtsev, S. V. Adichtchev, E. R\"{o}ssler,
and M. A. Ramos, J. Phys.: Cond. Matt. {\bf 16} , 223 (2004).

\item O. Pilla, S. Caponi, A. Fontana, J. R. Concalves, M. Montagna,
F. Rossi, G. Viliani, L. Angelani, G. Ruocco, G. Monaco, and F. Sette, J. Phys.:
Cond. Matt. {\bf 16}, 8519 (2004).

\item W. Schirmacher, G. Diezemann, and C. Ganter, Phys. Rev. Lett. {\bf 81}
, 136 (1998).

\item W. Schirmacher, G. Diezemann, and C. Ganter, Physica (Amsterdam)
{\bf 284-288B} ,1147 (2000).

\item V. L. Gurevich, D. A. Parshin, and H. R. Schober,
JETP Lett. {\bf 76}, 553 (2002).

\item V. L. Gurevich, D. A. Parshin, and H. R. Schober, Phys. Rev. B
{\bf 67}, 094203 (2003).

\item T. S. Grigera, V. Mart\'{i}n-Mayor, G. Parisi, and 
P. Verrocchio, Phys. Rev. Lett. {\bf 87} , 085502 (2001).

\item T. S. Grigera, V. Mart\'{i}n-Mayor, G. Parisi, and P. Verrochio,Nature(London) 
{\bf 422} , 289 (2003).

\item T. S. Grigera, V. Mart\'{i}n-Mayor, G. Parisi, and P. Verrochio,
J. Phys.: Cond. Matt. {\bf 14}, 2167 (2002).

\item S. N. Taraskin, Y. L. Loh, G. Natarajan, and S. R. Elliott, Phys. Rev. Lett.
{\bf 86} , 1255 (2001).

\item S. N. Taraskin and S. R. Elliott, Phys. Rev. B {\bf 56},
8605 (1997).

\item C. A. Angell, Y. Yue, L. -M. Yang, J. R. D. Copley, S. Borick, and
S. Mossa, J. Phys.: Cond. Matt. {\bf 15} , S1051 (2003).

\item A. I. Chumakov, I. Sergueev, U. van B\"{u}rck, W. Schirmacher, T. Asthalter,
R. R\"{u}ffer, O. Leupold and W. Petry, Phys. Rev. Lett. {\bf 92}, 245508 (2004).

\item S. Sastry, N. Deo, and S. Franz, Phys. Rev. E {\bf 64}, 016305 (2001).

\item S. K. Sarkar, G. S. Matharoo, and A. Pandey, Phys. Rev. Lett. {\bf 92}
, 215503 (2004).

\item G. S. Matharoo, S. K. Sarkar, and A. Pandey, Phys. Rev. B
{\bf 72}, 075401 (2005).

\item M. L. Mehta, {\it Random Matrices} (Academic Press, New York, 1991).

\item T. A. Brody, J. Flores, J. B. French, P. A. Mello, A. Pandey, and S. S. M. Wong,
Rev. Mod. Phys. {\bf 53} , 385 (1981).

\item O. Bohigas and M. J. Giannoni, in
{\it Mathematical and Computational Methods in Nuclear Physics, Proceedings
of the Sixth Granada Workshop, Granada, Spain, 1983}, Lecture Notes in Physics
 Vol. 209 (Springer, Berlin, 1984).

\item T. Guhr, A. M\"{u}ller-Groeling, and
H. A. Weidenm\"{u}ller, Phys. Rep. {\bf 299} , 189 (1998).

\item C. W. J. Beenakker, Rev. Mod. Phys. {\bf 69} , 731 (1997).

\item O. Bohigas, M. J. Giannoni, and C. Schmit, Phys. Rev. Lett. {\bf 52},
1 (1984).

\item T. H. Seligman, J. J. M. Verbaarschot, and M. R. Zirnbauer,
Phys. Rev. Lett. {\bf 53}, 215 (1984).

\item D. Wintgen and H. Marxer, Phys. Rev.
Lett. {\bf 60}, 971 (1988).

\item O. Bohigas, R. U. Haq, and A. Pandey, Phys. Rev. Lett. {\bf 54} , 1645
(1985).

\item G. Fagas, V. I. Fal'ko, and C. J. Lambert, Physica B {\bf 263-264} , 136
(1999).

\item F. Cleri and V. Rosato, Phys. Rev. B {\bf 48} , 22 (1993).

\item A. P. Sutton and J. Chen, Philos. Mag. Lett. {\bf 61} , 139 (1990).

\item R. P. Gupta, Phys. Rev. B {\bf 23} , 6265 (1981).

\item V. Rosato, M. Guillope and B. Legrand, Philos. Mag. A {\bf 59} , 321 (1989).

\item J. S. Hunjan, S. Sarkar and R. Ramaswamy, Phys. Rev. E {\bf 66}, 046704
(2002).

\item M Parrinello and A. Rahman, Phys. Rev. Lett. {\bf 45}, 1196 (1980).

\item D. C. Mattis, Phys. Lett. A {\bf 117} , 297 (1986).

\item D. C. Mattis, Phys. Lett. A {\bf 120} , 349 (1987).

\item M. I. Molina and D. C. Mattis, Phys. Lett. A {\bf 159} , 337 (1991).

\item M. Molina, Phys. Lett. A {\bf 161}, 145 (1991).

\item M. Molina, Ph.D. thesis, University of Utah (1991).

\item J. M. Y\'{a}\~{n}ez, M. I. Molina and D. C. Mattis, Phys. Lett. A {\bf 288},
277 (2001).

\end{enumerate}
\end{document}